\documentclass[twocolumn,showpacs,showkeys,preprintnumbers,amsmath,amssymb,prd,letterpaper,floatfix,superscriptaddress,10pt,aps,nofootinbib]{revtex4-1}
\usepackage[margin=1in,footskip=0.25in]{geometry}
\usepackage[utf8]{inputenc}
\usepackage{graphicx}
\usepackage{epsfig}
\usepackage{color}
\usepackage{longtable}
\usepackage[breaklinks=true]{hyperref}
\usepackage{latexsym}
\usepackage{amsmath}
\usepackage{amssymb}
\usepackage{dsfont}
\usepackage{verbatim}
\usepackage{bm}
\usepackage{bbm}
\usepackage{placeins}
\usepackage{adjustbox}
\usepackage{tabularx}
\usepackage[capitalize]{cleveref}
\usepackage{newunicodechar}
\usepackage{pdfpages}

\makeatletter
\AtBeginDocument{\let\LS@rot\@undefined}
\makeatother

\inputencoding{utf8}
\newunicodechar{γ}{$\gamma$}
\newunicodechar{ρ}{$\rho$}
\newunicodechar{π}{$\pi$}
\newunicodechar{Δ}{$\Delta$}
\newunicodechar{η}{$\eta$}
\newunicodechar{ψ}{$\psi$}

\hypersetup{colorlinks, linkcolor = [rgb]{0,0.0,0.75}, citecolor = [rgb]{0,0.0,0.75}, urlcolor = [rgb]{0,0.0,0.75}}

\newcommand{\I}{i}

\newcommand{\brackets}[1]{\langle #1 \rangle}

\newcommand{\qbar}{\bar{q}}

\newcommand{\pvec}{\vec{p}}
\newcommand{\xvec}{\vec{x}}

\newcommand{\pkpi}{\vec{P}}

\newcommand{\En}{E_n^{\vec{P},\,\Lambda}}
\newcommand{\sqrts}{\sqrt{s}_n^{\vec{P},\,\Lambda}}
\newcommand{\mpi}{m_{\pi}}
\newcommand{\mK}{m_{K}}
\newcommand{\ltwopi}{\frac{L}{2\pi}}
\newcommand{\twopiL}{\frac{2\pi}{L}}

\newcommand{\Pvec}{\vec{P}}

\newcommand{\Kmatrix}{$K$-matrix~}
\newcommand{\Tmatrix}{$T$-matrix~}

\def\MIT{Center for Theoretical Physics, Laboratory for Nuclear Science and Department of Physics, Massachusetts Institute of Technology, Cambridge, MA 02139, USA}
\def\RFWUB{Helmholtz-Institut f\"ur Strahlen- und Kernphysik, Rheinische Friedrich-Wilhelms-Universit\"at Bonn, Nu{\ss}allee 14-16, 53115 Bonn, Germany}

\def\RIKEN{RIKEN BNL Research Center, Brookhaven National Laboratory, Upton, NY 11973, USA}

\def\UOA{Department of Physics, University of Arizona, Tucson, AZ 85721, USA}
\def\SBU{Department of Physics and Astronomy, Stony Brook University, Stony Brook, NY 11794, USA}
\def\BNL{Department of Physics, Brookhaven National Laboratory, Upton, NY 11973, USA}
\def\JLAB{Thomas Jefferson National Accelerator Facility, Newport News, VA 23606, USA}
\def\JSC{Forschungszentrum J\"ulich GmbH, J\"ulich Supercomputing Centre, 52425 J\"ulich, Germany}
\def\ODU{Department of Physics, Old Dominion University, Norfolk, VA 23529, USA}
\def\MAINZ{Institut f\"ur Kernphysik, Johannes Gutenberg-Universit\"at Mainz, 55099 Mainz, Germany}

\begin{document}

\author{Gumaro Rendon}
  \email{jrendonsu@bnl.gov}
  \affiliation{\BNL}

\author{Luka Leskovec}
\affiliation{\JLAB}
\affiliation{{\ODU}}

\author{Stefan Meinel}
  \affiliation{\UOA}

\author{John Negele}
  \affiliation{\MIT}

\author{\mbox{Srijit Paul}}
\affiliation{\MAINZ}

\author{Marcus Petschlies}
  \affiliation{\RFWUB}

\author{Andrew Pochinsky}
  \affiliation{\MIT}

\author{Giorgio Silvi}
  \affiliation{\JSC}

\author{Sergey Syritsyn}
  \affiliation{\SBU}
  \affiliation{\RIKEN}


\date{\today}


\title{\texorpdfstring{$\bm{I=1/2}$ $\bm{S}$-wave and $\bm{P}$-wave $\bm{K\pi}$ scattering and the $\bm{\kappa}$ and $\bm{K^*}$ resonances \\ from lattice QCD}{}}


\vspace{0.2cm}

\begin{abstract}
  We present a lattice-QCD determination of the elastic isospin-$1/2$ $S$-wave and $P$-wave $K\pi$ scattering amplitudes as a function of the center-of-mass energy using L\"uscher's method. We perform global fits of \Kmatrix parametrizations to the finite-volume energy spectra for all irreducible representations with total momenta up to $\sqrt{3}\frac{2\pi}{L}$; this includes irreps that mix the $S$- and $P$-waves. Several different parametrizations for the energy dependence of the \Kmatrix are considered. We also determine the positions of the nearest poles in the scattering amplitudes, which correspond to the broad $\kappa$ resonance in the $S$-wave and the narrow $K^*(892)$ resonance in the $P$-wave. Our calculations are performed with $2+1$ dynamical clover fermions for two different pion masses of $317.2(2.2)$ and $175.9(1.8)$ MeV. Our preferred $S$-wave parametrization is based on a conformal map and includes an Adler zero; for the $P$-wave we use a standard pole parametrization including Blatt-Weisskopf barrier factors. The $S$-wave $\kappa$-resonance pole positions are found to be $\left[0.86(12) - 0.309(50)\,\I\right]\:{\rm GeV}$ at the heavier pion mass and $\left[0.499(55)- 0.379(66)\,\I\right]\:{\rm GeV}$ at the lighter pion mass. The $P$-wave $K^*$-resonance pole positions are found to be $\left[ 0.8951(64) - 0.00250(21)\,\I \right]\:{\rm GeV}$ at the heavier pion mass and $\left[0.8718(82) - 0.0130(11)\,\I\right]\:{\rm GeV}$ at the lighter pion mass, which corresponds to couplings of $g_{K^* K\pi}=5.02(26)$ and $g_{K^* K\pi}=4.99(22)$, respectively.
\end{abstract}

\maketitle

\section{Introduction}\label{sec_introduction}

As the simplest two-meson system with unequal mass and carrying strangeness, the $K\pi$ system plays an important role in particle and nuclear physics. A review of the early history of $K\pi$ scattering and the associated resonances can be found in Ref.~\cite{Lang:1978fk}. The $K\pi$ system also occurs in heavy-meson weak decay processes that are used to search for physics beyond the Standard Model \cite{Bevan:2014iga,Kou:2018nap,Bediaga:2018lhg,Asner:2008nq,Ablikim:2019hff}. This includes multibody nonleptonic decays such as $B\to K\pi\pi$, in which large $CP$-violating effects have been observed and two-body resonant sub-structures are seen \cite{Aaij:2014iva}, and semileptonic decays such as $B\to K\pi \ell^+ \ell^-$, which currently provides hints for new fundamental physics \cite{Aaij:2013iag,Aaij:2013qta,Descotes-Genon:2013wba,Horgan:2013pva,Aaij:2017vbb,Aebischer:2019mlg,Alguero:2019ptt,Aaij:2020nrf}.

$K\pi$ scattering has been studied in fixed-target scattering experiments with charged-kaon beams \cite{Estabrooks:1977xe, Aston:1987ir}, and, at low energies, through the formation and breakup of electromagnetically bound $K\pi$ atoms \cite{Adeva:2014xtx, Gorchakov:2016thm}. Further detailed investigations are planned using neutral kaon beams at the GlueX experiment \cite{Amaryan:2017ldw}.

The $I=1/2$ $S$-wave scattering amplitude is observed to be elastic up to approximately $1.3$ GeV \cite{Estabrooks:1977xe, Aston:1987ir}. The results for the elastic scattering phase shift slowly but monotonically increase and reach $60^{\circ}$ at approximately $1.1$ GeV. The rise in the phase shift is likely due to a very wide resonance, the $\kappa$ [also referred to as $K_0^*(800)$, or, more recently, $K_0^*(700)$]. However, since the phase shift does not cross $90^{\circ}$, the existence of the $\kappa$ remains a topic of discussion \cite{Cherry:2000ut}; the latest update of the Particle Data Group database \cite{Tanabashi:2018oca} still lists the $\kappa$ as ``requires confirmation''. Because the $\kappa$ is such a wide resonance
with total decay width $\Gamma_{\rm total}\approx 600$ MeV, its description is more involved. The search for a proper description and explanation for this elusive resonance is a hot topic in hadronic physics. The basic idea is to construct a parametrization of the scattering amplitude and fit it to the experimental data; by analytically continuing the amplitude into the complex plane, one searches for a pole attributed to the $\kappa$ resonance. The experimental data \cite{Estabrooks:1977xe,Aston:1987ir} have been described by effective lagrangians incorporating chiral symmetry \cite{Gasser:1984gg,Bernard:1990kw,Ishida:1997wn,Oller:1997ng,Oller:1998hw,Oller:1998zr,Black:1998zc,Roessl:1999iu,Jamin:2000wn,GomezNicola:2001as,Bijnens:2004bu,Nebreda:2010wv,Guo:2011pa,Magalhaes:2011sh,Wolkanowski:2015jtc} and models of meson interactions \cite{vanBeveren:1986ea,vanBeveren:2006ua}. The $\kappa$ was also studied in the large $N_c$ limit of QCD \cite{Pelaez:2004xp,Ledwig:2014cla} and with the inverse amplitude method \cite{Dobado:1992ha,Dobado:1996ps}. 
The authors of Ref.~\cite{Buettiker:2003pp, DescotesGenon:2006uk} used Roy-Steiner equations to determine the pole of the $\kappa$ resonance. Less rigorous, but similar studies using dispersion relations \cite{Zhou:2006wm,Pelaez:2020uiw} are able to consistently find the $\kappa$ pole. The relations between chiral perturbation theory and dispersion relations were explored in Ref.~\cite{Ananthanarayan:2000cp}. Fits of various models \cite{Pelaez:2016klv} and Pad\'e approximants \cite{Pelaez:2016tgi} to the experimental data led to similar results. Furthermore, it was observed that a $\kappa$ resonance is necessary to explain an enhancement \cite{Bugg:2005xx,Bugg:2005ni} in other production channels in the experiments \texttt{E791}~\cite{Aitala:2002kr} and BES~\cite{Bai:1994zm,Bai:2001dw}.

In the $P$-wave, the $I=1/2$ $K\pi$ scattering amplitude at energies below the $K\eta$ threshold is well described by a simple Breit-Wigner form with a single resonance, the $K^*(892)$\footnote{In the remainder of the text, we will denote the $K^*(892)$ in short as $K^*$.}, as observed in various processes ranging from kaon beam experiments to $\tau$ decays and $D$-meson decays \cite{Tanabashi:2018oca}. The total decay width of the $K^*(892)$ is approximately $50.8(0.9)$ MeV \cite{Tanabashi:2018oca} with branching ratios to $K\pi$ being $99.9\%$, to $K\gamma$ at the order of $10^{-3}$ and less than $10^{-5}$ to $K\pi\pi$.

In lattice QCD, scattering amplitudes can be determined from finite-volume energy spectra using L\"uscher's method \cite{Luscher:1990ux,Rummukainen:1995vs,Kim:2005gf,Davoudi:2011md,Fu:2011xz,Hansen:2012tf,Leskovec:2012gb,Gockeler:2012yj,Briceno:2014oea,Briceno:2017max}. For $S$-wave $K\pi$ scattering, the early lattice QCD calculations focused on scattering lengths describing low-energy scattering. The first such calculation, published in 2004, was performed for $I=3/2$ only and in the quenched approximation \cite{Miao:2004gy}. This was followed by a calculation in 2006 that included $N_f=2+1$ staggered sea quarks but employed a domain-wall valence action \cite{Beane:2006gj}; the authors determined the $I=3/2$ $S$-wave scattering length directly from the lattice and used chiral symmetry to extract also the $I=1/2$ scattering length at several pion masses. The $S$-wave scattering lengths have also been determined from scalar form factors for semileptonic decays \cite{Flynn:2007ki}. Reference \cite{Nagata:2008wk} contains the first direct lattice QCD calculation of the $S$-wave scattering length for both $I=1/2$ and $I=3/2$, albeit in the quenched approximation. The $K\pi$ system was also investigated using a staggered action for both the valence and sea quarks in Refs.~\cite{Fu:2011xb,Fu:2011xw,Fu:2011wc,Fu:2013sua}. Note that the presence of extra non-degenerate fermion tastes when using a staggered action
introduces complications for the L\"uscher method that are not yet fully understood. More recent dynamic lattice QCD calculations of $K\pi$ $S$-wave the scattering lengths employed valence Wilson fermions with either $N_f=2$ \cite{Lang:2012sv}, $N_f=2+1$ \cite{Sasaki:2013vxa} or $N_f=2+1+1$~\cite{Helmes:2018nug} dynamical Wilson fermions.

Early attempts to investigate the $\kappa$ resonance on the lattice focused on the energy spectrum and involved searching for additional energy levels that could be associated with the resonance. Finite-volume energies were investigated for the $\kappa$ system in Refs.~\cite{Alexandrou:2012rm} and Ref.~\cite{Prelovsek:2010kg}. In the latter reference, the quark-disconnected contributions were neglected. The authors later found that this leads to the wrong spectrum \cite{Lang:2012sv}, as also discussed in Ref.~\cite{Guo:2013nja} from a perturbative point of view. In the early 2010's it became clear that the $\kappa$ does not behave like the typical resonance on the lattice. Using unitarized chiral perturbation theory models, Refs.~\cite{Doring:2011nd,Bernard:2010fp,Doring:2012eu,Doring:2011nd,Xiao:2012vv,Doring:2013wka,Zhou:2014ana} determined the finite-volume energies and investigated what can be expected in lattice QCD calculations.

To date, there have been few fully-fledged determinations of the energy dependence of $K\pi$ scattering amplitudes with dynamical lattice QCD. The first such studies focused on the $P$-wave in the $K^*$ resonance region. In Ref.~\cite{Fu:2012tj}, $N_f=2+1$ staggered quarks were used to determine the $K^*$ phase shift from the rest frame spectra. A similar calculation with $N_f=2$ Wilson quarks included also moving frames \cite{Prelovsek:2013ela} and determined the scattering phase shift and the $K^*$ width. The authors of Ref.~\cite{Bali:2015gji} repeated the calculation for the $\rho$ and the $K^*$ at an almost physical pion mass on two large $N_f=2$ gauge ensembles. A more comprehensive study was published in Refs.~\cite{Wilson:2014cna,Dudek:2014qha}, where the authors calculated the scattering amplitudes in $S$-, $P$- and $D$-waves with $I=1/2$ and $I=3/2$ and determined their resonance content. They employed anisotropic gauge ensembles with $N_f=2+1$ Wilson fermions, similarly to Ref.~\cite{Brett:2018jqw}. Recently, Ref.~\cite{Wilson:2019wfr} reported a calculation of $I=1/2$ $S$- and $P$-wave scattering amplitudes at an unprecedented number of quark masses.

In the following, we present a new detailed analysis of $I=1/2$ $K\pi$ scattering using lattice QCD. This work provides further information on the interactions and resonances in this system, and also constitutes our first step toward a future lattice calculation of semileptonic form factors with $K\pi$ final states based on the formalism developed in Ref.~\cite{Briceno:2014uqa}. We simultaneously determine the energy dependence of both the $S$-wave and $P$-wave scattering amplitudes below the $K\eta$ threshold, and investigate several different parametrizations with and without an Adler zero for the $S$-wave amplitude. We determine the pole locations corresponding to the $\kappa$ and $K^*$ resonances, and also present the $K^*K\pi$ couplings. Our calculation is performed on two different gauge-field ensembles with $N_f=2+1$ dynamical clover fermions; the first has a lattice size of $32^3 \times 96$ with a spacing of $a\approx 0.114$ fm and a pion mass of $m_\pi\approx 317$ MeV, while the second has a lattice size of $48^3 \times 96$ with $a\approx 0.088$ fm and $m_\pi\approx 176$ MeV.

The paper is organized as follows: we begin by overviewing the continuum description of elastic $K\pi$ scattering in Sec.~\ref{sec_about_Kpi}. In Sec.~\ref{sec_gauge_ensembles}, we briefly describe the lattice gauge field ensembles, while Sec.~\ref{sec_corr_mat} gives details on the construction of the interpolating operators and correlation matrices. Our spectrum results are shown in Sec.~\ref{sec_spectrum}, and the finite-volume methods for the determination of the scattering amplitudes are discussed in Sec.~\ref{sec_luscher}. We present our results for the energy dependence of the phase shifts and the pole locations in Sec.~\ref{sec_phases}. Our conclusions, including a comparison with previous work, are given in Sec.~\ref{sec_summary}.

\section{\texorpdfstring{Parametrizations of the scattering amplitudes}{}}\label{sec_about_Kpi}

In this section we briefly review the $K$-matrix formalism describing $2\to2$ scattering \cite{Chung:1995dx}, and then discuss the specific parametrizations we use
to describe $K\pi$ scattering with $I(J^{P}) = 1/2(0^+)$ and $I(J^{P}) = 1/2(1^-)$. In general, the multi-channel scattering matrix can be expressed as
\begin{align}
\label{eq:TfromS}
S^{(\ell)}(s) = 1 + 2 \I \;T^{(\ell)}(s),
\end{align}
where $T^{(\ell)}$ is the $T$-matrix (also known as the scattering amplitude), which depends
on the invariant mass $s$ of the system, and $\ell$ is the partial wave of the scattering process.
From the unitarity of $S^{(\ell)}$ one gets
\begin{align}
\frac{1}{2\I}[T^{(\ell)}_{ij}-T^{(\ell)*}_{ji}] &= \operatorname{Im}{T^{(\ell)}_{ij}} \cr
&= \sum_k T^{(\ell)*}_{ik} \theta(s-s^{(k)}_{\text{thr}}) T^{(\ell)}_{kj},
\end{align}
where we used that due to time-reversal invariance of the strong interaction the \Tmatrix is symmetric.
Here, the indices $i$, $j$, ... label the scattering channels and $s_{thr}^{(i)}$ denotes the threshold in channel $i$. Equivalently,
\begin{align}
\operatorname{Im}{\{\,{T^{(\ell)}}^{-1}\}_{ij}}= - \theta(s-s^{(i)}_{\text{thr}})\delta_{ij}.
\end{align}
That means that we can split the real and imaginary contributions to 
${T^{(\ell)}}^{-1}$ 
in the following way:
\begin{align}
\{\,{T^{(\ell)}}^{-1}\}_{ij} = \{\,{K^{(\ell)}}^{-1}\}_{ij} - \I\, \theta(s-s^{(i)}_{\text{thr}})\delta_{ij},
\end{align}
where $K^{(\ell)}$ has to be real and symmetric to ensure unitarity and time-reversal invariance.

In order to incorporate the correct analytic structure from the $K\pi$ threshold, we define the phase-space factor $\rho$, which is a diagonal matrix in channel space:
\begin{align}
\rho_{ii} = \sqrt{\left(1-\left(\frac{m_a^i + m_b^i}{\sqrt{s}}\right)^2\right)\left(1-\left(\frac{m_a^i - m_b^i}{\sqrt{s}}\right)^2\right)}.
\end{align}
Above, $a$ and $b$ label the two mesons undergoing elastic scattering in channel $i$. 
For example, at scattering energies above the $K\eta$ threshold we have two scattering channels, $i=0,1$, 
which correspond to the scattering of $K\pi$ and $K\eta$, respectively.
However, the $K\eta$ channel is not relevant at the energies we consider here,
and our spectra can be described by purely elastic $K\pi$ scattering ($i=0$ only).
We omit the channel indices in the remainder of the paper.

Using the phase-space factor, we define the rescaled \Kmatrix $\hat{K}^{(\ell)}$ through
\begin{align}
K^{(\ell)} = \rho^{1/2} \hat{K}^{(\ell)} \rho^{1/2}.
\end{align}
The elastic scattering phase shift $\delta_\ell$ is related to the scattering amplitude as
\begin{align}
T^{(\ell)} &= e^{\I \delta_{\ell}} \sin(\delta_{\ell}) 
= \frac{1}{\cot{\delta_{\ell}}-\I}\,,
\label{eq:tlofs}
\end{align}
and to the \Kmatrix as
\begin{align}
  K^{(\ell)}=\tan(\delta_{\ell}) \quad \mathrm{~and~} \quad 
\hat{K}^{(\ell)}=\frac{1}{\rho}\tan(\delta_{\ell})\,.
\end{align}
We now proceed to the discussion of the parametrizations we use for the $s$ dependence. For the $P$-wave, which is governed by the narrow $K^*$ resonance, we find
that a simple one-pole $K$-matrix parametrization with Blatt-Weisskopf barrier factors describes the data well. For the $S$-wave, we also consider three
additional parametrizations: the effective range expansion \cite{Landau:1991wop,Krane:1987ky}, Bugg's parametrization \cite{Bugg:2003kj,Bugg:2009uk} that accounts for a zero in the scattering amplitude known as the Adler zero \cite{Adler:1965ga,Adler:1964um,Bessler:1974cb}, and the conformal-map-based parametrization used in Ref.~\cite{Pelaez:2016tgi}, which also has an Adler zero.

\subsection{Chung's parametrization}

Chung's parametrization is a raw \Kmatrix pole parametrization \cite{Chung:1995dx} combined with Blatt-Weisskopf barrier factors \cite{VonHippel:1972fg}. The latter describe a centrifugal barrier effect in the $P$-wave but are trivial for $S$-wave scattering. For both the $S$- and $P$-waves the \Kmatrix pole parametrization is
\begin{align}
\label{eq:ChungBad}
\hat{K}^{(\ell)} = \sum_{\alpha } \frac{g_{\ell,\alpha}(\sqrt{s})g_{\ell,\alpha}(\sqrt{s})}{(m_{\ell,\alpha }^2 - s)\rho},
\end{align}
where $\alpha$ labels the resonances present and
\begin{align}
g_{\ell,\alpha}(\sqrt{s}) = \sqrt{ m_{\ell,\alpha } \Gamma_{\ell,\alpha}(\sqrt{s}) }
\end{align}
with
\begin{align}\label{eq:gamma_Chung}
\Gamma_{\ell,\alpha}(\sqrt{s}) = \gamma_{\ell,\alpha}^2 \:\rho \left[ B^{\ell}_{\alpha}(k,k_{\alpha}) \right]^2 \Gamma_{\ell,\alpha }^0.
\end{align}
Here, the $\gamma_{\ell,\alpha}$ are the resonance couplings, $B^{\ell}_{\alpha}(k ,k_{\alpha})$ are the Blatt-Weisskopf barrier factors (defined further below),
and the parameters $\Gamma_{\ell,\alpha }^0$ are related to the widths of the resonances. Since only the product $\gamma_{\ell,\alpha}^2\Gamma_{\ell,\alpha }^0$ appears in $\Gamma_{\ell,\alpha}(\sqrt{s})$, we perform our fits in terms of new parameters $g^0_{\ell,\alpha}$ defined as
\begin{equation}
\label{eq:g0}
g^0_{\ell,\alpha} = \gamma_{\ell,\alpha} \sqrt{m_{\ell,\alpha} \Gamma^0_{\ell,\alpha}}.
\end{equation}
Inserting Eq.~\eqref{eq:g0} into Eq.~\eqref{eq:ChungBad} gives the final form
\begin{equation}
\label{eq:Chung}
\hat{K}^{(\ell)}= \sum_{\alpha} \frac{g^0_{\ell,\alpha}g^0_{\ell,\alpha}B^{\ell}_{\alpha}(k ,k_{\alpha})B^{\ell}_{\alpha}(k ,k_{\alpha})}{(m_{\ell,\alpha}^2 - s)}.
\end{equation}
The Blatt-Weisskopf factors are functions of $k$, the scattering momentum at the given $s$, and $k_\alpha$, the scattering momentum at $s=m_\alpha^2$.
The scattering momentum $k$ is defined via
\begin{equation}
\sqrt{s} = \sqrt{\mpi^2 + k^2} + \sqrt{\mK^2 + k^2},
\end{equation}
which gives
\begin{equation}
k^2 = \frac{s^2 + (\mpi^2 - \mK^2)^2 - 2 s (\mpi^2 + \mK^2) }{4 s}.
\end{equation}
The Blatt-Weisskopf factors are equal to $B^{\ell}_{\alpha}(k ,k_{\alpha}) = F_{\ell}(k)/F_{\ell}(k_{\alpha })$, where
\begin{align}
F_0(k) &= 1, \\
F_1(k) &= \sqrt{\frac{2(k\,R_{1,\alpha})^2}{1+(k\,R_{1,\alpha})^2}},
\end{align}
with $R_{1,\alpha}$ the characteristic range for $\ell=1$, which we also take to be a fit parameter. Since we include only one resonance in each partial wave, we omit the index $\alpha$ in the following and denote the fit parameters as
\begin{equation}
m_\ell,\:\: g^0_\ell\:\: (\text{for }\ell=0,1),\:\: R_1. 
\end{equation}

\subsection{Effective-range expansion}

The effective range expansion (ERE) to order $k^2$ for $\ell=0$ is given by \cite{Landau:1991wop,Krane:1987ky}
\begin{equation}\label{eq:ERE}
\hat{K}^{(\ell=0)^{-1}} = \frac{\rho}{k}\left(\frac{1}{a}+\frac{1}{2}r_0 k^2\right),
\end{equation}
where $a$ is the zero-energy scattering length and $r_0$ represents the effective range of the interaction. The actual fit parameters we use are
\begin{align}
c_{0}=\frac{1}{a}, \qquad c_{1}=\frac{1}{2}r_0.
\end{align}

\subsection{Bugg's parametrization}

The author of Ref.~\cite{Bugg:2009uk} performed a fit to FOCUS and E791 data for the $D\to K \pi \pi$ decay \cite{Link:2009ng,Aitala:2005yh}, using a modified version of the parametrization of Ref.~\cite{Jamin:2000wn} to accommodate the broad nature of the $\kappa$ resonance. This is a \Kmatrix pole parametrization where the width is taken to have a zero of the form $s-s_{A}$, intended to account for the prediction from chiral perturbation theory that the \Tmatrix has a zero at $s=s_A$ (the Adler zero), where \cite{Pelaez:2016tgi}
\begin{equation}
 s_A=\frac15\left(m_K^2 + m_\pi^2 + 2\sqrt{4(m_K^2-m_\pi^2)^2+m_K^2 m_\pi^2}\right).
\end{equation}
The parametrization enables the scattering amplitude to reproduce the experimental phase shift near the threshold much better. A model explanation for this form of the amplitude is discussed in Ref.~\cite{vanBeveren:2006ua}. We implement this parametrization by multiplying the \Kmatrix pole with an enveloping term of the form $s-s_{A}$ so that the \Tmatrix also becomes zero as predicted:
\begin{align}\label{eq:Bugg}
\hat{K}^{(\ell=0)}&=\frac{[G_0(s)]^2}{m_0^2-s}, \\
G_0(s)&=G_0^{0} \sqrt{\frac{s-s_{A}}{s_A-m_0^2}}.
\end{align}

\subsection{Conformal map parametrization\label{subsec_conformal}}

The final parametrization we consider for the $S$-wave is that of Ref.~\cite{Pelaez:2016tgi}, which involves a power series
in a new variable $\omega(s)$. The regions of analyticity in the complex-$s$ plane are conformally mapped to the interior of the unit disk,
while the elastic, inelastic, left-hand, and circular cuts are mapped to the circle $|\omega|=1$.
The parametrization also includes an Adler zero and is given by
\begin{equation}
{K^{(\ell=0)}}^{-1}=\frac{\sqrt{s}}{2k}F(s)\sum_n B_n\, \omega^n(s)
\label{eq:generalconformal}
\end{equation}
with
\begin{equation}
F(s)=\frac{1}{s-s_{A}}.
\end{equation}
The conformal map is defined as
\begin{equation}
\omega(y)=\frac{\sqrt{y}-\alpha \sqrt{y_0-y}}{\sqrt{y}+\alpha \sqrt{y_0-y}}, \quad y(s)=\left(\frac{s-\Delta_{K\pi}}{s+\Delta_{K\pi}}\right)^2,
\label{eq:conformalvars}
\end{equation}
where $\Delta_{K\pi}=m^2_{K}-m^2_{\pi}$ and $y_0\equiv y(s_0)$. The constant $s_0$ determines the maximum value of $s$ for which the map is applicable, while the constant $\alpha$ determines the origin of the expansion. We set $\sqrt{s_0}$ equal to the $K\eta$ threshold, using leading-order chiral perturbation theory to express $m_\eta$ in terms of $m_K$ and $m_\pi$:
\begin{equation}\label{eq:GMO}
 \sqrt{s_0} = m_K + \sqrt{\frac{4 m_K^2 - m_\pi^2}{3}}.
\end{equation}
We choose $\alpha=1.3$ for both ensembles so that the origin of the expansion is around the middle of the data points. We found it sufficient to expand to first order in $\omega$.

\section{Gauge Ensembles and Single-Meson Energies}\label{sec_gauge_ensembles}

In this work, we utilize two different gauge-field ensembles, labeled \texttt{C13} and \texttt{D6}, with parameters given in \cref{tab:kpi_lattices}.
These ensembles use the tadpole-improved tree-level Symanzik gluon action \cite{Symanzik:1983pq, Symanzik:1983dc, Symanzik:1983gh, Luscher:1985zq} and include $2+1$ flavors of clover fermions \cite{Wilson:1974sk, Sheikholeslami:1985ij}. The gauge links in the fermion action are Stout smeared \cite{Morningstar:2003gk} with
a staple weight of $\rho=0.125$. We use the same clover action also for the valence quarks. The lattice spacings, $a$, were determined using the $\Upsilon(2S)-\Upsilon(1S)$ splitting~\cite{Davies:2009tsa,Meinel:2010pv} computed with improved lattice NRQCD \cite{Lepage:1992tx}.

\begin{table}[b]
\centering
\begin{tabular}{|c|c|c|}
\hline
 & \texttt{C13} & \texttt{D6} \cr
\hline
$N_s^3\times N_t$ & $32^3\times 96$ & $48^3\times 96$ \cr
$\beta$           & $6.1$           & $6.3$ \cr
$a m_{u,d}$       & $-0.285$        & $-0.2416$  \cr
$a m_{s}$         & $-0.245$        & $-0.205$  \cr
$c_{\rm SW}$      & $1.2493$        & $1.2054$  \cr
$a$ [fm] & $0.11403(77)$ & $0.08766(79)$ \cr
$L$ [fm] & $3.649(25)$ & $4.208(38)$ \cr
$a m_\pi$ & $0.18332(29)$ & $0.07816(35)$ \cr
$a m_K$   & $0.30475(17)$ & $0.22803(15)$ \cr
$m_\pi$ [MeV] & $317.2(2.2)$ & $175.9(1.8)$ \cr
$m_K$ [MeV] & $527.4(3.6)$ & $513.3(4.6)$ \cr
$N_{\text{config}}$ & $896$ & $328$ \cr
\hline
\end{tabular}
\caption{Parameters of the gauge-field ensembles.} \label{tab:kpi_lattices}
\end{table}

\begin{figure}[t]
  \includegraphics[width=\columnwidth]{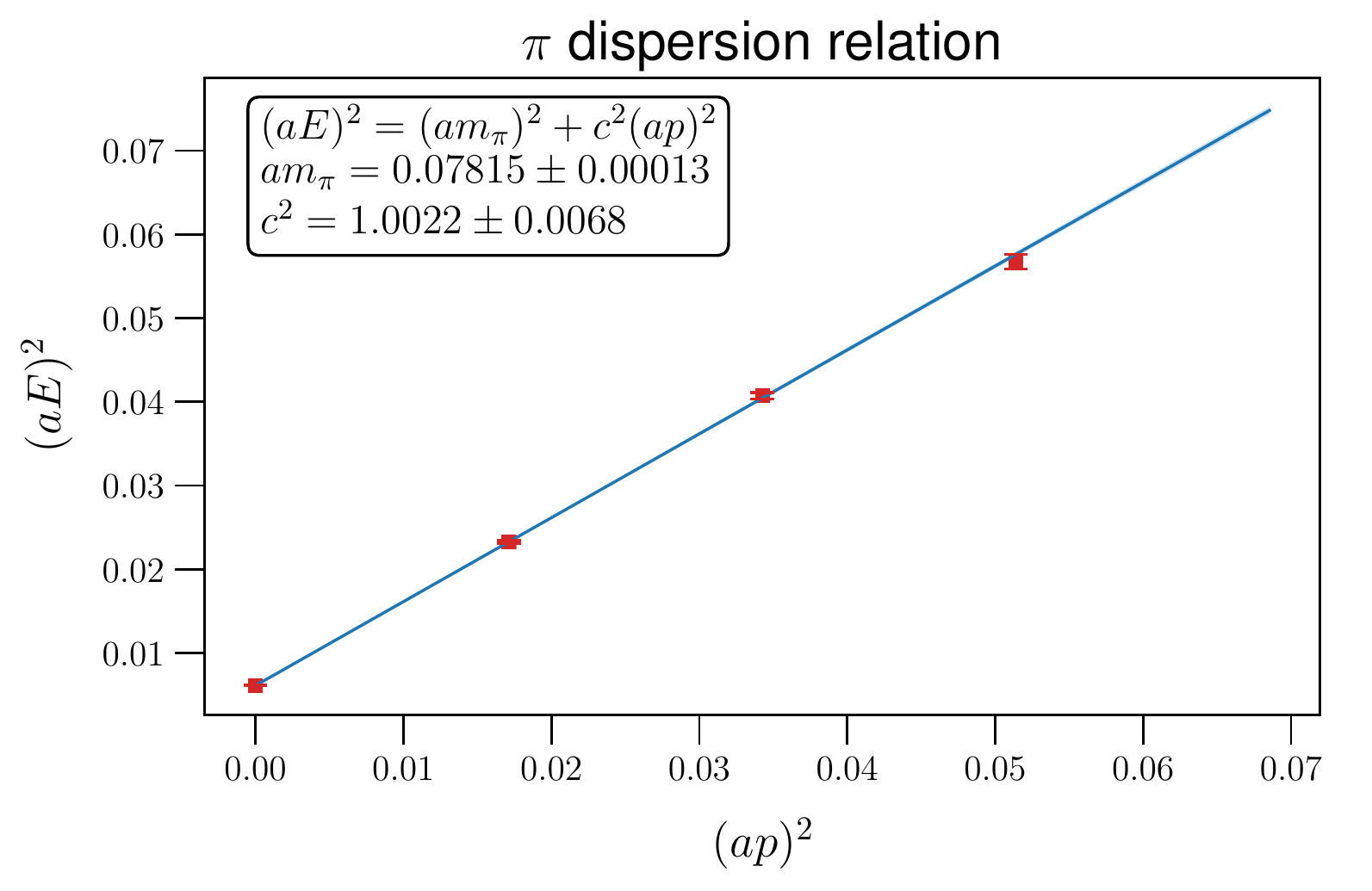}
  \caption{Pion dispersion relation for \texttt{D6} ensemble. The mass of the $\pi$ and the speed of light determined from the multiple-momenta simultaneous fit matches the relativistic dispersion relation with the rest frame $\pi$ mass fit.\label{fig:dispersion_rel_pi}}
\end{figure}

\begin{figure}[t]
  \includegraphics[width=\columnwidth]{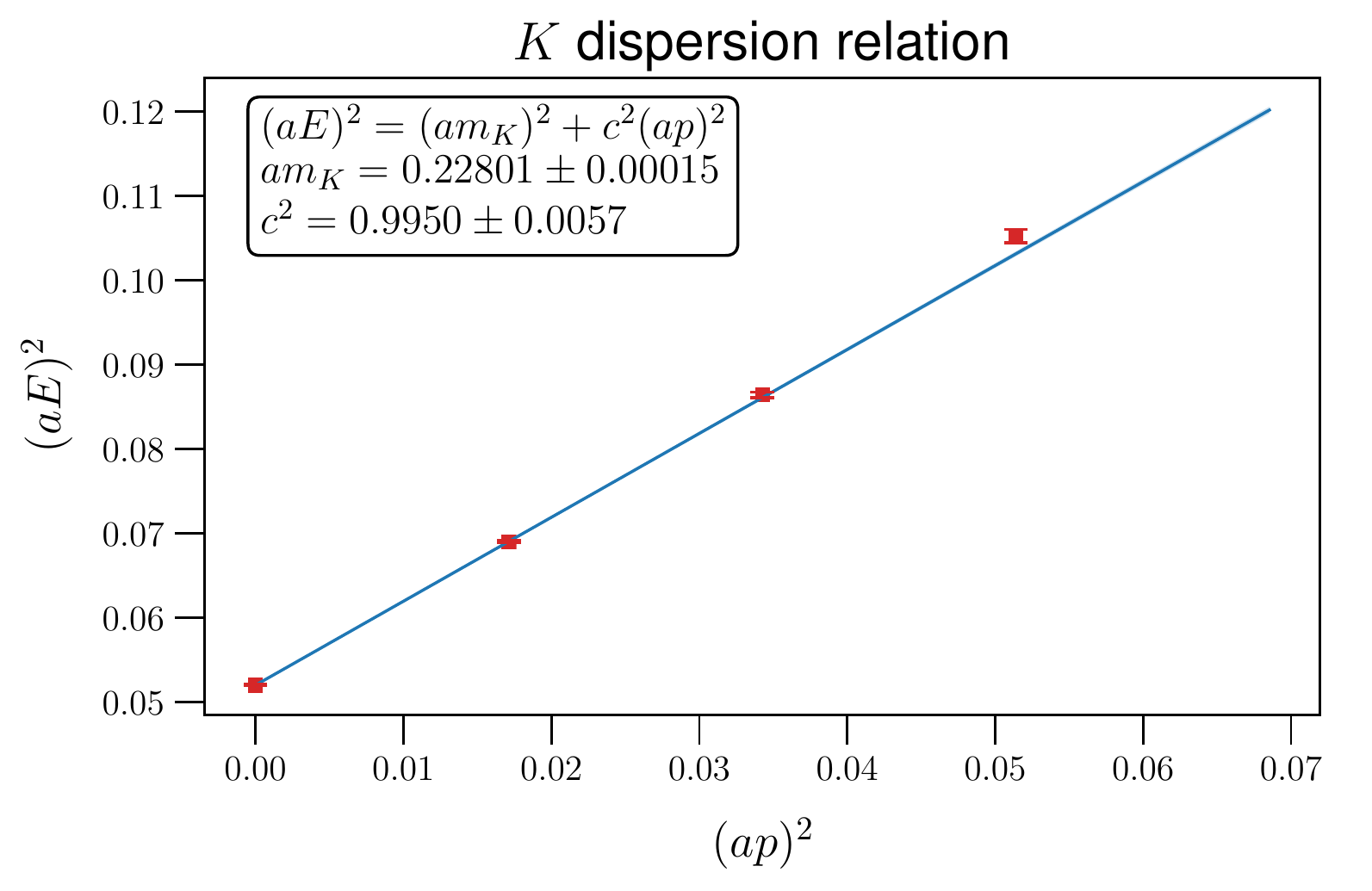}
  \caption{Like Fig.~\ref{fig:dispersion_rel_pi}, but for the kaon.\label{fig:dispersion_rel_K}}
\end{figure}

When using L\"uscher's method, the pion and kaon dispersion relations are needed to relate energies to scattering momenta.
To study the dispersion relations on the lattice, we computed pion and kaon two-point correlation functions projected to different momenta.
Fits of the dispersion relations on the \texttt{D6} ensemble are shown in Figs.~\ref{fig:dispersion_rel_pi} and~\ref{fig:dispersion_rel_K}.
We find that the data with $|\vec{p}|\leq \sqrt{3}\cdot 2\pi/L$ are consistent with the relativistic continuum dispersion relation on both ensembles,
and we therefore use this form in the further analysis.

%

\section{Interpolating operators and correlation matrix construction}\label{sec_corr_mat}

\begin{table*}
\centering
\begin{tabular}{ |c|c|c|c|c|c|  }
 \hline
 $\ltwopi\vec{P}$& Little Group $LG$ & Irrep $\Lambda$ & Ang.~mom.~content & Operator number & Operator structure \\
\hline
(0,0,0)   & $O_h$       & $A_{1g}$     &   $J=0,4,...$   & 1 & $K_0^+$  \\
          &             &              &                 & 2 & $K\pi$ with $|\vec{p}_1|=|\vec{p}_2|=0$ \\
          &             &              &                 & 3 & $K\pi$ with $|\vec{p}_1|=|\vec{p}_2|=\frac{2\pi}{L}$ \\
          &             &              &                 & 4 & $K\pi$ with $|\vec{p}_1|=|\vec{p}_2|=\sqrt{2}\frac{2\pi}{L}$ \\
          &             &              &                 & 5 & $K\pi$ with $|\vec{p}_1|=|\vec{p}_2|=\sqrt{3}\frac{2\pi}{L}$ \\          
\hline          
(0,0,0)   & $O_h$       & $T_{1u}$     &   $J=1,3,...$   & 1 & $K^{*+}_{i}$ \\
          &             &              &                 & 2 & $K^{*+}_{ti}$ \\
          &             &              &                 & 3 & $K\pi$ with $|\vec{p}_1|=|\vec{p}_2|=\frac{2\pi}{L}$ \\
          &             &              &                 & 4 & $K\pi$ with $|\vec{p}_1|=|\vec{p}_2|=\sqrt{2}\frac{2\pi}{L}$ \\
          &             &              &                 & 5 & $K\pi$ with $|\vec{p}_1|=|\vec{p}_2|=\sqrt{3}\frac{2\pi}{L}$ \\
\hline          
(0,0,1)   & $C_{4v}$    & $A_1$        &   $J=0,1,...$   & 1 & $K^{*+}_{i}$ \\
          &             &              &                 & 2 & $K^{*+}_{ti}$ \\
          &             &              &                 & 3 & $K_0^+$  \\
          &             &              &                 & 4 & $K\pi$ with $|\vec{p}_1|=0$ and $|\vec{p}_2|=\frac{2\pi}{L}$ \\
          &             &              &                 & 5 & $K\pi$ with $|\vec{p}_1|=\frac{2\pi}{L}$ and $|\vec{p}_2|=0$ \\
          &             &              &                 & 6 & $K\pi$ with $|\vec{p}_1|=\frac{2\pi}{L}$ and $|\vec{p}_2|=\sqrt{2}\frac{2\pi}{L}$ \\
          &             &              &                 & 7 & $K\pi$ with $|\vec{p}_1|=\sqrt{2}\frac{2\pi}{L}$ and $|\vec{p}_2|=\frac{2\pi}{L}$ \\
          &             &              &                 & 8 & $K\pi$ with $|\vec{p}_1|=\sqrt{2}\frac{2\pi}{L}$ and $|\vec{p}_2|=\sqrt{3}\frac{2\pi}{L}$ \\
          &             &              &                 & 9 & $K\pi$ with $|\vec{p}_1|=\sqrt{3}\frac{2\pi}{L}$ and $|\vec{p}_2|=\sqrt{2}\frac{2\pi}{L}$ \\                                   
\hline          
(0,0,1)   & $C_{4v}$    & $E$          &   $J=1,2,...$   & 1 & $K^{*+}_{i}$ \\
          &             &              &                 & 2 & $K^{*+}_{ti}$ \\
          &             &              &                 & 3 & $K\pi$ with $|\vec{p}_1|=\sqrt{2}\frac{2\pi}{L}$ and $|\vec{p}_2|=\frac{2\pi}{L}$ \\
          &             &              &                 & 4 & $K\pi$ with $|\vec{p}_1|=\frac{2\pi}{L}$ and $|\vec{p}_2|=\sqrt{2}\frac{2\pi}{L}$ \\
          &             &              &                 & 5 & $K\pi$ with $|\vec{p}_1|=\sqrt{3}\frac{2\pi}{L}$ and $|\vec{p}_2|=\sqrt{2}\frac{2\pi}{L}$ \\
          &             &              &                 & 6 & $K\pi$ with $|\vec{p}_1|=\sqrt{2}\frac{2\pi}{L}$ and $|\vec{p}_2|=\sqrt{3}\frac{2\pi}{L}$ \\                                
\hline          
(0,1,1)   & $C_{2v}$    & $A_1$        &   $J=0,1,...$   & 1 & $K^{*+}_{i}$ \\
          &             &              &                 & 2 & $K^{*+}_{ti}$ \\
          &             &              &                 & 3 & $K_0^+$  \\
          &             &              &                 & 4 & $K\pi$ with $|\vec{p}_1|=0$ and $|\vec{p}_2|=\sqrt{2}\frac{2\pi}{L}$ \\
          &             &              &                 & 5 & $K\pi$ with $|\vec{p}_1|=\sqrt{2}\frac{2\pi}{L}$ and $|\vec{p}_2|=0$ \\
          &             &              &                 & 6 & $K\pi$ with $|\vec{p}_1|=\frac{2\pi}{L}$ and $|\vec{p}_2|=\sqrt{3}\frac{2\pi}{L}$ \\
          &             &              &                 & 7 & $K\pi$ with $|\vec{p}_1|=|\vec{p}_2|=\frac{2\pi}{L}$ \\          
          &             &              &                 & 8 & $K\pi$ with $|\vec{p}_1|=\sqrt{3}\frac{2\pi}{L}$ and $|\vec{p}_2|=\frac{2\pi}{L}$ \\
          &             &              &                 & 9 & $K\pi$ with $|\vec{p}_1|=|\vec{p}_2|=\sqrt{2}\frac{2\pi}{L}$ \\                                                  
\hline          
(0,1,1)   & $C_{2v}$    & $B_1$        &   $J=1,2,...$   & 1 & $K^{*+}_{i}$  \\
          &             &              &                 & 2 & $K^{*+}_{ti}$ \\
          &             &              &                 & 3 & $K\pi$ with $|\vec{p}_1|=\sqrt{3}\frac{2\pi}{L}$ and $|\vec{p}_2|=\frac{2\pi}{L}$ \\
          &             &              &                 & 4 & $K\pi$ with $|\vec{p}_1|=|\vec{p}_2|=\sqrt{2}\frac{2\pi}{L}$ \\
          &             &              &                 & 5 & $K\pi$ with $|\vec{p}_1|=\frac{2\pi}{L}$ and $|\vec{p}_2|=\sqrt{3}\frac{2\pi}{L}$ \\          
\hline          
(0,1,1)   & $C_{2v}$    & $B_2$        &   $J=1,2,...$   & 1 & $K^{*+}_{i}$  \\
          &             &              &                 & 2 & $K^{*+}_{ti}$ \\
          &             &              &                 & 3 & $K\pi$ with $|\vec{p}_1|=|\vec{p}_2|=\frac{2\pi}{L}$ \\
          &             &              &                 & 4 & $K\pi$ with $|\vec{p}_1|=|\vec{p}_2|=\sqrt{2}\frac{2\pi}{L}$ \\          
\hline          
(1,1,1)   & $C_{3v}$    & $A_1$        &   $J=0,1,...$   & 1 & $K^{*+}_{i}$ \\
          &             &              &                 & 2 & $K^{*+}_{ti}$ \\
          &             &              &                 & 3 & $K_0^+$  \\
          &             &              &                 & 4 & $K\pi$ with $|\vec{p}_1|=0$ and $|\vec{p}_2|=\sqrt{3}\frac{2\pi}{L}$ \\
          &             &              &                 & 5 & $K\pi$ with $|\vec{p}_1|=\sqrt{3}\frac{2\pi}{L}$ and $|\vec{p}_2|=0$ \\
          &             &              &                 & 6 & $K\pi$ with $|\vec{p}_1|=\frac{2\pi}{L}$ and $|\vec{p}_2|=\sqrt{2}\frac{2\pi}{L}$ \\
          &             &              &                 & 7 & $K\pi$ with $|\vec{p}_1|=\sqrt{2}\frac{2\pi}{L}$ and $|\vec{p}_2|=\frac{2\pi}{L}$ \\
\hline          
(1,1,1)   & $C_{3v}$    & $E$          &   $J=1,2,...$   & 1 & $K^{*+}_{i}$  \\
          &             &              &                 & 2 & $K^{*+}_{ti}$ \\
          &             &              &                 & 3 & $K\pi$ with $|\vec{p}_1|=\frac{2\pi}{L}$ and $|\vec{p}_2|=\sqrt{2}\frac{2\pi}{L}$ \\
          &             &              &                 & 4 & $K\pi$ with $|\vec{p}_1|=\sqrt{2}\frac{2\pi}{L}$ and $|\vec{p}_2|=\frac{2\pi}{L}$ \\
\hline
\end{tabular}
\caption{List of operators for all irreps that we use. The operators with structures labeled $K_0^+$, $K^{*+}_i$, and $K^{*+}_{ti}$ are quark-antiquark operators, while the operators with structures labeled $K\pi$ are two-meson operators. \label{tab:all_irreps}}
\end{table*}

To determine the $K\pi$ scattering amplitudes, we use the L\"uscher method; the first step of such a calculation is to determine the finite-volume spectra for different total momenta and irreducible representations. We determine the spectra in specific momentum frames and irreducible representations by calculating two-point correlation functions constructed from a set of interpolating operators.\\
\subsection{Interpolating operators}\label{subsec_ops}

We use two types of interpolating operators in this work. 
The single-hadron operators, built from local quark-antiquark bilinears, are constructed as follows:
\begin{align}
  K^{*+}_{i}(t,\vec{P})&= \sum_{\vec{x}} e^{\I \vec{P}\cdot \vec{x}}\,\overline{s}(t,\xvec)\, \gamma_i \, u(t,\xvec), \\
K^{*+}_{ti}(t,\vec{P})&= \sum_{\vec{x}} e^{\I\vec{P}\cdot \vec{x}}\,\overline{s}(t,\xvec)\, \gamma_t \gamma_i \, u(t,\xvec), \\
K^{+}_{0}(t,\vec{P})&= \sum_{\vec{x}} e^{\I \vec{P}\cdot \vec{x}}\,\overline{s}(t,\xvec)\,  u(t,\xvec). 
\end{align}
These operators have manifest isospin $I=1/2$; the projection to irreducible representations of the lattice symmetry group is discussed further below. The two-hadron operators are constructed from products of pseudoscalar $\pi$ and $K$ operators with definite individual momenta, combined to $I=1/2$ via $SU(2)$ Clebsch-Gordan coefficients:
\begin{align}
O_{K\pi}\left(t, \vec{p}_1,\vec{p}_2\right) &= \sqrt{\frac{2}{3}}\,\pi^{+}(t,\vec{p}_1) \, K^0(t,\vec{p}_2) \cr
&-\sqrt{\frac{1}{3}}\,\pi^{0}(t,\vec{p}_1)\, K^+(t,\vec{p}_2),
\end{align}
where
\begin{align}
 \pi^{+}(t,\vec{p}_1) &= \sum_{\vec{x}} e^{\I \vec{p}_1\cdot \vec{x}}\,\overline{d}(t,\xvec) \gamma_5  u(t,\xvec)\,, \\
 \pi^{0}(t,\vec{p}_1) &= \sum_{\vec{x}} e^{\I \vec{p}_1\cdot \vec{x}}\,
 \frac{1}{\sqrt{2}}\,
 \left[
 \overline{d}(t,\xvec) \,\gamma_5 \, d(t,\xvec) \right. 
 \nonumber \\
 & \left. \qquad -\overline{u}(t,\xvec)\, \gamma_5\,  u(t,\xvec) \right]\,, \\
 K^{+}(t,\vec{p}_2) &= \sum_{\vec{x}} e^{\I \vec{p}_2\cdot \vec{x}}\,\overline{s}(t,\xvec)\, \gamma_5\,  u(t,\xvec)\,, \\
 K^{0}(t,\vec{p}_2) &= \sum_{\vec{x}} e^{\I \vec{p}_2\cdot \vec{x}}\,\overline{s}(t,\xvec)\, \gamma_5\,  d(t,\xvec)\,.
\end{align}
All quark fields in the single-hadron and multi-hadron operators are Wuppertal-smeared \cite{Gusken:1989ad} with $\alpha_{\rm Wup}=3.0$ and 
$N_{\rm Wup}=20$ on the \texttt{C13} ensemble and $N_{\rm Wup}=55$ on the \texttt{D6} ensemble, using APE-smeared gauge links \cite{Albanese:1987ds} with $\alpha_{\rm APE}=2.5$ and $N_{\rm APE}= 25 ,\, 32 $ for \texttt{C13, D6} in the smearing kernel.

Since the finite-volume box in which we perform our calculation reduces the symmetry with respect to the infinite volume, we project the operators to the irreducible representations that respect the symmetry of the lattice:

\begin{align}
  O_{K\pi}^{\Lambda,\vec{P}} &= \dfrac{\mathrm{dim} (\Lambda)}{\mathrm{order}(LG(\vec{P}))} \sum_{\vec{p}}\sum\limits_{R\in LG(\vec{P})} \chi^{\Lambda}(R) \nonumber \\
  &\quad\times\:O_{K\pi}(R\,\vec{p}, \vec{P}-R\,\vec{p}), \\
  O_{K^{*}\!,i}^{\Lambda,\vec{P}} &= \dfrac{\mathrm{dim} (\Lambda)}{\mathrm{order}(LG(\vec{P}))} \sum\limits_{R\in LG(\vec{P})} \chi^{\Lambda}(R)\nonumber \\
  &\quad\times\sum_j R_{ij}\:K_j^{*+}(\vec{P}\,).
\end{align}
Above, $LG(\vec{P})$ denotes the Little Group on the lattice for total momentum $\vec{P}$ (i.e., the subgroup of the cubic group that remains a symmetry for the given momentum), and $\chi^{\Lambda}(R)$ are the characters for irrep $\Lambda$, which can be found for example in Ref.~\cite{Dresselhaus:2008}. In the sum over $\vec{p}$ (with components being integer multiples of $2\pi/L$), we fix the magnitudes $|\vec{p}_1|=|R\,\vec{p}\,|=|\vec{p}\,|$ and $|\vec{p}_2|=|\vec{P}-R\,\vec{p}\,|$, and different choices for these magnitudes yield different operators in the same irrep.

In the following we will denote the irrep-projected operators as
 $O_A^{\Lambda,\vec{P}}$,
where the operator index $A$ counts the different internal structures within a given irrep, as detailed in \cref{tab:all_irreps}.

\subsection{Wick contractions}\label{kpi_subsec_wick}

\begin{figure*}
  \centering
  \includegraphics[width=0.9\textwidth]{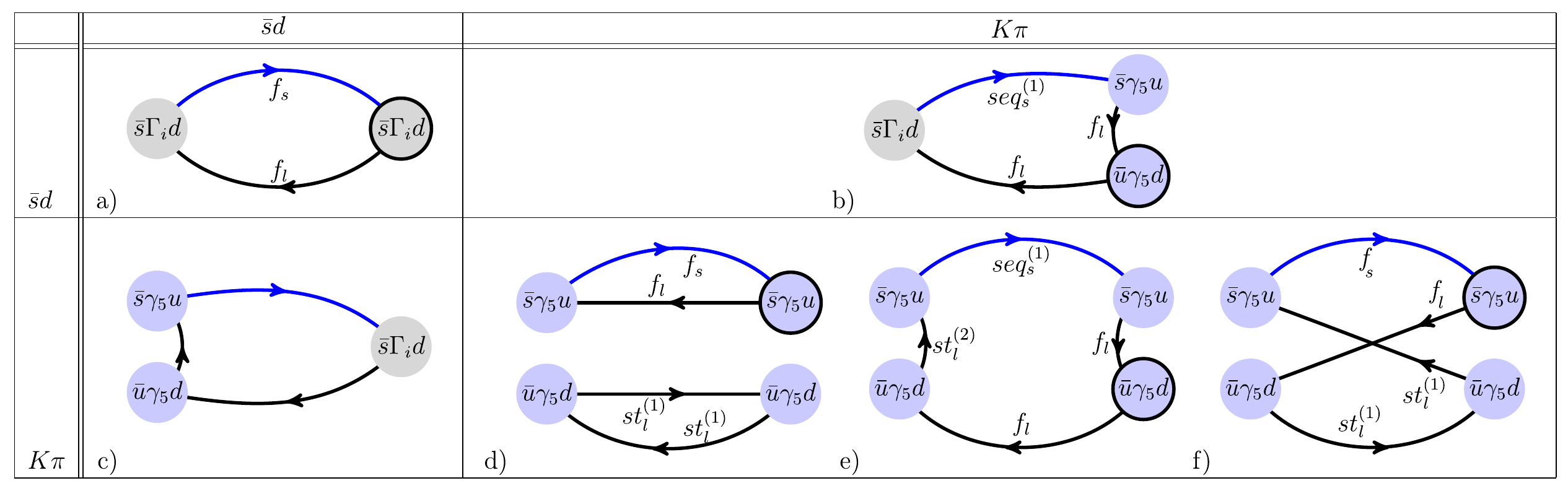}
  \caption{\label{fig:kpi_wick2pt} The Wick contractions corresponding to the correlation matrix elements of type $C_{\qbar q - \qbar q}$ (a),
  $C_{K\pi - \qbar q}$ (b,c), 
  $C^{\mathrm{direct}}_{K\pi - K\pi}$ (d), 
  $C^{\mathrm{box}}_{K\pi - K\pi}$ (e)
  and $C^{\mathrm{cross}}_{K\pi - K\pi}$ (f). We do not directly compute diagram (c), since it can be obtained as the complex-conjugate of diagram (b). The black circles outlining one of the interpolating fields in each diagram indicate the location of the point-to-all-propagator source. }
\end{figure*}

The correlation matrix $C^{\Lambda,\vec{P}}(t)$ for irreducible representation $\Lambda$ of Little Group $LG(\Pvec)$ is obtained from the interpolators
defined above as
\begin{align}
  C_{AB}^{\Lambda,\vec{P}}(t_{\rm snk} - t_{\rm src}) &= \brackets{O_A^{\Lambda,\vec{P}}(t_{\rm snk})\: O_B^{\Lambda,\vec{P}}(t_{\rm src})^\dagger}\,,
  \label{eq:.kpi_correlation_matrix_element}
\end{align}
where $t_{\rm src}$ is the source time and $t_{\rm snk}$ is the sink time. The correlation matrix elements
are expressed in terms of quark propagators by performing the Wick contractions (i.e., by
performing the path integral over the quark fields in a given gauge-field configuration).
The resulting quark-flow diagrams are shown in Fig.~\ref{fig:kpi_wick2pt} (for the case $I=1/2$
considered here).
The diagrams in Fig.~\ref{fig:kpi_wick2pt} are obtained from point-to-all propagators (labeled
$f$), sequential propagators (labeled $seq$) and stochastic timeslice-to-all propagators
(labeled $st$), as in Ref.~\cite{Alexandrou:2017mpi}. One summation over $\vec{x}$ at the source is eliminated using
translational symmetry. In the following description of the different types of propagators
we omit the smearing kernels for brevity, but we note that all propagators are smeared at source and sink with the parameters given in Sec.~\ref{subsec_ops}.

The point-to-all propagator for quark flavor $q$ with full spin-color-dilution is given by
\begin{align}
 & S_{f_q}\left( t_{\rm snk},\xvec_{\rm snk}; t_{\rm src},\xvec_{\rm src} \right)^{ab}_{\alpha\beta}  
  \label{eq:point-to-all-propagator}\\
 & \quad= \sum\limits_{x,\beta',b'}\,
  D^{-1}_q\left( t_{\rm snk},\xvec_{\rm snk}; x \right)^{a,b'}_{\alpha,\beta'}\,\eta^{(t_{i},\xvec_{\rm src},\beta,b)}\left( x \right)_{\beta'}^{b'}, 
  \nonumber
\end{align}
where
\begin{align}
  & \eta^{(t_{i},\xvec_{\rm src},\beta,b)}\left( t,\xvec \right)_{\beta'}^{b'} = \delta_{t,t_{\rm src}}\,\delta^{(3)}_{\xvec,\xvec_{\rm src}}\,\delta_{b,b'}\,\delta_{\beta,\beta'}.
\end{align}

The sequential propagator $S_{seq}$ follows from the solution of the lattice Dirac equation with right-hand side given by (\ref{eq:point-to-all-propagator})
restricted to the source time $t_{\rm src}$ and dressed with a vertex,
\begin{align}
  & S_{seq_q}\left( t_{\rm snk},\xvec_{\rm snk}; \Gamma(\pvec); t_{\rm src},\xvec_{\rm src} \right)^{ab}_{\alpha\beta}
  \nonumber \\
  & \quad = \sum\limits_{\xvec,\beta',b'}\,
  D^{-1}_q\left( t_{\rm snk},\xvec_{\rm snk}; t_{\rm src}, \xvec \right)^{a,b'}_{\alpha,\alpha'}\,  \cr
  &\quad\quad\times \Gamma_{\alpha'\beta'}\,e^{\I\pvec\cdot \xvec}\,S_{l}\left( t_{\rm src}, \xvec; t_{\rm src},\xvec_{\rm src} \right)^{b^\prime b}_{\beta',\beta}. \label{eq:sequential-propagator}
\end{align}
For the purpose of this calculation we require only $\Gamma = \gamma_5$ to realize the pseudoscalar pion or kaon interpolators 
and $q = l / s$ for the second inversion of the light / strange Dirac operator.

In addition we use timeslice-to-all stochastic propagators for the sink-to-sink quark propagation in diagram (e) in Fig. \ref{fig:kpi_wick2pt}.
These follow from solving the Dirac equation with a fully time-diluted stochastic source
\begin{align}
  & \nonumber S_{st^{(2)}}\left( t_t,\xvec_{\rm snk}; t_{\rm src}, \xvec_{\rm src} \right)
  \\
  & \quad  = \sum\limits_{\xvec,a',\alpha'}\, 
  D^{-1}_{l}\left( t_{\rm snk}, \xvec_{\rm snk}; t_{\rm src},\xvec \right)^{aa'}_{\alpha\alpha'}\cr
  & \quad\quad\times\eta^{(t_{\rm src})}\left( \xvec \right)^{a'}_{\alpha'}\,\eta^{(t_{\rm src})}\left( \xvec_{\rm src} \right)^{b *}_{\beta}. \label{eq:stochastic-timeslice-to-all-propagator} 
\end{align}
The space-spin-color components of $\eta^{(t_{\rm src})}$ are independent, $\mathbb{Z}_2 \times \mathbb{Z}_2$-distributed random numbers with 
zero mean and unit variance.

Finally, the all-source-to-all-sink quark propagation in diagrams (d) (lower fermion loop) and diagram (f) are realized with 
stochastic source-timeslice-to-all propagators from spin-diluted noise sources based on the one-end-trick \cite{McNeile:2006bz},
\begin{align}
  & S_{st^{(1)}}\left( t_{\rm snk}, \xvec_{\rm snk}; t_{\rm src}, \pvec_{\rm src} \right)^{a}_{\alpha \beta} 
  \label{eq:stochastic-timeslice-to-all-oet-propagator} \\
  & \quad= \sum\limits_{x,a',\alpha'}\, D^{-1}_{l}\left( t_{\rm snk},\xvec_{\rm snk}; x \right)^{aa'}_{\alpha\alpha'}\,\eta^{(t_{\rm src},\pvec_{\rm src},\beta)}\left( x \right)^{a'}_{\alpha'} 
  \nonumber
\end{align}
where
\begin{align}
&  \eta^{(t_{\rm src},\pvec_{\rm src},\beta)}\left( x \right)^{a'}_{\alpha'} = \delta_{\alpha',\beta}\,\delta_{t,t_{\rm src}}\,\eta(\xvec)^{a'}\,e^{\I\pvec_{\rm src}\cdot \xvec}\,,
\end{align}
with components $\eta\left( \xvec \right)^{a'}$ again $\mathbb{Z}_2 \times \mathbb{Z}_2$ noise.

\section{Spectrum Results}\label{sec_spectrum}

\begin{figure*}
  \includegraphics[width=\columnwidth]{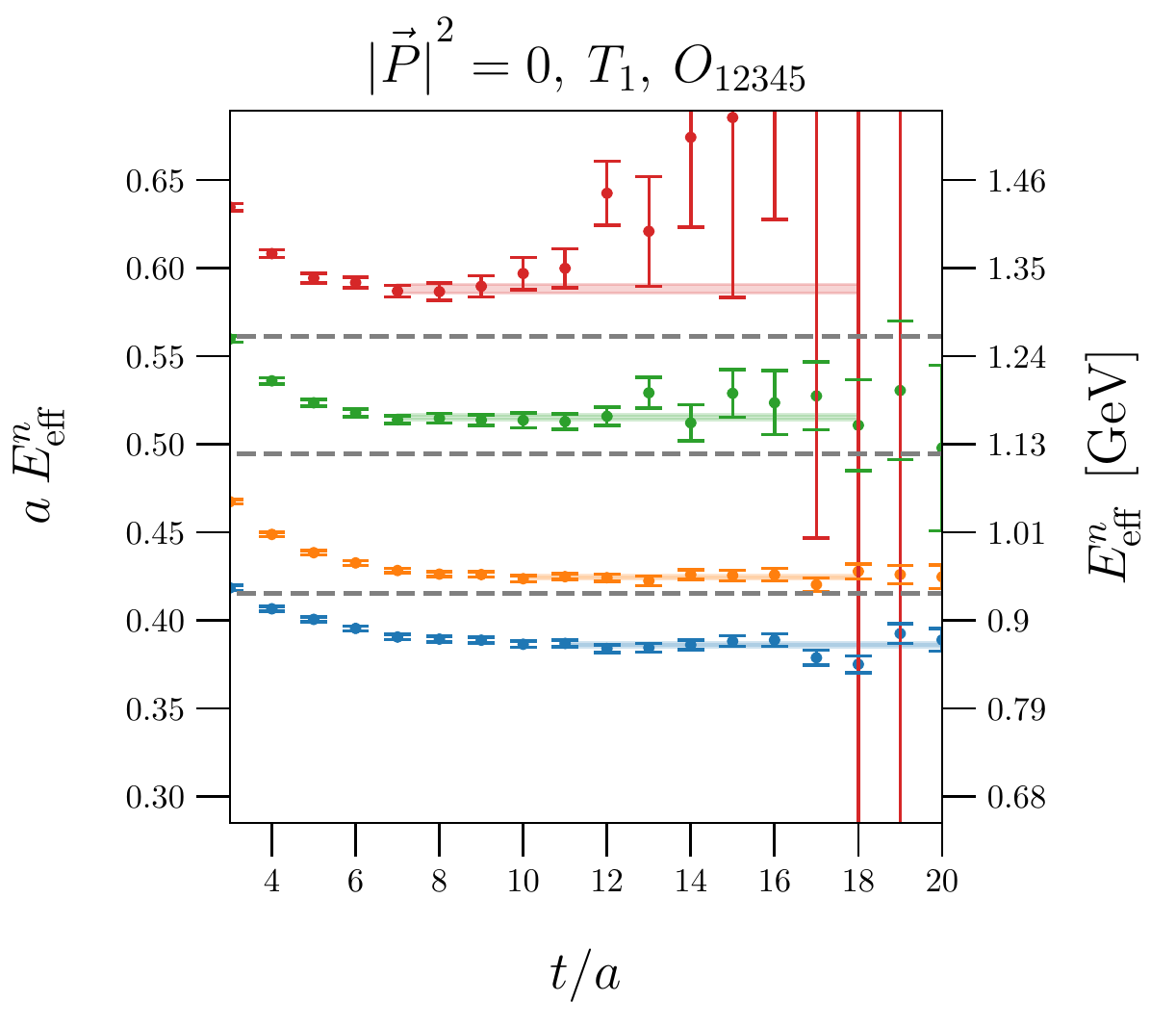}
  \includegraphics[width=\columnwidth]{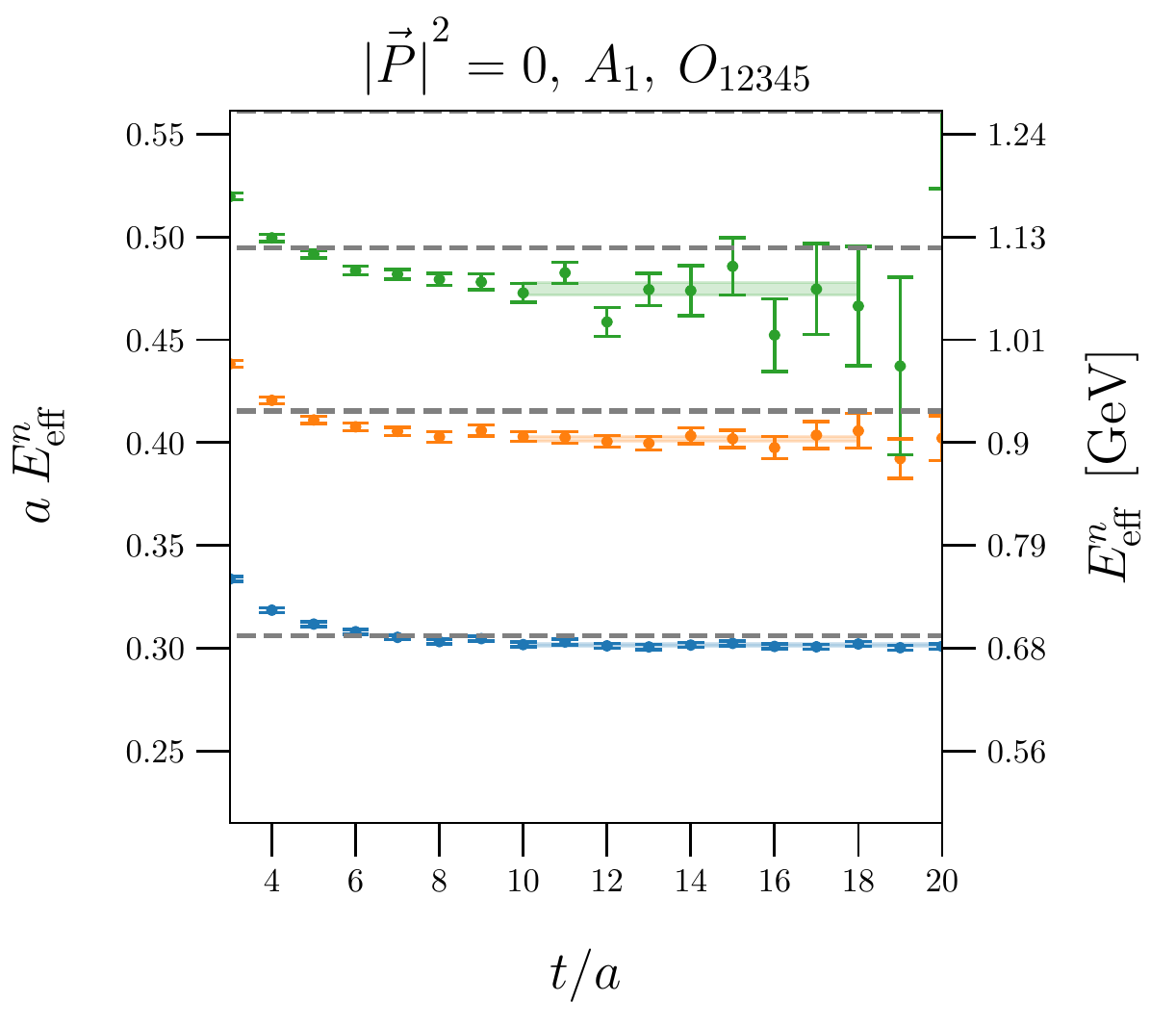}  
  \includegraphics[width=\columnwidth]{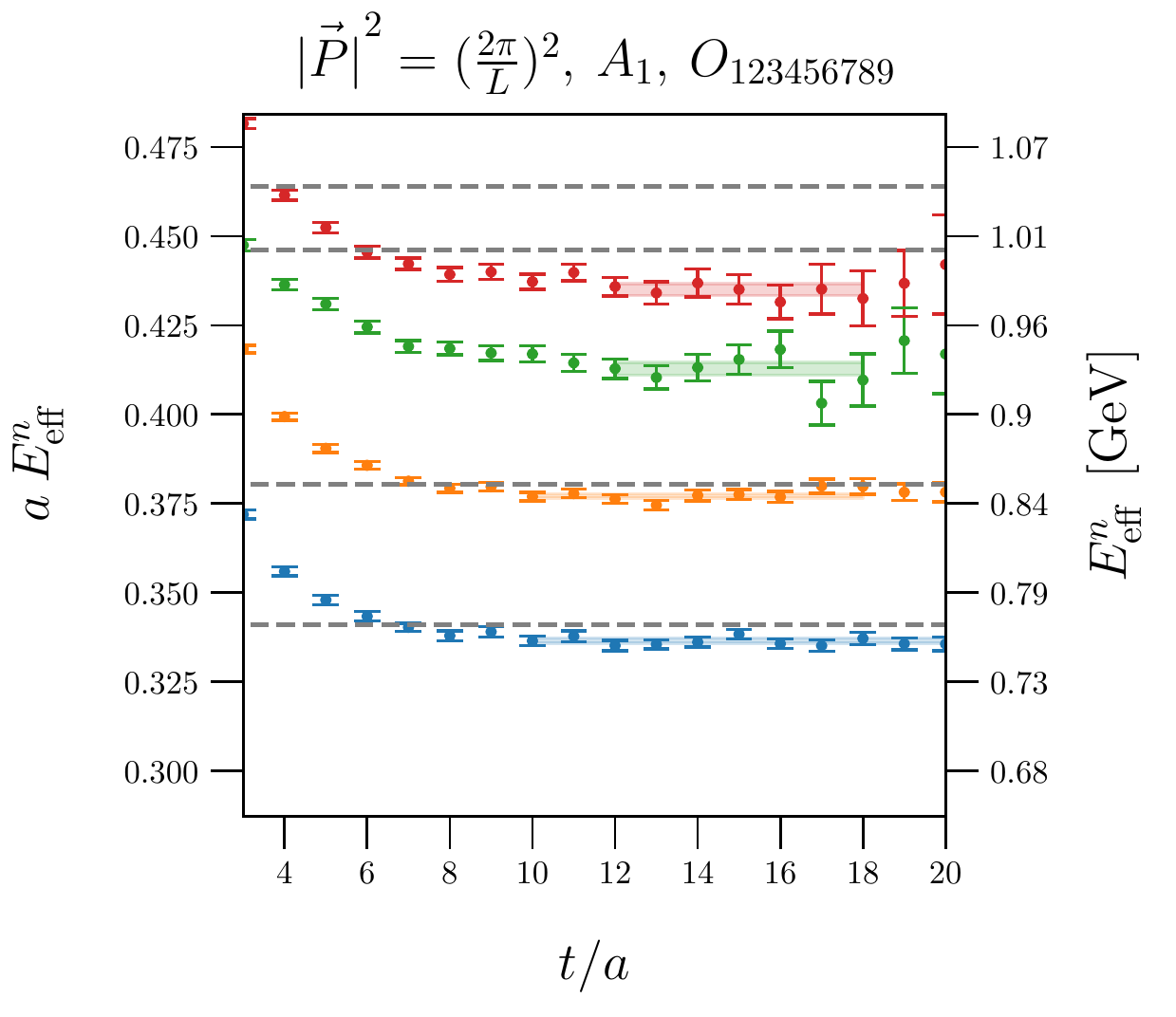}
  \includegraphics[width=\columnwidth]{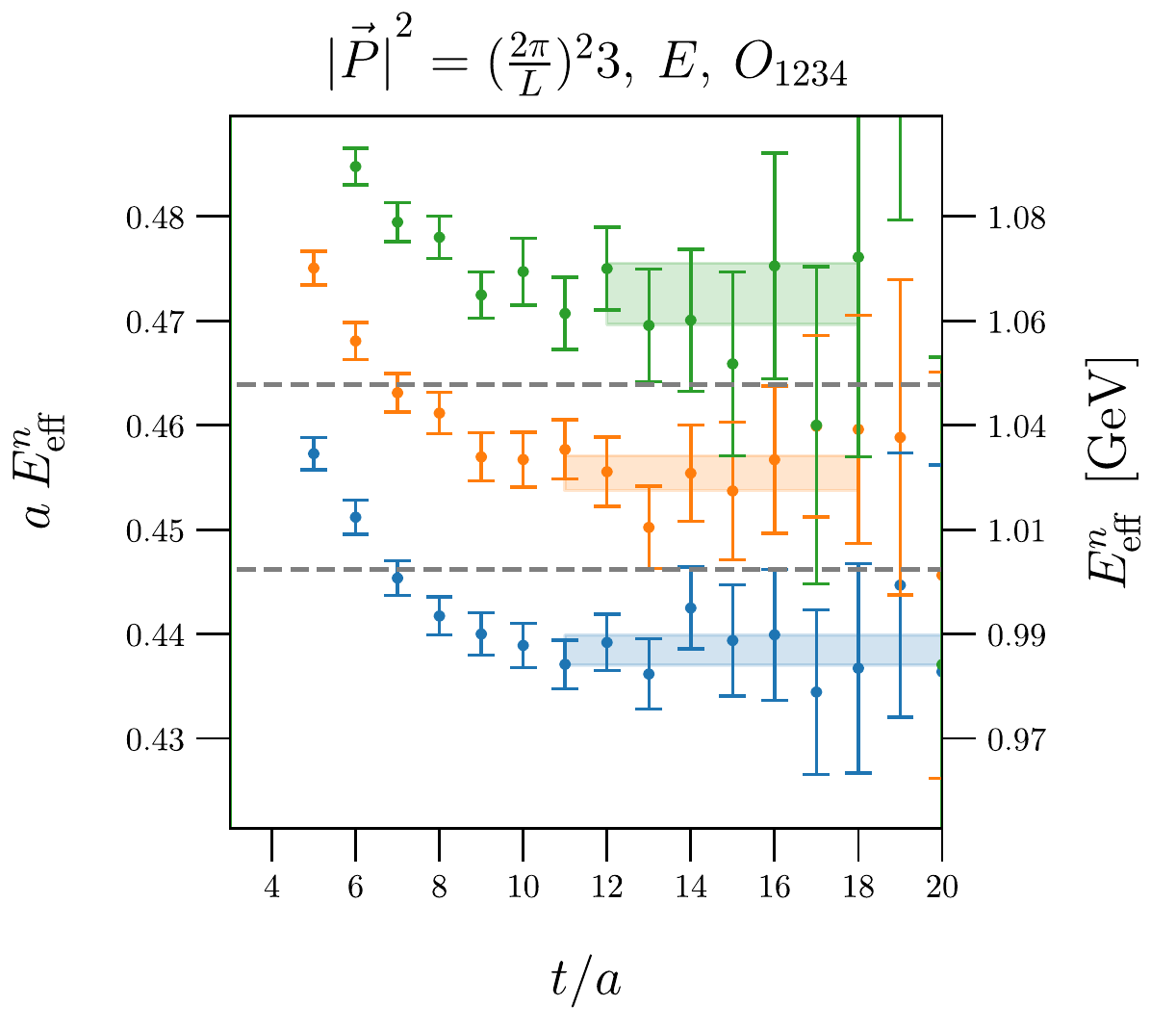}
  \caption{Sample plots of effective energies [defined in Eq.~(\ref{eq:meff})] from the \texttt{D6} ensemble. The noninteracting energy levels are indicated with dashed lines. The results from single-exponential fits are shown as the shaded bands, indicating the $\pm1\sigma$ energy range and the fit range.\label{fig:meff_plot}}
\end{figure*}

\begin{table*}
\begin{tabular}{|c|c l | c c c | l l|}
\hline
Ensemble & $\frac{L}{2\pi}|\vec{P}|$ & $\Lambda$ & $n$ & Fit Range & $\frac{\chi^2}{\text{dof}}$ & $aE^{\vec{P},\Lambda}_n$ & $a\sqrt{s^{\vec{P},\Lambda}_n}$ \\
\hline
\texttt{C13} & $0$ & $T_1$ & $1$ & $10-20$ & $0.76$ & $0.5189(17)$ & $0.5189(17)$\\
\hline
\texttt{C13} & $0$ & $A_1$ & $1$ & $9-20$ & $1.14$ & $0.48318(63)$ & $0.48318(63)$\\
\hline
\texttt{C13} & $1$ & $A_1$ & $1$ & $8-20$ & $0.68$ & $0.53809(74)$ & $0.50099(79)$\\
\texttt{C13} & $1$ & $A_1$ & $2$ & $8-20$ & $0.29$ & $0.5544(10)$ & $0.5184(11)$\\
\texttt{C13} & $1$ & $A_1$ & $3$ & $8-20$ & $1.37$ & $0.57660(89)$ & $0.54214(94)$\\
\hline
\texttt{C13} & $1$ & $E$ & $1$ & $8-20$ & $0.97$ & $0.5547(13)$ & $0.5188(14)$\\
\hline
\texttt{C13} & $\sqrt{2}$ & $A_1$ & $1$ & $9-20$ & $0.48$ & $0.5809(16)$ & $0.5103(18)$\\
\texttt{C13} & $\sqrt{2}$ & $A_1$ & $2$ & $8-18$ & $0.83$ & $0.5977(16)$ & $0.5292(18)$\\
\texttt{C13} & $\sqrt{2}$ & $A_1$ & $3$ & $8-18$ & $0.44$ & $0.6242(12)$ & $0.5590(14)$\\
\hline
\texttt{C13} & $\sqrt{2}$ & $B_2$ & $1$ & $8-20$ & $1.12$ & $0.5866(19)$ & $0.5167(21)$\\
\texttt{C13} & $\sqrt{2}$ & $B_2$ & $2$ & $9-20$ & $0.84$ & $0.6374(12)$ & $0.5737(13)$\\
\hline
\texttt{C13} & $\sqrt{2}$ & $B_1$ & $1$ & $9-20$ & $0.90$ & $0.5871(23)$ & $0.5172(26)$\\
\hline
\texttt{C13} & $\sqrt{3}$ & $A_1$ & $1$ & $7-20$ & $0.81$ & $0.6183(24)$ & $0.5163(29)$\\
\texttt{C13} & $\sqrt{3}$ & $A_1$ & $2$ & $7-18$ & $1.18$ & $0.6397(22)$ & $0.5418(26)$\\
\texttt{C13} & $\sqrt{3}$ & $A_1$ & $3$ & $7-18$ & $0.42$ & $0.6686(20)$ & $0.5757(23)$\\
\hline
\texttt{C13} & $\sqrt{3}$ & $E$ & $1$ & $8-20$ & $0.94$ & $0.6189(32)$ & $0.5171(38)$\\
\texttt{C13} & $\sqrt{3}$ & $E$ & $2$ & $8-18$ & $1.01$ & $0.6843(13)$ & $0.5938(14)$\\
\hline
\texttt{D6} & $0$ & $T_1$ & $1$ & $11-20$ & $1.54$ & $0.3861(11)$ & $0.3861(11)$\\
\texttt{D6} &  $0$ & $T_1$ & $2$ & $10-18$ & $0.38$ & $0.4244(10)$ & $0.4244(10)$\\
\hline
\texttt{D6} & $0$ & $A_1$ & $1$ & $10-20$ & $0.71$ & $0.30162(71)$ & $0.30162(71)$\\
\texttt{D6} & $0$ & $A_1$ & $2$ & $10-18$ & $0.22$ & $0.4018(13)$ & $0.4018(13)$\\
\hline
\texttt{D6} & $1$ & $A_1$ & $1$ & $10-20$ & $0.79$ & $0.33657(73)$ & $0.31007(80)$\\
\texttt{D6} & $1$ & $A_1$ & $2$ & $10-18$ & $0.96$ & $0.37714(58)$ & $0.35369(62)$\\
\texttt{D6} & $1$ & $A_1$ & $3$ & $12-18$ & $1.15$ & $0.4128(18)$ & $0.3915(19)$\\
\texttt{D6} & $1$ & $A_1$ & $4$ & $12-18$ & $0.19$ & $0.4350(17)$ & $0.4148(18)$\\
\hline
\texttt{D6} & $1$ & $E$ & $1$ & $12-20$ & $1.51$ & $0.4072(15)$ & $0.3856(16)$\\
\texttt{D6} & $1$ & $E$ & $2$ & $9-18$ & $0.63$ & $0.45134(94)$ & $0.43194(98)$\\
\texttt{D6} & $1$ & $E$ & $3$ & $12-16$ & $0.03$ & $0.4694(17)$ & $0.4508(18)$\\
\hline
\texttt{D6} & $\sqrt{2}$ & $A_1$ & $1$ & $9-20$ & $0.62$ & $0.36787(82)$ & $0.31789(95)$\\
\texttt{D6} & $\sqrt{2}$ & $A_1$ & $2$ & $10-18$ & $1.23$ & $0.41041(82)$ & $0.36629(92)$\\
\texttt{D6} & $\sqrt{2}$ & $A_1$ & $3$ & $10-18$ & $0.75$ & $0.42372(97)$ & $0.3811(11)$\\
\texttt{D6} & $\sqrt{2}$ & $A_1$ & $4$ & $9-18$ & $0.77$ & $0.4412(13)$ & $0.4005(15)$\\
\hline
\texttt{D6} & $\sqrt{2}$ & $B_2$ & $1$ & $11-18$ & $0.36$ & $0.41064(76)$ & $0.36655(85)$\\
\texttt{D6} & $\sqrt{2}$ & $B_2$ & $2$ & $9-18$ & $0.54$ & $0.4414(15)$ & $0.4007(16)$\\
\hline
\texttt{D6} & $\sqrt{2}$ & $B_1$ & $1$ & $10-20$ & $0.89$ & $0.4294(14)$ & $0.3874(16)$\\
\texttt{D6} & $\sqrt{2}$ & $B_1$ & $2$ & $9-18$ & $0.47$ & $0.4775(13)$ & $0.4402(14)$\\
\hline
\texttt{D6} & $\sqrt{3}$ & $A_1$ & $1$ & $8-20$ & $0.98$ & $0.3966(12)$ & $0.3254(15)$\\
\texttt{D6} & $\sqrt{3}$ & $A_1$ & $2$ & $8-18$ & $1.42$ & $0.4397(13)$ & $0.3767(15)$\\
\texttt{D6} & $\sqrt{3}$ & $A_1$ & $3$ & $8-18$ & $1.26$ & $0.4497(15)$ & $0.3884(18)$\\
\texttt{D6} & $\sqrt{3}$ & $A_1$ & $4$ & $9-18$ & $0.82$ & $0.4659(20)$ & $0.4070(23)$\\
\hline
\texttt{D6} & $\sqrt{3}$ & $E$ & $1$ & $11-20$ & $0.30$ & $0.4385(14)$ & $0.3753(17)$\\
\texttt{D6} & $\sqrt{3}$ & $E$ & $2$ & $11-18$ & $0.38$ & $0.4554(17)$ & $0.3950(19)$\\
\texttt{D6} & $\sqrt{3}$ & $E$ & $3$ & $12-18$ & $0.39$ & $0.4726(29)$ & $0.4146(34)$\\
\hline
\end{tabular}
\caption{Results of single-exponential fits to the generalized eigenvalues, for the two different ensembles, the different total momenta $\vec{P}$, and the different irreps $\Lambda$. We set $t_0/a=3$ on the \texttt{C13} ensemble and $t_0/a=4$ on the \texttt{D6} ensemble. Ancillary files with the central values of $a\sqrt{s^{\vec{P},\Lambda}_n}$ and their covariances for each ensemble are provided (\texttt{D6\_spectrum.dat} and \texttt{C13\_spectrum.dat}). \label{tab:spectrum}}
\end{table*}

\begin{figure*}
  \includegraphics[width=0.8\linewidth]{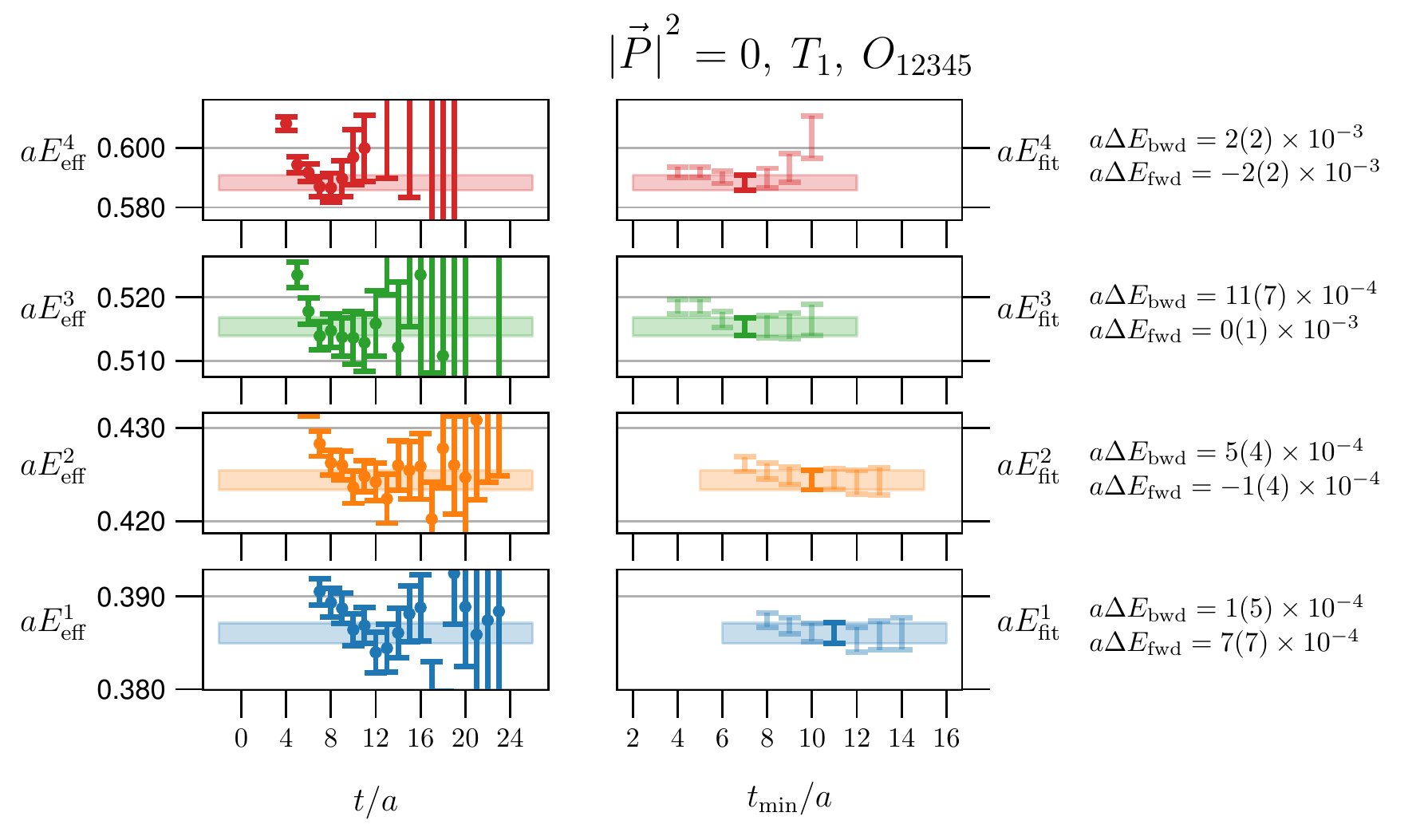}
  
  \vspace{0.05\textheight}
  
  \includegraphics[width=0.8\linewidth]{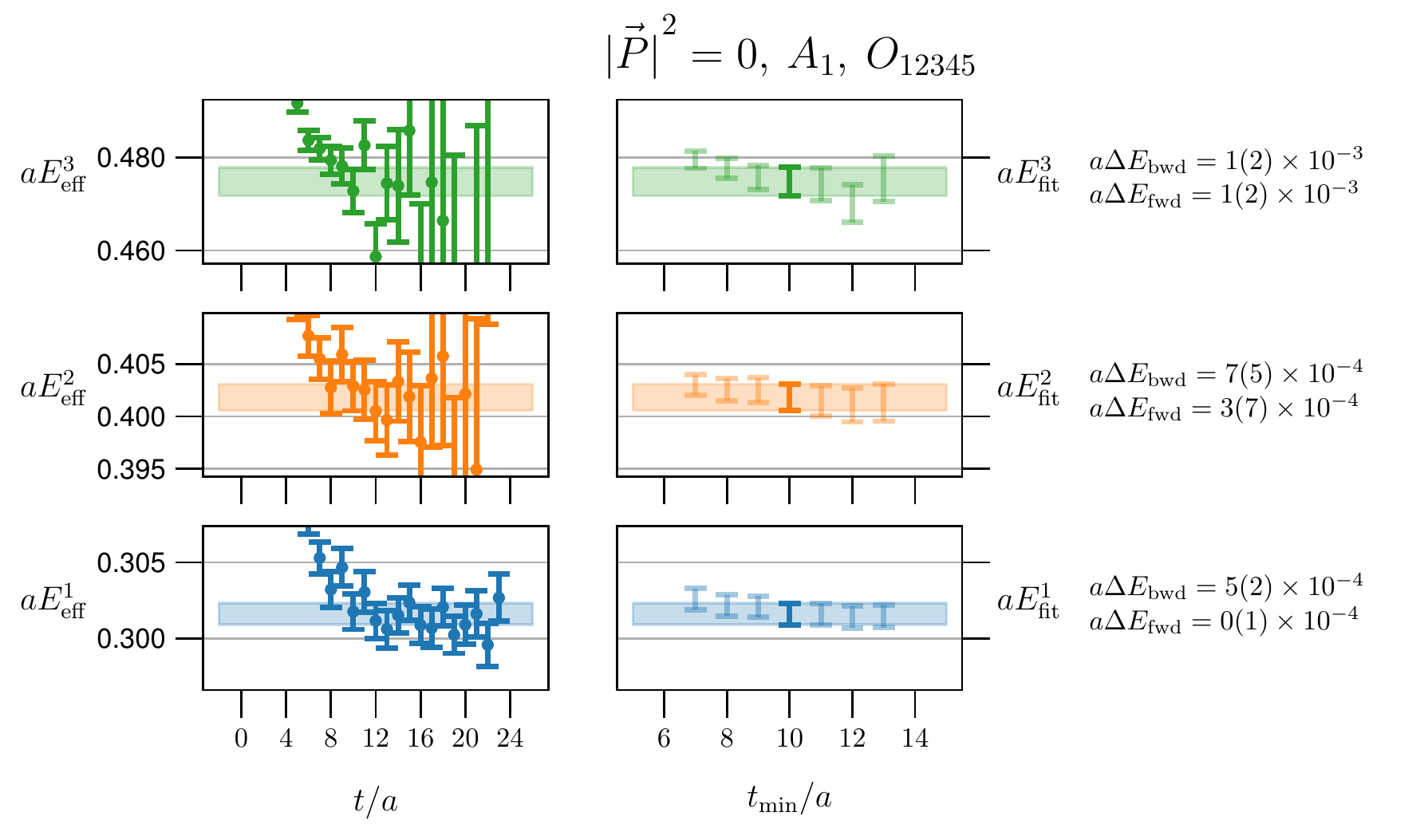}  

  \caption{Sample plots illustrating our tests of the stability of the fitted energies under variations of the lower bound of the fit range, $t_{\rm min}$. The left panels show the effective-energy plots for the generalized eigenvalues. The center panels show the fitted energies as a function of $t_{\rm min}$. On the right, we give the changes in the fitted energies when shifting $t_{\rm min}$ by one lattice step: $\Delta E_{\rm bwd}=E|_{t_{\rm min}-a}-E|_{t_{\rm min}}$ and $\Delta E_{\rm fwd}=E|_{t_{\rm min}}-E|_{t_{\rm min}+a}$, for our nominal choice of $t_{\rm min}$. The results shown here are from  the \texttt{D6} ensemble for the irreps $A_{1g}$ and $T_{1u}$ of the Little Group $O_{h}$.\label{fig:stability_plot1}
}
\end{figure*}
\begin{figure*}
  \includegraphics[width=0.8\linewidth]{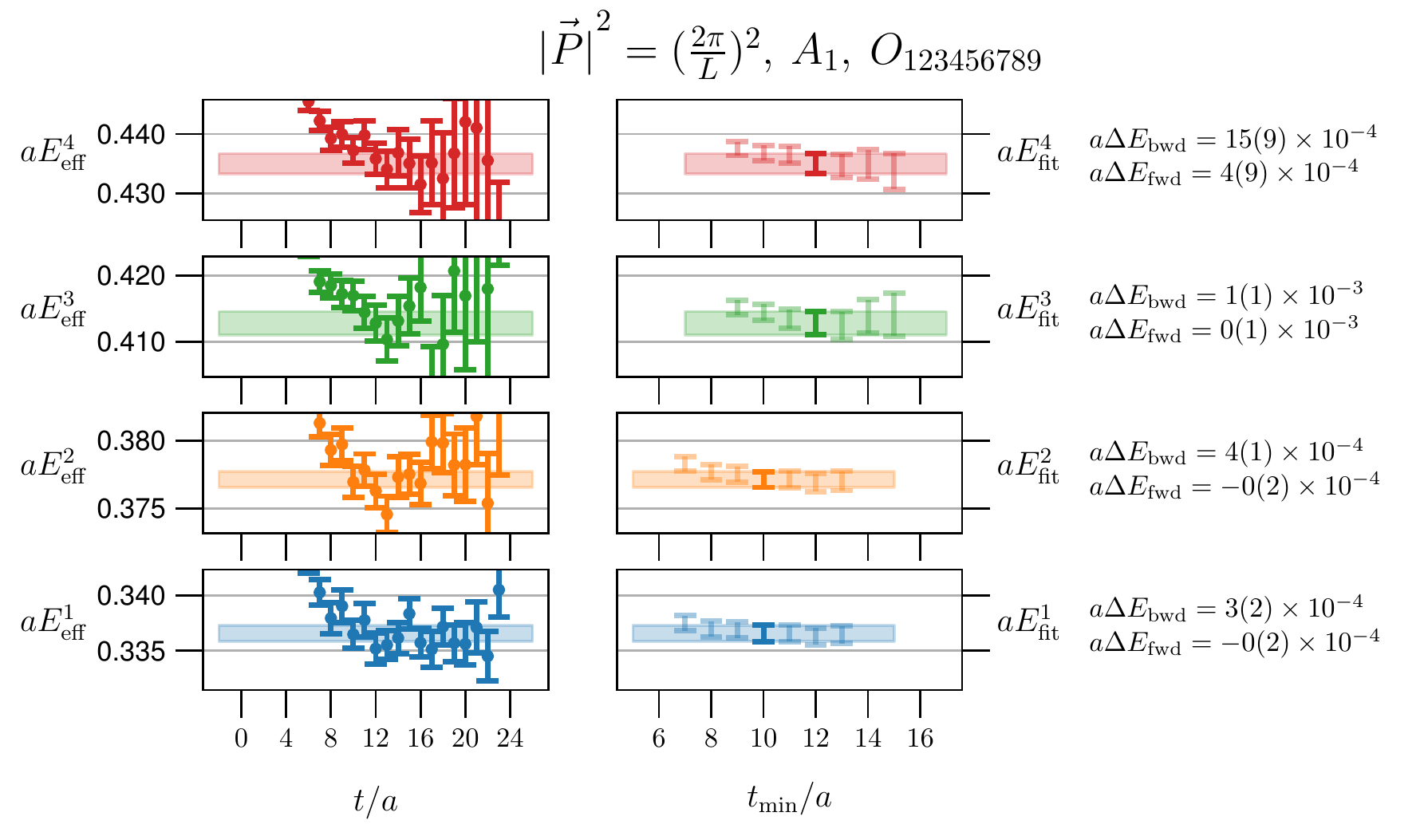}
  
  \vspace{0.05\textheight}
  
  \includegraphics[width=0.8\linewidth]{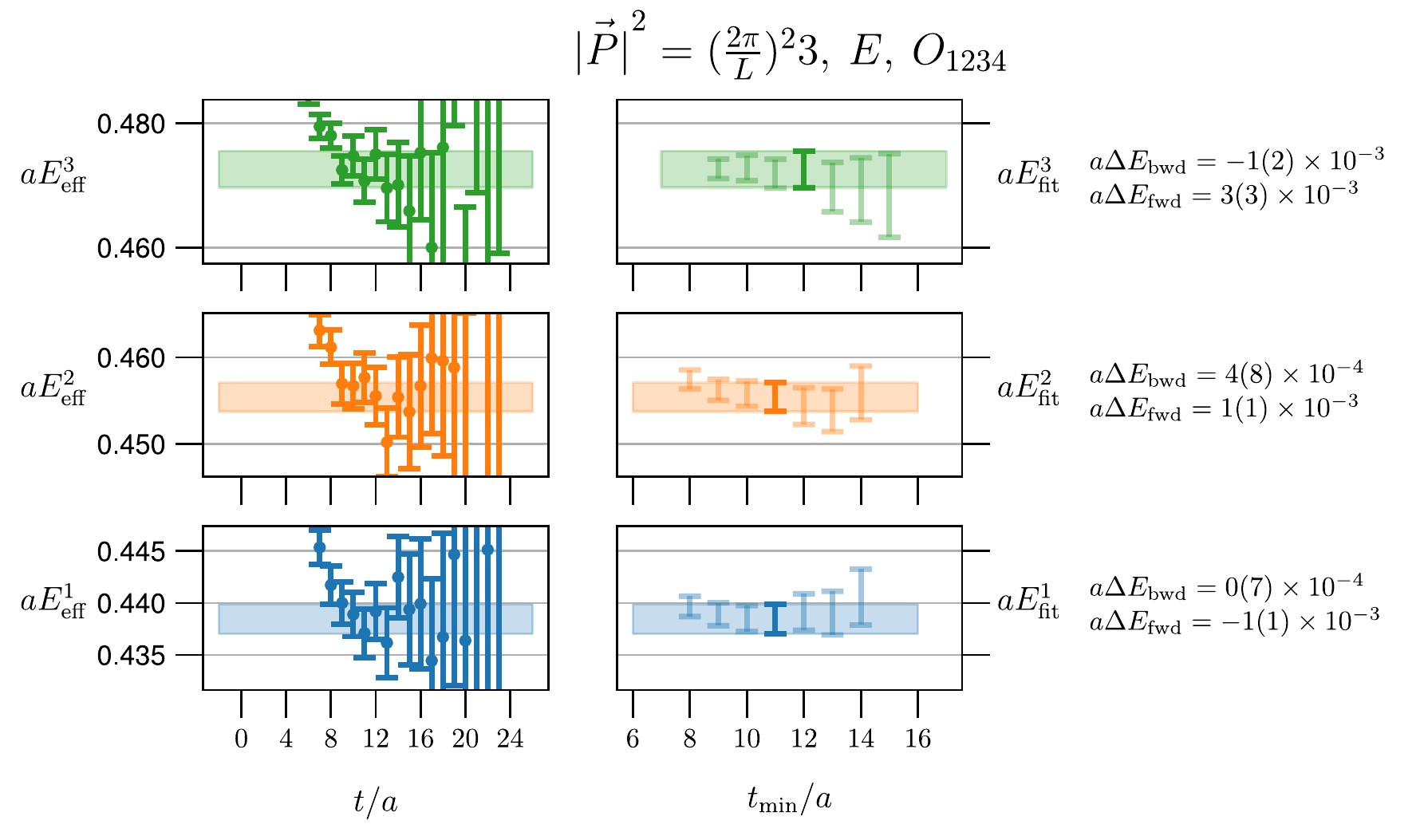}
  
  \caption{Like Fig.~\protect\ref{fig:stability_plot1}, but for irreps $A_1$ and $E$ of the Little Groups $C_{4v}$ and $C_{3v}$, respectively.\label{fig:stability_plot2}}
\end{figure*}

We extracted the energy levels $E^{\Lambda,\vec{P}}_n$ from the correlation matrices using the generalized eigenvalue problem (GEVP) \cite{Luscher:1990ck,Blossier:2009kd}
\begin{equation}
\label{eq:GEVP}
\sum_B C^{\Lambda,\vec{P}}_{AB}(t)\, u^n_B(t) = \lambda^n(t,t_0)\sum_B C^{\Lambda,\vec{P}}_{AB}(t_0)\, u_B^n(t),
\end{equation}
where $n$ labels the eigenpair. 
Here, $t_0$ is a reference timeslice whose variation does not affect noticeably the large-$t$ behavior~\cite{Alexandrou:2017mpi}. At large enough values of $t$ and $t_0$, the eigenvalues $\lambda^n(t,t_0)$ take the form of a single exponential
\begin{align}
\label{oneexp}
\lambda^{n}(t,t_{0}) = e^{-E_{n}^{\Lambda,\vec{P}}(t-t_{0})}.
\end{align}
We can make some initial observations by looking at the effective energies
\begin{equation}\label{eq:meff}
  aE_{\mathrm{eff}}^{n}(t) = \ln \frac{\lambda_n(t,t_0)}{\lambda_n(t+a,t_0)},
\end{equation}
shown in Fig.~\ref{fig:meff_plot} for four different irreps of three different Little Groups. Dashed lines indicate the noninteracting energy levels of the $K\pi$ system with the pion and kaon momenta $|\vec{p}_1|$ and $|\vec{p}_2|$ used for the corresponding operator basis. In the plot for irrep $T_{1u}$, note that the highest three energy levels are shifted upward relative to the noninteracting energies. These states overlap most strongly (in relative terms) with the states created by the corresponding multi-hadron operators. The lowest energy level is an extra energy level whose occurrence is related to the presence of a narrow resonance (the $K^*$). This state overlaps most strongly with the states created by the quark-antiquark operators. Similarly, on the top-right plot ($\Lambda=A_{1g}$, $|\vec{P}|^2=0$), there is a downward shift with respect to the noninteracting energies from which we can expect an attractive interaction and a positive $S$-wave scattering phase shift. With the absence of an extra energy level, one can not straightforwardly identify the presence of a resonance in the depicted energy range.

Our main results for the energies $E_{n}^{\Lambda, \vec{P}}$ are obtained directly from single-exponential fits to the generalized eigenvalues $\lambda^{n}(t,t_{0})$ and are given in \cref{tab:spectrum}. Also shown in the table are the center-of-momentum-frame energies $\sqrts$, which are related to the lattice-frame energy $E_{n}^{\Lambda, \vec{P}}$ through
\begin{equation}
\sqrts = \sqrt{(\En)^2 - (\pkpi)^2}.
\end{equation}

We have chosen the fit ranges such that the contributions from higher excited states are negligible compared to statistical uncertainty. We ensured this by varying the lower bound of the fit range, $t_{\rm min}$, as shown in Figs.~\ref{fig:stability_plot1} and \ref{fig:stability_plot2}. In each case, the nominal value for $t_{\rm min}/a$ is chosen such that $\Delta E_{\rm fwd}=E|_{t_{\rm min}}-E|_{t_{\rm min}+a}$ is consistent with zero.

\FloatBarrier

\section{L\"uscher Analysis}\label{sec_luscher}

Assuming elasticity and neglecting exponential finite-volume effects, the energy levels of a two-particle system with total momentum $\vec{P}$ in a cubic box
with periodic boundary conditions are given by the solutions of the L\"uscher quantization condition
\begin{align}
\label{eq:QC}
{\rm det} \left( \mathbbm{1} + \I  T \, (\mathbbm{1} + \I {\cal M}^{\vec{P}}) \right) = 0.
\end{align}
The object in parentheses is a matrix with indices $\ell m,\ell' m'$. 
The \Tmatrix introduced in Sec.~\ref{sec_about_Kpi} is diagonal,
\begin{align}
  T_{\ell m,\ell' m'} &=  T^{(\ell)} \,\delta_{\ell \ell'} \delta_{m m'}\, ,
  \label{eq:T_gen}
\end{align}
and for a single scattering channel as considered here, 
$T^{(\ell)}$ denotes the scattering amplitude for partial wave $\ell$.
The amplitudes $T^{(\ell)}$ depend
only on the center-of-mass energy, or, equivalently, the scattering momentum $k$. The elements of ${\cal M}^{\vec{P}}$ for $\ell,\ell^\prime \leq 1$ are given by \cite{Leskovec:2012gb}
\begin{widetext}
\begin{align}
\label{M_gen}
&\left({\cal M}^{\vec{P}}_{\ell m, \ell' m'}\right) = \bordermatrix{~ & 0 \phantom{\mbox{-}}0        & 1 \phantom{\mbox{-}}0                    & 1 \phantom{\mbox{-}}1                       & 1\mbox{-}1\cr
  0 \phantom{\mbox{-}}0 & w_{00} & \I\sqrt{3}w_{10} & \I \sqrt{3} w_{11} & \I\sqrt{3} w_{1 \mbox{-}1} \cr
  1 \phantom{\mbox{-}}0 & -\I\sqrt{3}w_{10} & w_{00}+2w_{20} & \sqrt{3} w_{21} & \sqrt{3}w_{2\mbox{-}1} \cr
  1 \phantom{\mbox{-}}1 & \I \sqrt{3}w_{1\mbox{-}1} & -\sqrt{3}w_{2\mbox{-}1} & w_{00}-w_{20} & -\sqrt{6} w_{2\mbox{-}2} \cr
  1 \mbox{-}1 & \I \sqrt{3} w_{11} & -\sqrt{3}w_{21} & -\sqrt{6}w_{22} & w_{00}-w_{20} \cr}\ ,
\end{align}
\end{widetext}
where the functions $w_{\ell m}$ depend on the scattering momentum $k$, the box size $L$, and the total momentum $\vec{P}$,
\begin{align}
&w_{\ell m}=w_{\ell m}^{\pkpi}(k,\,L) = \frac{ Z_{\ell m}^{\pkpi}\left(1;(k \ltwopi )^2\right) }{ \gamma \pi^{3/2} \sqrt{2 \ell+1} (k \ltwopi)^{\ell+1}}\,.
\end{align}
Here, $Z_{\ell m}^{\pkpi}\left(1;(k \ltwopi )^2\right)$ is the generalized zeta function  and $\gamma = E^{\vec{P}}/\sqrt{s}$ is the Lorentz boost factor.
The matrix ${\cal M}^{\vec{P}}$ can be further simplified by taking into account the symmetries for a given Little Group $LG(\vec{P})$ and irreducible representation
$\Lambda$ \cite{Leskovec:2012gb}. Each irrep in principle contains infinitely many partial waves, the first few of which are listed in Table \ref{tab:all_irreps} (since the $K$ and $\pi$ are both spinless, we have $J=\ell$). However, the contributions from higher partial waves are increasingly suppressed, and we neglect the contributions from $\ell\geq 2$ in this work. This then leads to the following quantization conditions, where we write the scattering amplitudes $T^{(0)}$ and $T^{(1)}$ in terms of the phase shifts $\delta_0$ and $\delta_1$, respectively:
\begin{align}
\nonumber &\pkpi = \twopiL (0,0,0), \;\;\,\;LG=O_h,\;\;\,\; \Lambda = A_{1g} \text{: } \\
&\; \cot{\delta_0} =  w_{00}, \\
\nonumber&\pkpi = \twopiL (0,0,0), \;\;\,\;LG=O_h,\;\;\,\; \Lambda = T_{1u} \text{: } \\
&\; \cot{\delta_1} =  w_{00}, 
\end{align}
\begin{align}
& \pkpi = \twopiL (0,0,1), \;\;\,\;LG=C_{4v}, \;\;\, \;\Lambda = A_{1} \text{: } \cr 
& \left( \cot{\delta_0 }-  w_{00}  \right) \left(\cot{\delta_1}-w_{00}-2 w_{20}\right) -3 w_{10}^2=0, \cr
\label{eq:cantdis1} \\
& \pkpi = \twopiL (0,0,1), \;\;\,\;LG=C_{4v}, \;\;\,\; \Lambda = E \text{: }\cr
& \cot{\delta_1} = w_{00} - w_{20}, \label{eq:cantdis2} \\
& \pkpi = \twopiL (0,1,1), \;\;\,\;LG=C_{2v}, \;\;\, \;\Lambda = A_{1} \text{: } \cr 
& \left(\cot{\delta_0}-w_{00}\right) \left(\cot{\delta_1}-w_{00}+w_{20}+\I \sqrt{6}\, w_{22}\right) \cr
&+6 i \,w_{11}^2=0 , \\
& \pkpi = \twopiL (0,1,1), \;\;\,\;LG=C_{2v}, \;\;\,\; \Lambda = B_{1} \text{: }\cr
& \cot{\delta_1}  = w_{00} + 2 w_{20} , \\
& \pkpi = \twopiL (0,1,1), \;\;\,\;LG=C_{2v}, \;\;\,\; \Lambda = B_{2} \text{: }\cr
& \cot{\delta_1}  = w_{00} - w_{20} -\sqrt{6}\, {\rm Im}\left[ w_{22} \right]  , \\
& \pkpi = \twopiL (1,1,1), \;\;\,\;LG=C_{3v}, \;\;\, \;\Lambda = A_{1} \text{: } \cr 
& \left(\cot{\delta_0}-w_{00}\right) \left(-\cot{\delta_1}+w_{00}-2 \I \sqrt{6}\, w_{22}\right) \cr
& + 9 \,w_{10}^2=0, \\
& \pkpi = \twopiL (1,1,1), \;\;\,\;LG=C_{3v}, \;\;\,\; \Lambda = E \text{: }\cr
& \cot{\delta_1} = w_{00}  + \I \sqrt{6}\, w_{22} .
\end{align}
Note that at nonzero momenta, the quantization conditions in the $A_1$ irreps depend on both the $S$-wave and the $P$-wave phase shifts. This mixing between even and odd
partial waves occurs because the reciprocal space of momenta in the unequal-mass $K\pi$ system lacks inversion symmetry \cite{Leskovec:2012gb}. Traditionally, the L\"uscher method has often been used to map individual energy levels on the lattice to individual phase shift values at the corresponding center-of-mass energies. However, this is no longer possible in the $A_1$ irreps with partial-wave mixing. \Cref{eq:cantdis1}, for example, has the two unknowns $\delta_0(k_n^{\vec{P},A_1})$ and $\delta_1(k_n^{\vec{P},A_1})$, and it does not help to combine Eqs.~(\ref{eq:cantdis1}) and (\ref{eq:cantdis2}) either, because the solutions of Eq.~(\ref{eq:cantdis2}) occur at different values of the scattering momentum, $k_n^{\vec{P},E}$. Since we want to use all irreps, we follow a different approach \cite{Guo:2012hv}, in which we parametrize the energy dependence of the phase shifts $\delta_0$ and $\delta_1$ using the models discussed in Sec.~\ref{sec_about_Kpi}, and then perform a global fit of the model parameters for both the $S$- and $P$-waves 
to all energy levels by minimizing the following $\chi^2$ function:
\begin{align}
\label{eq:tmatchi2}
\chi^2 &= \sum_{\vec{P},\Lambda,n}\:\: \sum_{\vec{P}^\prime,\Lambda^\prime,n^\prime}\:\:[C^{-1}]_{\vec{P},\Lambda,n;\vec{P}',\Lambda',n'}\cr
&\times  \:\: \bigg( \sqrt{s_n^{\Lambda, \vec{P}}}^{[{\rm data}]} - \sqrt{s_n^{\Lambda, \vec{P}}}^{[{\rm model}]} \bigg) \cr
&\times \:\: \bigg( \sqrt{s_{n'}^{\Lambda', \vec{P}'}}^{[{\rm data}]} - \sqrt{s_{n'}^{\Lambda', \vec{P}'}}^{[{\rm model}]}  \bigg).
\end{align}
Here, $[C^{-1}]$ is the data covariance matrix of the spectrum determined on the lattice and $\sqrt{s_n^{\Lambda, \vec{P}}}^{[{\rm model}]}$ is obtained from the parametrized scattering amplitudes using the L\"uscher quantization conditions\footnote{In Ref.~\cite{Alexandrou:2017mpi}, we demonstrated that the results of this approach are consistent with those from the traditional two-step approach (when applicable) of first extracting individual phase shifts followed by a fit of a model to the phase shifts.}. We fit $17$ energy levels on the \texttt{C13} ensemble and $26$ levels on the \texttt{D6} ensemble, as listed in \cref{tab:spectrum}. In choosing these energy levels, we have stayed further below the $K\eta$ threshold, $\sim 0.95\sqrt{s_{K\eta}}$, determined through \cref{eq:GMO} in order to safely avoid effects from the $K\eta$ threshold, the $K^\star(1410)$ resonance, or re-scattering from three particle channels \cite{Tanabashi:2018oca}. In practice, we found it helpful to obtain initial guesses for the $P$-wave model parameters using an initial
fit to only those irreps without $S$-wave contributions, followed by the full fit to all irreps. The results for the $P$-wave parameters from the full fits are consistent
with the results from the reduced fits, but are about 10\% more precise. Moreover, we performed a combined fit with a reduced set of irreps ($T_{1u}$, $A_{1g}$, and $C_{4v}$'s $A_1$) which, apart from increasing the uncertainty in some parameters compared to using the full list, has also proven to worsen the $\chi^2/dof$  minimum to an unacceptable value. This further justifies the use of data from higher momentum frames. On each ensemble, we performed four different full fits that differ in the type of parametrization used for the $S$-wave amplitude: Chung's parametrization [Eq.~(\ref{eq:Chung})], effective-range expansion [Eq.~(\ref{eq:ERE})], Bugg's parametrization [Eq.~(\ref{eq:Bugg})], and conformal-map parametrization [Eq.~(\ref{eq:generalconformal})]. The parametrization for the $P$-wave amplitude was always of the form given in Eq.~(\ref{eq:Chung}). The best-fit parameters and $\chi^2$ values of the full fits are given in Tables \ref{tab:C13fits} and \ref{tab:D6fits} in the Appendix. 

\section{Results for the phase shifts and pole positions}\label{sec_phases}

\begin{figure*}
  \includegraphics[width=\columnwidth]{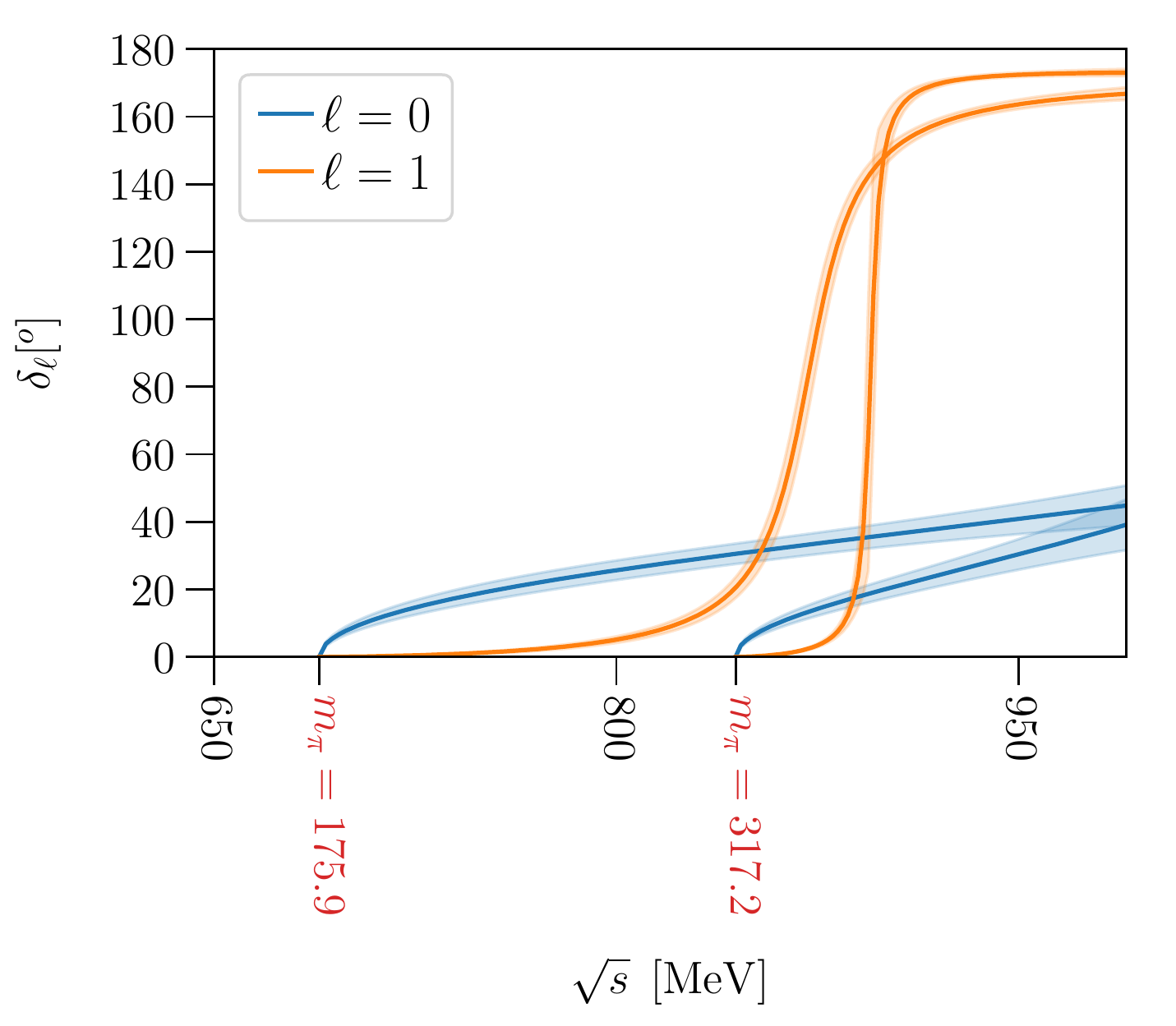}
  \includegraphics[width=\columnwidth]{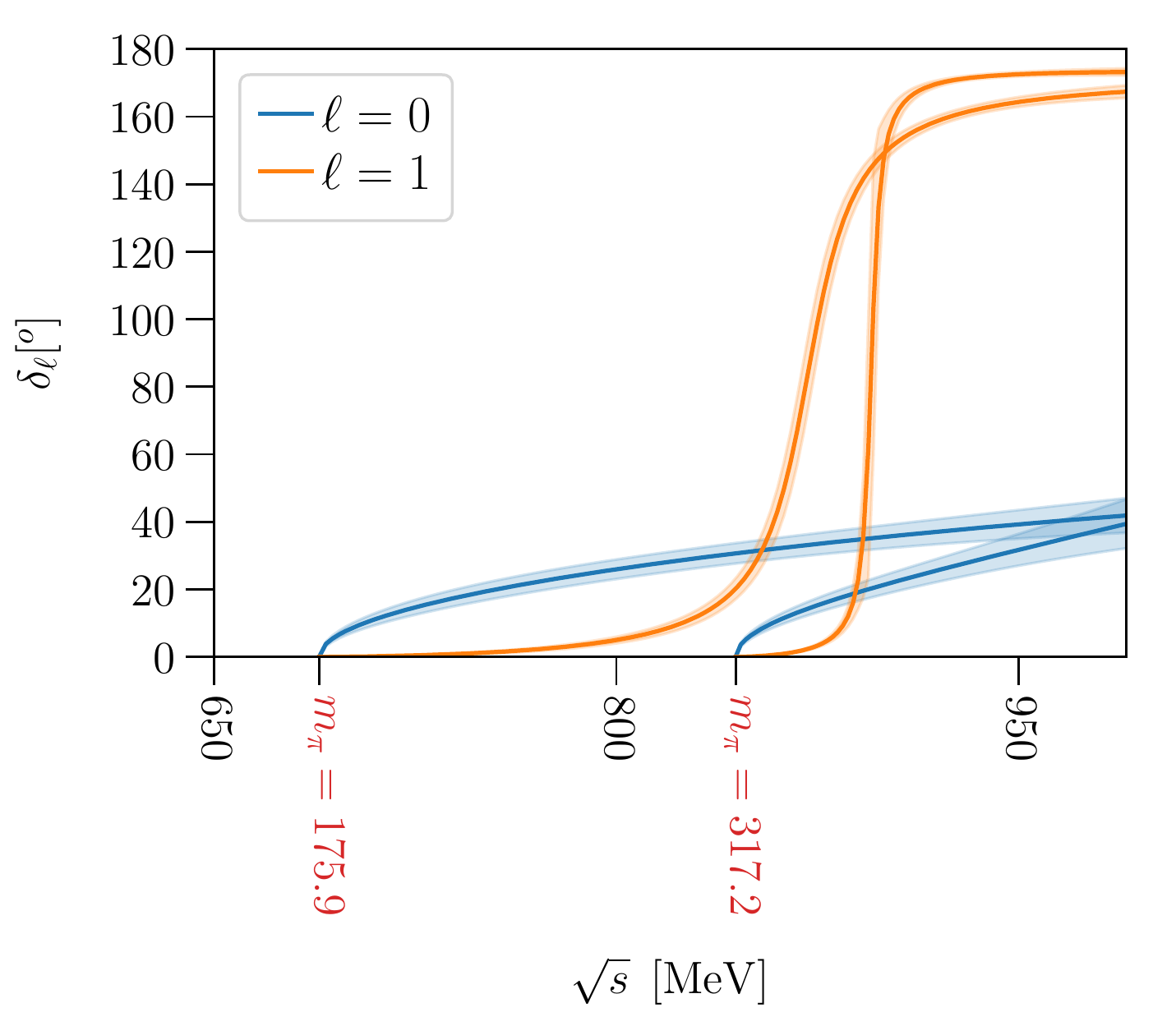}
  \includegraphics[width=\columnwidth]{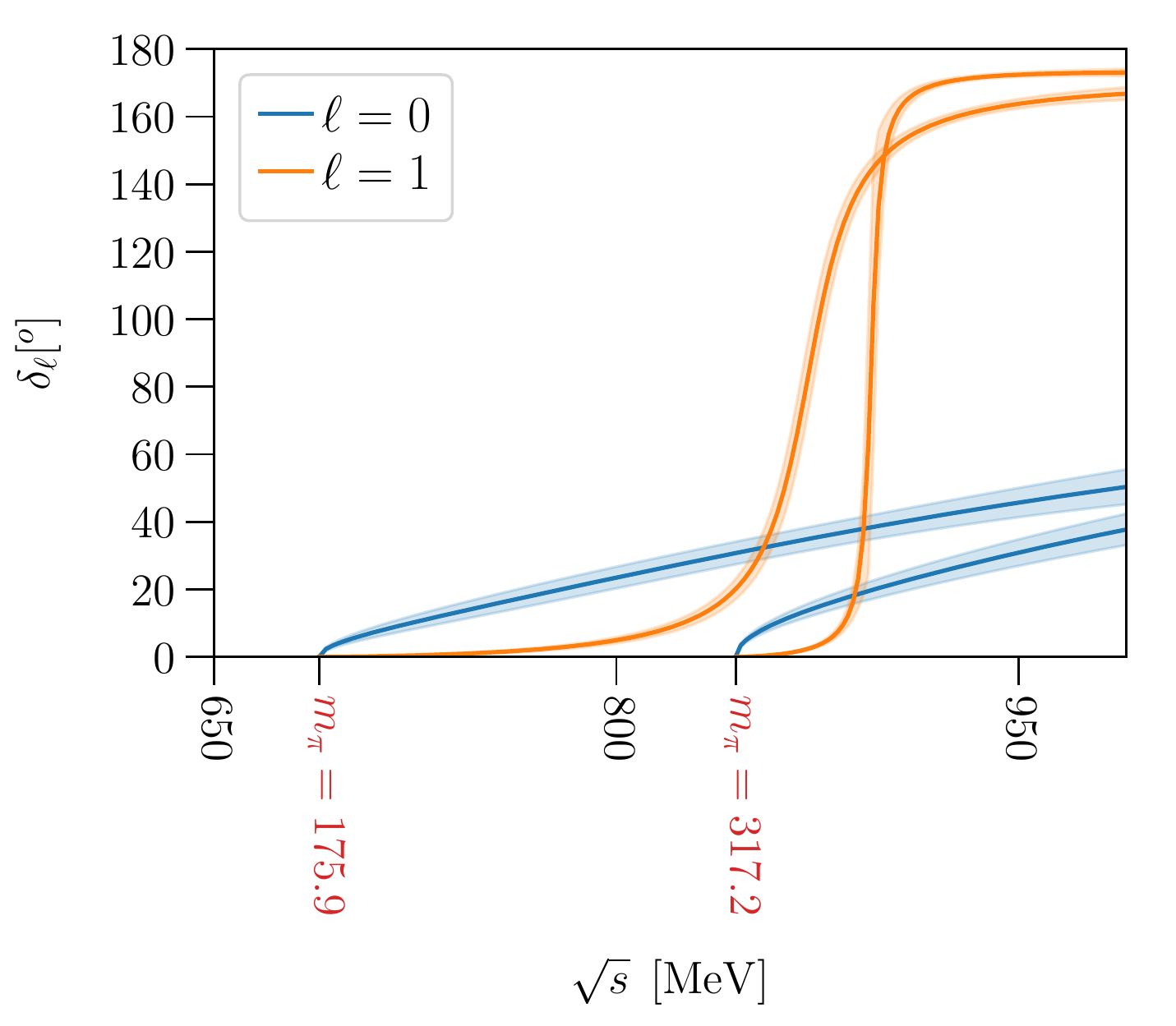}
  \includegraphics[width=\columnwidth]{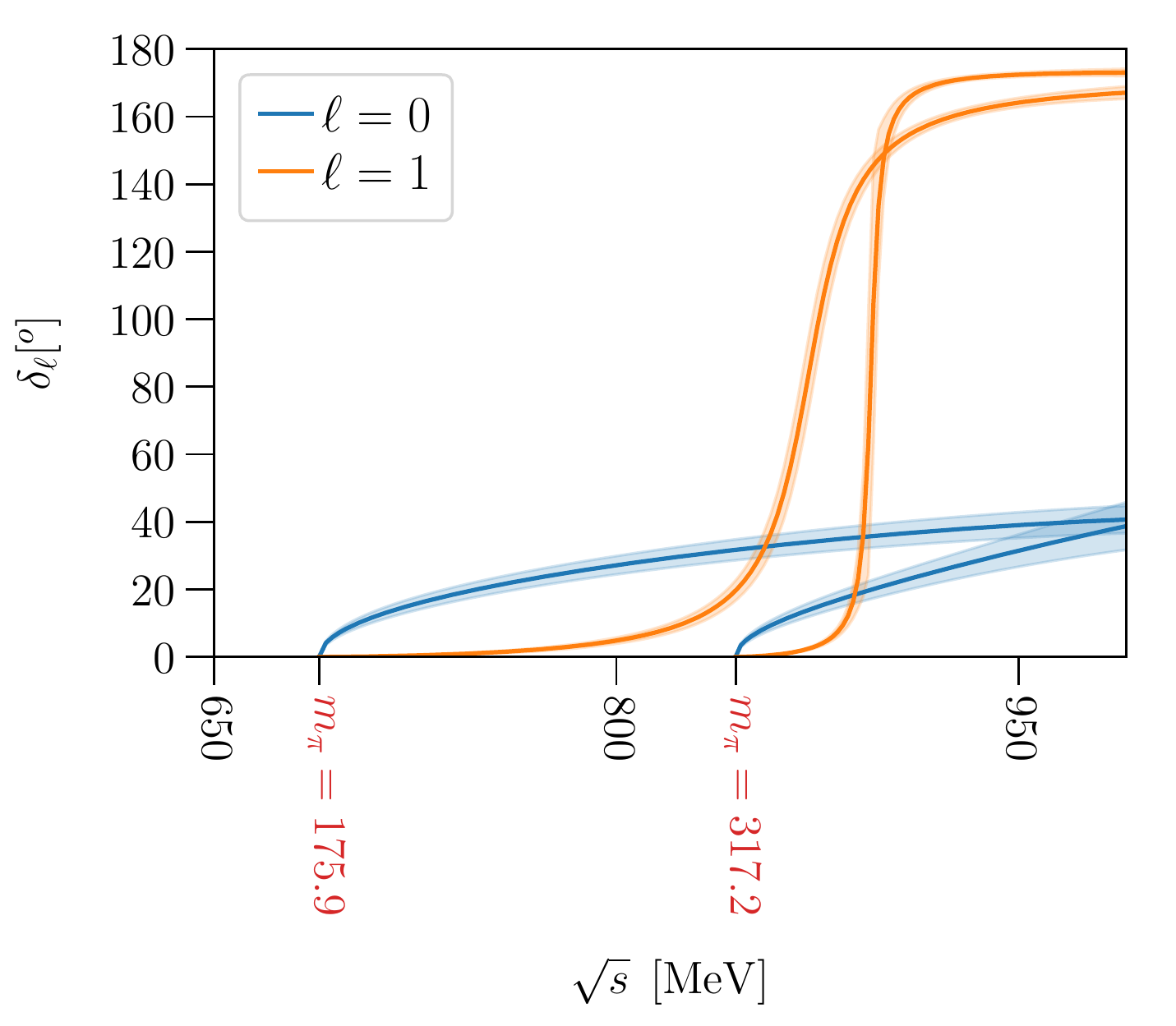}
  \caption{$S$- and $P$-wave phase shift results from both ensembles, labeled here by to their pion masses. The four different plots differ in the type of parametrization used for the $S$-wave amplitude. From top-left to right-bottom: Chung's parametrization [Eq.~(\ref{eq:Chung})], effective-range expansion [Eq.~(\ref{eq:ERE})], Bugg's parametrization [Eq.~(\ref{eq:Bugg})], and conformal-map parametrization [Eq.~(\ref{eq:generalconformal})]. The parametrization for the $P$-wave amplitude was always of the form given in Eq.~(\ref{eq:Chung}).
}\label{fig:phase_shifts}
\end{figure*}

\begin{figure*}
\includegraphics[width=0.99\linewidth]{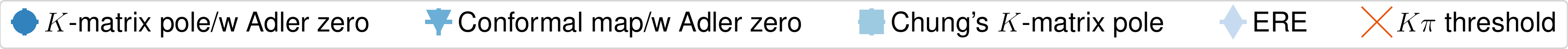}

\includegraphics[width=0.5\linewidth]{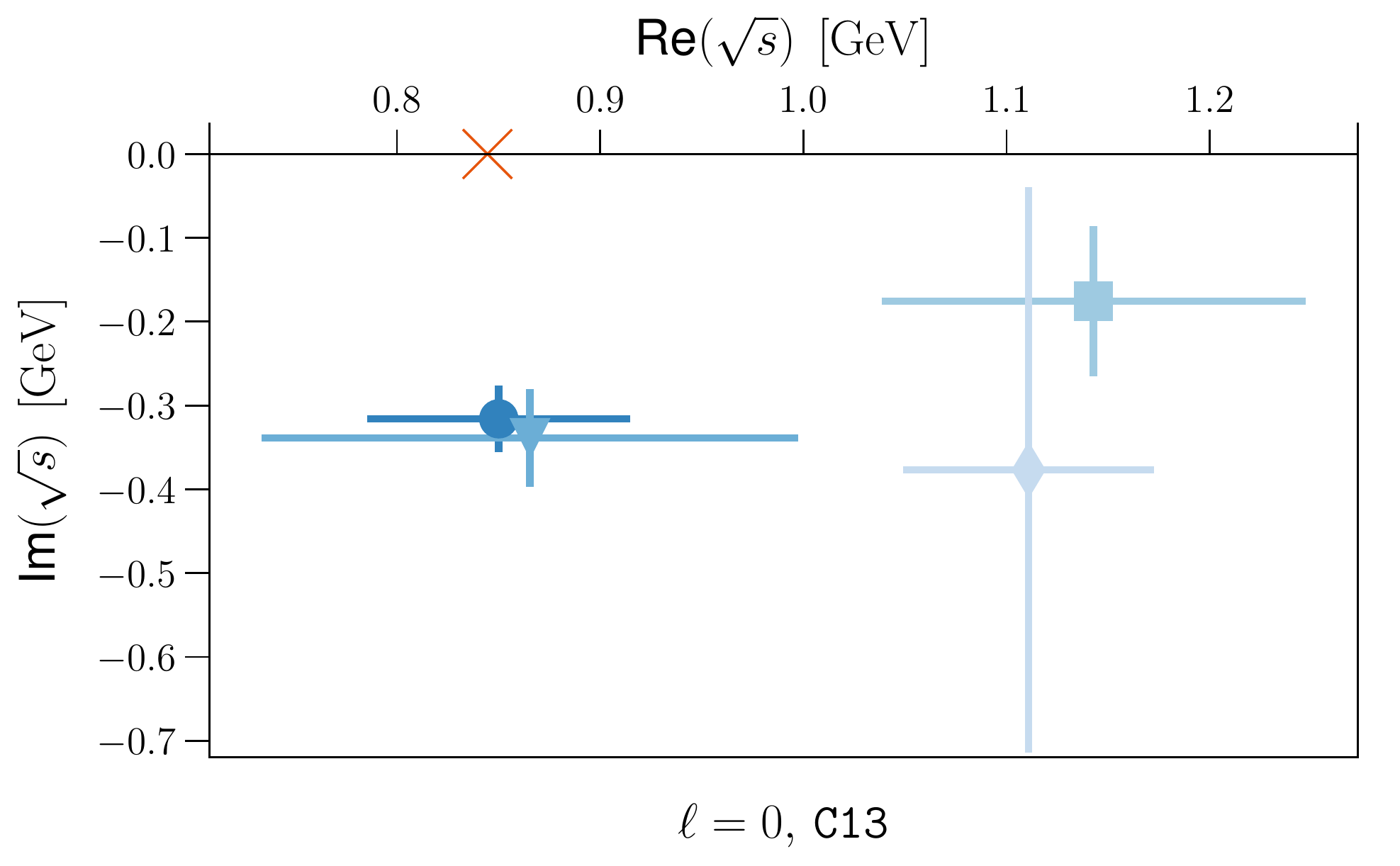}\includegraphics[width=0.5\linewidth]{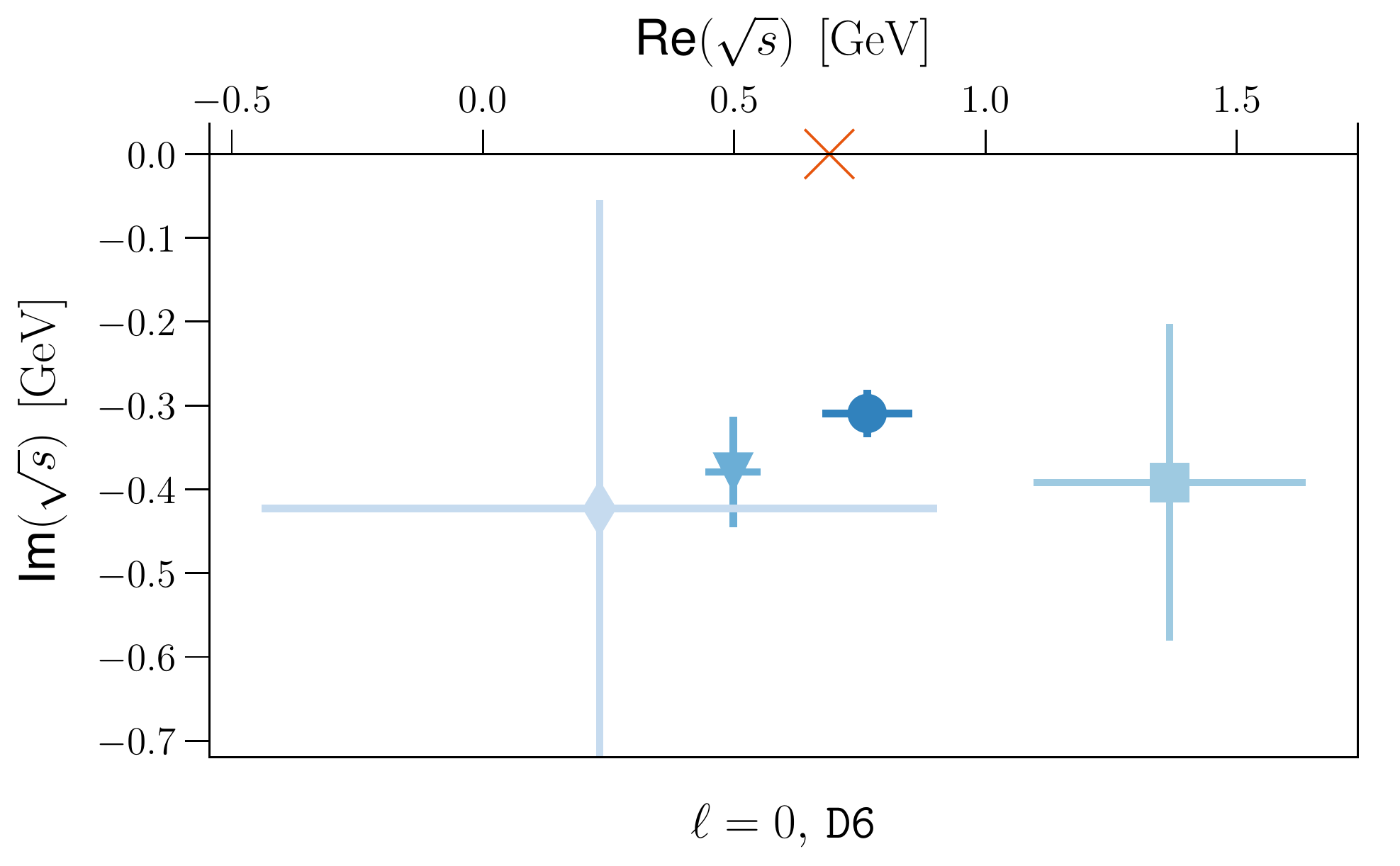}

\vspace{2ex}

\includegraphics[width=0.5\linewidth]{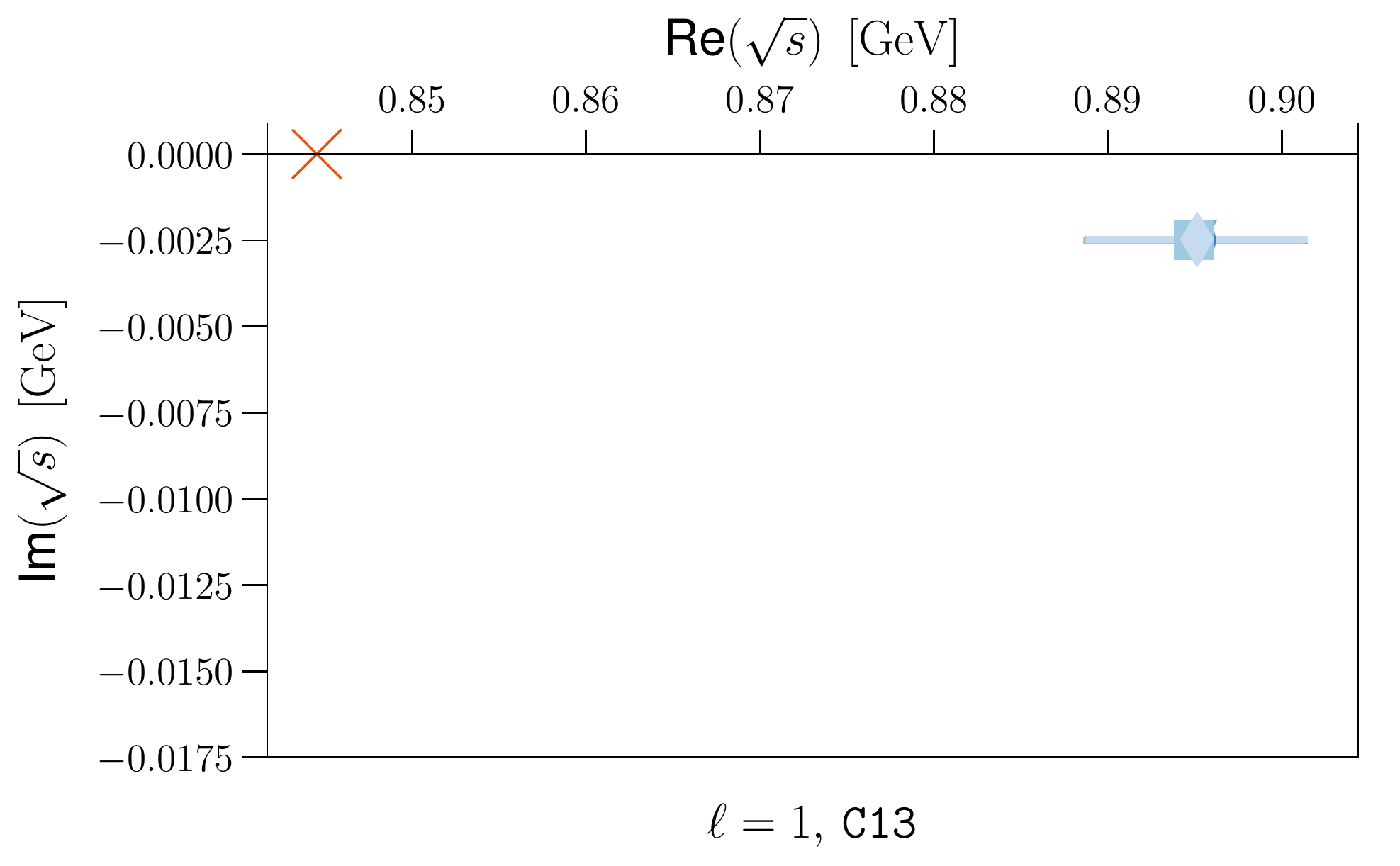}\includegraphics[width=0.5\linewidth]{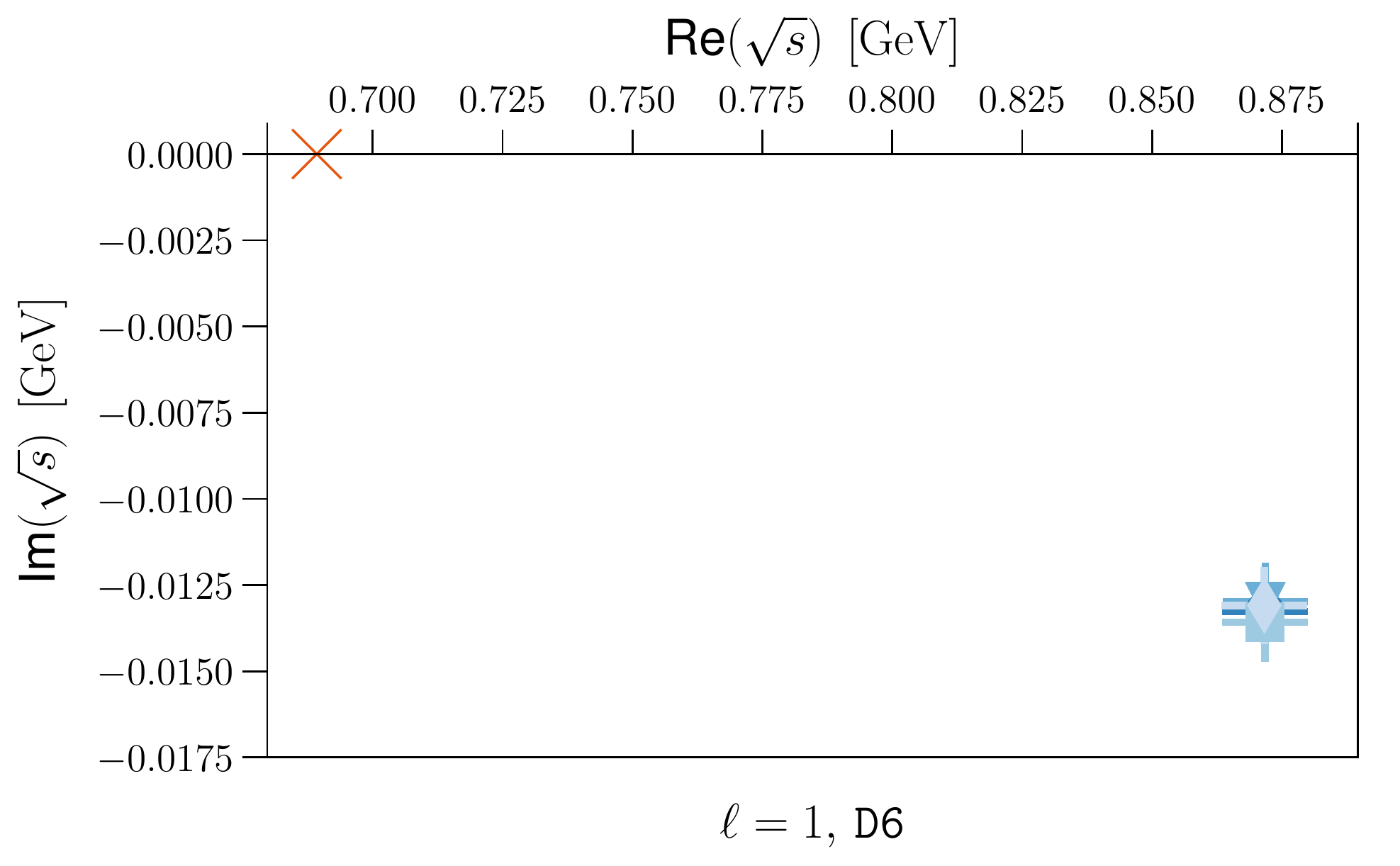}

\caption{ \Tmatrix pole positions for the $S$-wave (top) and $P$-wave (bottom). The plots on the left show the results from the \texttt{C13} ensemble [$m_\pi=317.2(2.2)~\text{MeV}$], while the plots on the right show the results from the \texttt{D6} [$m_\pi=175.9(1.8)~\text{MeV}$] ensemble. In each plot,
the four different data points correspond to four different parametrizations of the $S$-wave amplitude.
We can see a better stability of the $S$-wave pole position for the parametrizations with an Adler zero.}\label{fig:poles}
\end{figure*}

The phase-shift curves obtained from the four different fits and the two ensembles are presented in Fig.~\ref{fig:phase_shifts}. In addition,
we determined the positions of the closest $T$-matrix poles in the complex $\sqrt{s}$ plane, which are associated with the $\kappa$ and $K^*$ resonances. The pole positions are shown in Fig.~\ref{fig:poles} and are listed in Table \ref{tab:poles}. All poles are located on the second Riemann sheet. In the following, we discuss our observations separately for the $S$- and $P$-waves.

  \begin{table*}
  \begin{tabular}{|l|l|c|c|}
  \hline
  $S$-wave parametrization & Ensemble & $S$-wave \Tmatrix poles [GeV] & $P$-wave \Tmatrix poles [GeV]    \cr
  \hline
Conformal map & \texttt{C13} &  $0.86(12)- 0.309(50)\,\I$ & $0.8951(64)- 0.00250(21)\,\I $ \cr
                  & \texttt{D6}  &  $0.499(55)- 0.379(66)\,\I$  & $0.8718(82)- 0.0130(11)\,\I$  \cr
  \hline
  Bugg's parametrization & \texttt{C13} &  $0.850(65)- 0.315(40)\,\I$ & $0.8951(64)- 0.00250(21)\,\I$ \cr
                          & \texttt{D6}  &  $0.765(90)- 0.310(28)\,\I$ & $0.8717(82)- 0.0133(11)\,\I$  \cr
  \hline
  Effective-range expansion & \texttt{C13} &  $1.111(62)- 0.38(34)\,\I$     & $0.8951(64)- 0.00248(21)\,\I$ \cr
                            & \texttt{D6}  &  $0.23(67)- 0.42(37)\,\I$    & $0.8716(82)- 0.0131(11)\,\I$ \cr
  \hline
  Chung's parametrization & \texttt{C13} &  $1.14(10)- 0.176(89)\,\I $     & $0.8949(64) - 0.00250(21)\,\I$ \cr
                           & \texttt{D6}  &  $1.37(27)- 0.39(19)\,\I$  &  $0.8718(82)- 0.0136(11)\,\I$ \cr
  \hline
  \end{tabular}
\caption{Pole positions the $S$-wave and $P$-wave scattering amplitudes on the \texttt{C13} [$m_\pi=317.2(2.2)~\text{MeV}$] and \texttt{D6} [$m_\pi=175.9(1.8)~\text{MeV}$] ensembles.} \label{tab:poles}
\end{table*}

\subsection{\texorpdfstring{$\bm{S}$-wave scattering}{}}

The $S$-wave phase-shift curves from the four different parametrizations are in reasonable agreement with each other, given the uncertainties. We observe that the phase shifts remain below $80^{\circ}$ in the energy region considered. Even though there is little model dependence in the phase-shift curves for real-valued $\sqrt{s}$, the positions of the resulting poles of the scattering amplitude vary widely between the different parametrizations. Moreover, some of the parametrizations
lead to a much stronger dependence on the pion mass than others:

\begin{itemize}
  \item The ERE parametrization [Eq.~(\ref{eq:ERE})] yields a pole at $[1.11(6)-0.38(34)\,\I]$ GeV on the \texttt{C13} ensemble and at $[0.33(23)- 0.35(22)\,\I] $ GeV on the \texttt{D6} ensemble. This change is significantly larger than expected from the relatively minor change in the quark masses. Note that the ERE is meant to describe the behavior near the threshold and likely becomes unreliable in the upper range of our energy region.

  \item  The poles for Chung's $K$-matrix parametrization without an Adler zero [Eq.~(\ref{eq:Chung})] appear above $1.1$ GeV -- in the region where we do not have  data points to fully constrain the amplitude -- with only mild dependence on the quark masses. What may be happening is that this fit is sensing the $K_0^*(1430)$ resonance located not that far away \cite{Tanabashi:2018oca}.
  \item For the conformal-map-based parametrization [Eq.~(\ref{eq:generalconformal})] and Bugg's parametrization [Eq.~(\ref{eq:Bugg})], which both include an Adler zero,
  the poles appear near $[0.7- 0.3\,\I]$ GeV consistently for the two pion masses, and consistently for the two parametrizations. On the lower-pion-mass ensemble in particular, these parametrizations also yield much smaller statistical uncertainties for the pole locations.
\end{itemize}

In summary, the poles of the $S$-wave amplitude are significantly more stable for parametrizations incorporating an Adler zero. Because the conformal-map parametrization describes the data on the lower-pion-mass ensemble (\texttt{D6}) better than Bugg's parametrization, we choose the conformal-map parametrization as our nominal parametrization.

\begin{figure*}
  \includegraphics[width=0.6\linewidth]{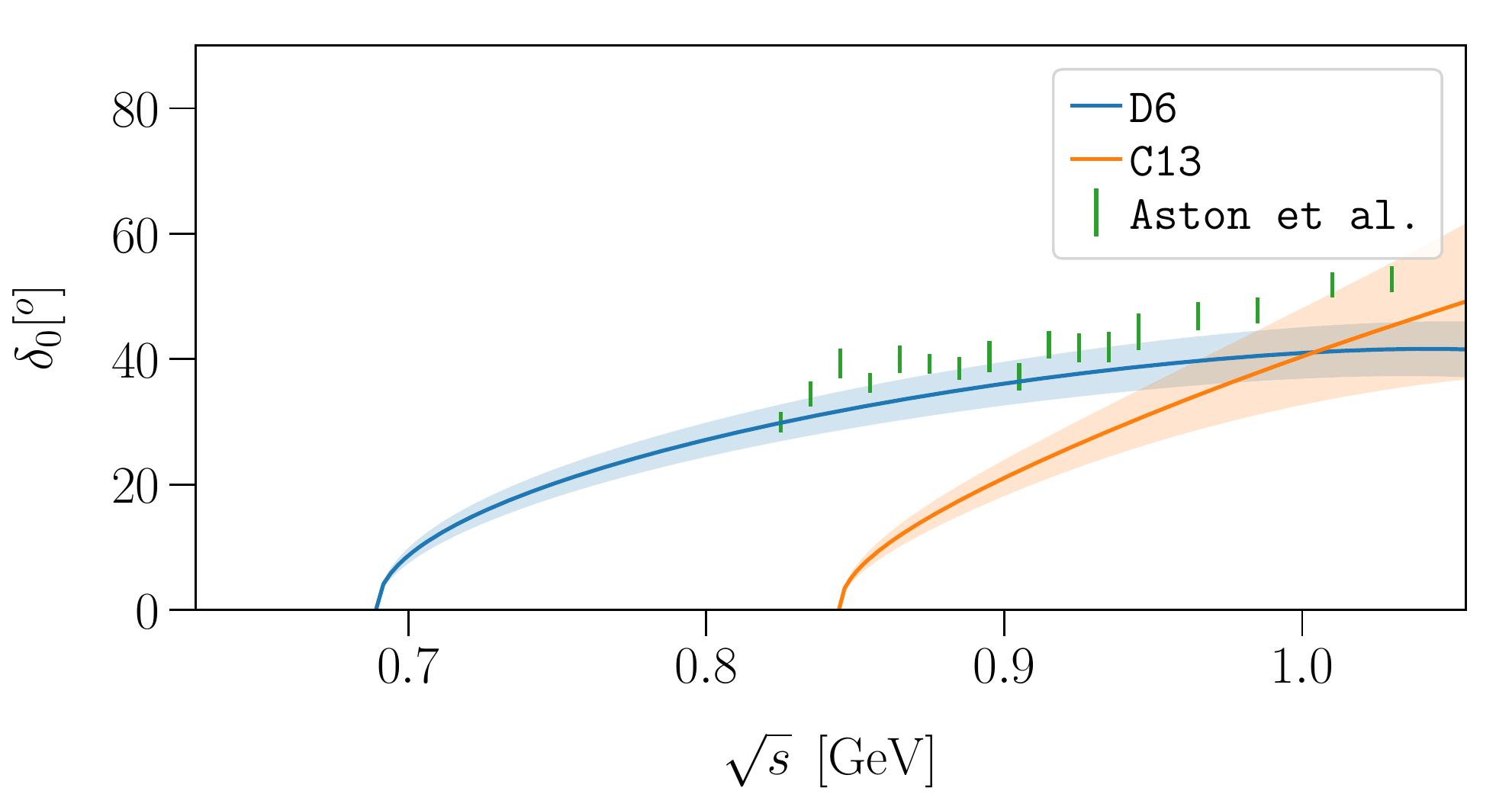}
  
  \includegraphics[width=0.61\linewidth]{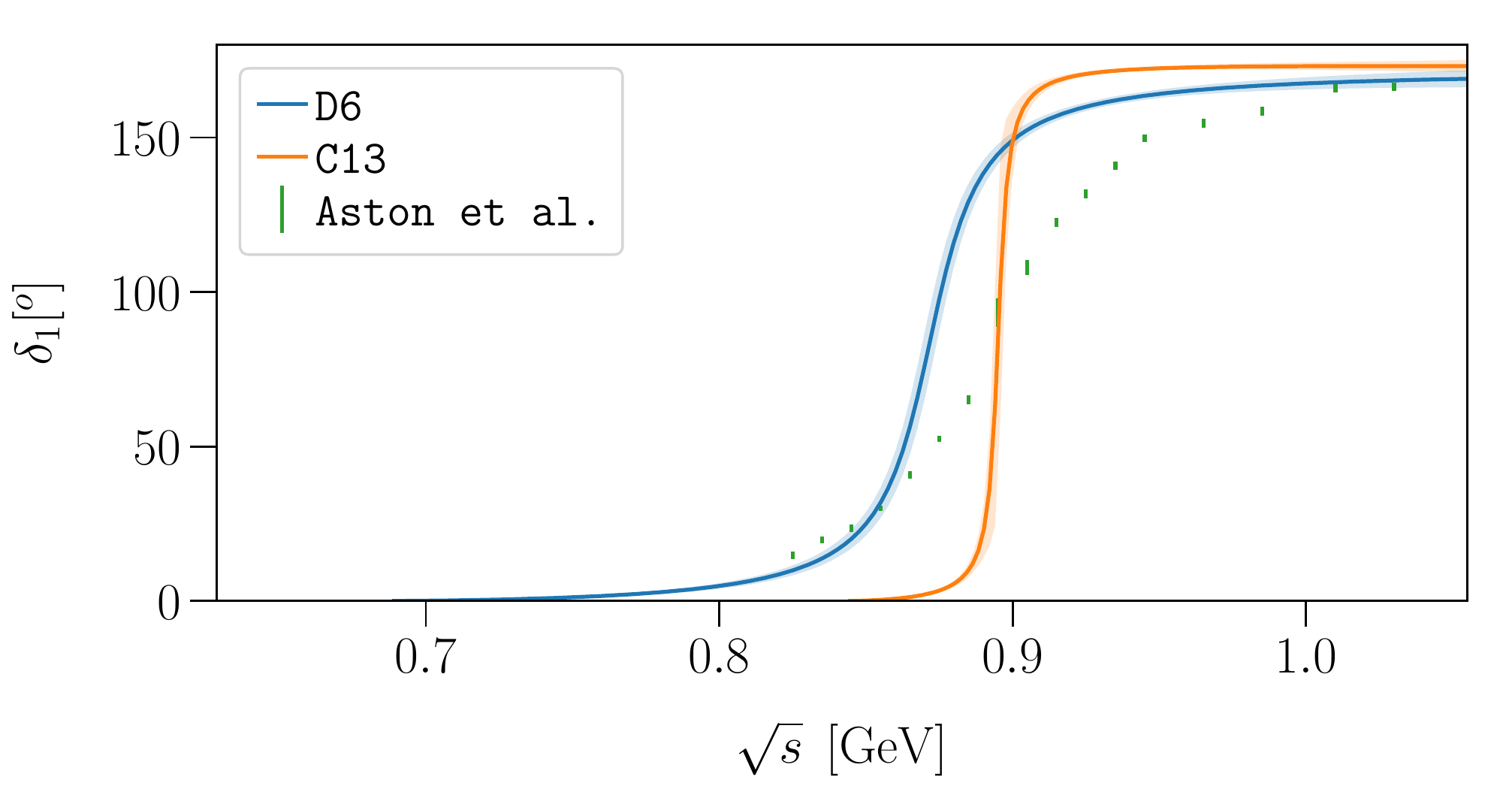}
  
  \caption{Phase shift results, using the conformal-map parametrization for the $S$-wave and Chung's parametrization for the $P$-wave, compared to experimental results from Ref.~\cite{Aston:1987ir}. The fitted parameters with the corresponding covariances used to produce these plots are included in the ancillary files as \texttt{D6\_fit\_parameters.dat} and \texttt{C13\_fit\_parameters.dat}.}\label{fig:nominal_phase_shifts}
\end{figure*}

In Fig~\ref{fig:nominal_phase_shifts}, we compare the phase-shift curves from this parametrization to experimental results from Ref.~\cite{Aston:1987ir}. We see that the $S$-wave phase-shift curves approach the experimental data as the pion mass is lowered toward its physical value. The pion mass of the \texttt{D6} ensemble is nearly physical, and the resulting curve is very close to the data. The behavior of the curve as a function of the pion mass seen here is also consistent with the lattice results in Ref.~\cite{Wilson:2019wfr} at a higher pion mass.

\subsection{\texorpdfstring{$\bm{P}$-wave scattering}{}}

The results presented in Figs.~\ref{fig:phase_shifts}, \ref{fig:poles} and Table \ref{tab:poles} were all obtained with Chung's \Kmatrix parametrization, Eq.~(\ref{eq:Chung}), for the $P$-wave. While we explored other $P$-wave parametrizations that do not include Blatt-Weisskopf barrier factors, we found little variation (an explicit comparison can be found for $I=1$ $\pi\pi$ scattering in Ref.~\cite{Alexandrou:2017mpi}, which also showed no significant variation). Furthermore, the $P$-wave phase-shift curves and pole locations do not significantly depend on the choice of the $S$-wave parametrization, which confirms that partial-wave mixing between $\ell=0$ and $\ell=1$ is under good control. We choose the same combination of parametrizations as above (Chung's parametrization for the $P$-wave combined with the conformal map for the $S$ wave) for our nominal results for the $P$ wave.

\begin{figure*}
  \includegraphics[width=0.9\textwidth]{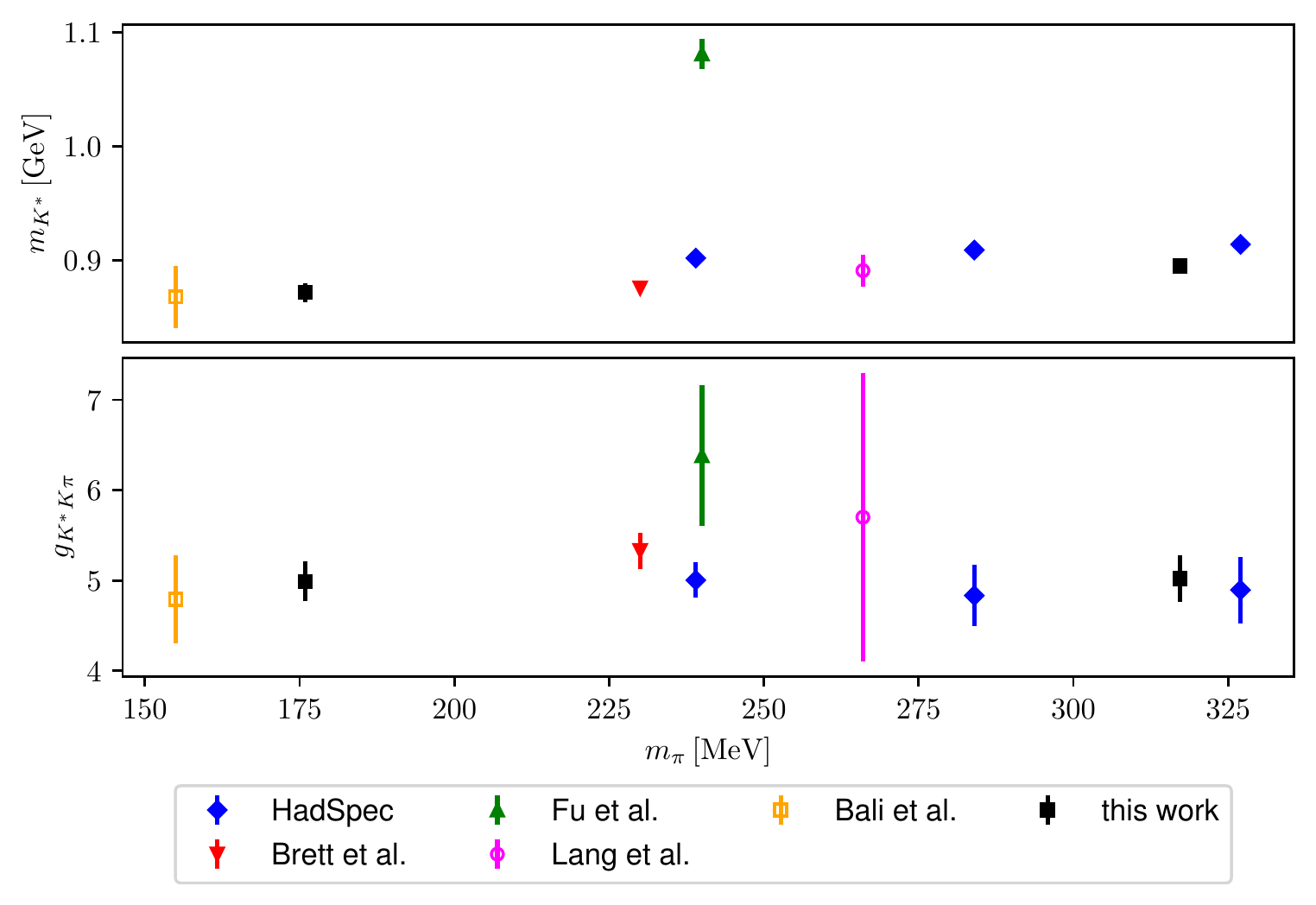}
  \caption{Comparison between different lattice-QCD results for the $K^*$ mass and its coupling to the $K\pi$ channel, plotted as a function of the pion mass. The references are: HadSpec \cite{Wilson:2019wfr}, Fu {\it et al.}~\cite{Fu:2012tj}, Bali {\it et al.}~\cite{Bali:2015gji}, Brett  {\it et al.}~\cite{Brett:2018jqw}, and Lang  {\it et al.}~\cite{Lang:2012sv}. Open symbols indicate $N_f=2$ results and the filled symbols indicate $N_f=2+1$ results.
}\label{fig:comparison_Kast}
\end{figure*}

A comparison of the phase-shift curves for the two different pion masses with experimental data \cite{Aston:1987ir} is shown in Fig.~\ref{fig:nominal_phase_shifts}. A clear resonance shape is observed for both pion masses. In this channel, the resonance width $\Gamma_{K^*\to K \pi} = -2\, \mathrm{Im}(\sqrt{s}_R)$, where $\sqrt{s}_R$ is the location of the pole, depends strongly on the pion mass due to the large change in available phase space. We find
\begin{align}
  \Gamma_{K^*\to K \pi}^{\texttt{C13}} &= (4.99 \pm 0.41)\:\:\text{MeV} , \\
  \Gamma_{K^*\to K \pi}^{\texttt{D6}} &= (26.0 \pm 2.2)\:\:\text{MeV}, 
\end{align}
while the value in nature is $50.8(0.9)$ MeV \cite{Tanabashi:2018oca}. Consequently, even at the close-to-physical pion mass of the \texttt{D6} ensemble, the phase shift curve is still noticeably steeper than in nature. In this situation, it is more appropriate to consider the $K^* K \pi$ coupling $g_{K^*K\pi}$, which can be obtained from the decay width through
\begin{align}
  \Gamma_{K^*\to K \pi} &= \frac{g_{K^*K\pi}^2}{6\pi}\frac{k_*^3}{
    \mathrm{Re}(\sqrt{s}_R)^2
  },
  \label{eq:decay_width_formula}
\end{align}
where $k_*$ is the scattering momentum for $\sqrt{s}=\mathrm{Re}(\sqrt{s}_R)$. This gives
\begin{align}
  g_{K^*K\pi}^{\texttt{C13}} &= 5.02(26) , \\
  g_{K^*K\pi}^{\texttt{D6}} &= 4.99(22) .
\end{align}
These values are consistent with each other, and also consistent with similar calculations \cite{Bali:2015gji,Wilson:2019wfr}. Our results are slightly below the experimental value of $g_{K^*K\pi}=5.603(4)$. A comparison of our results for $m_{K^*}$ and $g_{K^*K\pi}$ with previous lattice results is shown in Fig.~\ref{fig:comparison_Kast}.

\section{Conclusions}\label{sec_summary}

We have obtained precise results for the $I=1/2$ $S$- and $P$-wave $K\pi$ scattering phase shifts as functions of the center-of-mass energy up to 1.1 GeV, for
quark masses corresponding to $m_\pi\approx 176$ MeV and $m_\pi\approx 317$ MeV. We also determined the positions of the closest poles in the scattering amplitudes,
which we identify with the $K_0^*(700)$ (also referred to as $\kappa$) and $K^*(892)$ resonances.

For the $S$-wave amplitude, we investigated several different parametrizations proposed in the literature, some including an Adler zero and some without it. All parametrizations considered, including the effective-range expansion that is similar to the widely used LASS parametrization \cite{Aston:1987ir}, describe the phase shifts well for real $\sqrt{s}$ in the energy region considered. However, we found that the pole positions are stable only for those parametrizations that include an Adler zero. Using a conformal-map-based parametrization with an Adler zero, we found the poles in the $S$-wave scattering amplitude at $\left[0.86(12) - 0.309(50)\,\I\right]\:{\rm GeV}$ for $m_\pi\approx 317$ MeV and $\left[0.499(55) - 0.379(66)\,\I\right]\:{\rm GeV}$ for $m_\pi\approx 176$ MeV. Despite the unphysical pion masses and the lack of continuum extrapolations, these results are consistent with the $\kappa$ pole position extracted from experiments as reported by the Particle Data Group \cite{Tanabashi:2018oca}.

Earlier lattice calculations at a heavier pion mass of $m_\pi\approx 390$ MeV performed by the Hadron Spectrum Collaboration \cite{Wilson:2014cna,Dudek:2014qha} found the $\kappa$ as a bound state. More recently, the same collaboration reported results for a wider range of pion masses down to approximately $200$ MeV \cite{Wilson:2019wfr}. Investigating a large number of parametrizations, the authors did not find a sufficiently unique result to report numerically. For the parametrizations inspired by unitarized chiral perturbation theory, they did, however, find a $\kappa$ pole with a real part near the $K\pi$ threshold and a large imaginary part. This is consistent with our findings for parametrizations that include the Adler zero.

In the vector channel, our results for the $K^*$ mass and width have high statistical precision. Since the $K^*$ width depends strongly on the pion mass through kinematic effects, it is more appropriate to consider the $K^*K\pi$ coupling. Our results for this coupling and for the $K^*$ mass
are compared with previous lattice results \cite{Lang:2012sv,Fu:2012tj,Wilson:2014cna,Bali:2015gji,Brett:2018jqw,Wilson:2019wfr} in Fig.~\ref{fig:comparison_Kast}. Note that the calculations were performed with different numbers of flavors, different gluon and fermion discretizations, and with different procedures to set the lattice scale; none of the calculations included a continuum extrapolation. Keeping in mind these caveats, we note that our results for both $g_{K^*K\pi}$ and $m_{K^*}$ agree well with previous calculations, except for the higher mass obtained by Fu et al. using a staggered fermion action \cite{Fu:2012tj}. Apart from this outlier,
the results for $m_{K^*}$ show only very mild quark-mass dependence, while $g_{K^*K\pi}$ has no discernible quark-mass dependence, similar to $g_{\rho\pi\pi}$ \cite{Alexandrou:2017mpi}. Furthermore, the results from $N_f=2+1$ and $N_f=2$ ensembles appear to follow common staight lines.

The calculations performed here can also be used in future lattice determinations of $1\to 2$ transition matrix elements of external currents with the same $K\pi$ states. The scattering amplitudes are needed to map the finite-volume matrix elements to infinite-volume matrix elements via the formalism of Ref.~\cite{Briceno:2014uqa}, as has already been done for $\pi\gamma^* \to \pi\pi$ \cite{Briceno:2015dca, Alexandrou:2018jbt}. Such a calculation will be particularly important for rare $B\to K \pi \ell^+\ell^-$ decays \cite{Aaij:2013iag,Aaij:2013qta,Descotes-Genon:2013wba,Horgan:2013hoa,Horgan:2013pva,Aaij:2017vbb,Aebischer:2019mlg,Alguero:2019ptt,Aaij:2020nrf}.

\begin{acknowledgments}
We thank Kostas Orginos, Balint Joó, Robert Edwards, and their collaborators for providing the gauge-field configurations. Computations for this work were carried out in part on (1) facilities of the USQCD Collaboration, which are funded by the Office of Science of the U.S. Department
of Energy, (2) facilities of the Leibniz Supercomputing Centre, which is funded by the Gauss Centre for Supercomputing,  (3) facilities at the National Energy Research Scientific Computing Center, a DOE Office of Science User Facility supported by the Office of Science of the U.S. Department of Energy under Contract No. DE-AC02-05CH1123, (4) facilities of the Extreme Science and Engineering Discovery Environment (XSEDE) \cite{XSEDE}, which is supported by National Science Foundation grant number ACI-1548562, and (5) the Oak Ridge Leadership Computing Facility, which is a DOE Office of Science User Facility supported under Contract DE-AC05-00OR22725.

SM is supported by the U.S. Department of Energy, Office of Science, Office of High Energy Physics under Award Number D{E-S}{C0}009913.
MP gratefully acknowledges support by the Sino-German collaborative research center CRC-110.
JN, and AP are supported in part by the U.S. Department of Energy, Office of Nuclear Physics under grants DE-SC0011090 and DE-SC0018121, respectively.
GR is supported by the U.S. Department of Energy, Office of Science, Office of Nuclear Physics, under Contract No. DE-SC0012704 (BNL).
LL acknowledges support from the U.S. Department of Energy, Office of Science, through contracts DE-SC0019229 and DE-AC05-06OR23177 (JLAB).
SS is supported by the National Science Foundation under CAREER Award PHY-1847893 and by the RHIC Physics Fellow Program of the RIKEN BNL Research Center.
\end{acknowledgments}

\appendix
\section{Fit parameters of scattering amplitudes}

The fit parameters for the scattering amplitudes are presented in Tables \ref{tab:C13fits} and \ref{tab:D6fits}.

\begin{table*}[h]
  \begin{tabularx}{0.6\textwidth}{|X|X|c|}
  \hline
   $S$-wave parametrization & Fit parameters & $\chi^2/{\rm dof}$  \\
   \hline
  Conformal map &
  $g_1^0 a = 0.0682 \pm 0.0025$ \newline
  $(m_1 a)^2 = 0.2676 \pm 0.0012$ \newline
  $R_1 a^{-1} = 0 \pm 170$ \newline
  $B_{0} a^{2} = 0.174 \pm 0.030$ \newline
  $B_{1} a^{2} = -0.05 \pm 0.15$ & $0.764$ \\
   \hline
  Bugg's parametrization & 
  $g_1^0 a = 0.0683 \pm 0.0025$  \newline
  $(m_1 a)^2 = 0.2677 \pm 0.0012$ \newline
  $R_1 a^{-1} = 0 \pm 17$ \newline
  $G_0^0 a = 4.5 \pm 8.6$ \newline
  $(m_0 a)^2 = 2.1 \pm 3.6$ & $0.761$ \\
  \hline
  Effective-range expansion & 
  $g_1^0 a = 0.0680 \pm 0.0025$ \newline
  $(m_1 a)^2 = 0.2677 \pm 0.0012$ \newline
  $R_1 a^{-1} = 1 \pm 7.0$ \newline
  $c_{0} a = 0.248 \pm 0.039$ \newline
  $c_{1} a^{2} = -3.3 \pm 2.5$ & $0.773$ \\
  \hline
  Chung's parametrization &  
  $g_1^0 a = 0.0684 \pm 0.0025$ \newline
  $(m_1 a)^2 = 0.2676 \pm 0.0012$ \newline
  $R_1 a^{-1} = 0 \pm 20$ \newline
  $g_0^0 a = 0.44 \pm 0.10$ \newline
  $(m_0 a)^2 = 0.448 \pm 0.081$ & $0.753$ \\
  \hline
  \end{tabularx}
  \caption{Fit results for the \Kmatrix parameters from the \texttt{C13} ensemble with $m_\pi\approx 317~\text{MeV}$. The type of parametrization for the $P$-wave is always the same (Chung's parametrization), while the parametrization for the $S$-wave changes and is given in the leftmost column. Here, $a$ denotes the lattice spacing.} \label{tab:C13fits}
  \end{table*}

  \begin{table*}[h]
  \begin{tabularx}{0.6\textwidth}{|X|X|c|}
  \hline
   $S$-wave parametrization & Fit parameters & $\chi^2/{\rm dof}$  \\
  \hline
  Conformal map  &
  $g_1^0 a = 0.0894 \pm 0.0037$ \newline
  $(m_1 a)^2 = 0.15046\pm 0.00086$ \newline
  $R_1 a^{-1} = 0 \pm 23$ \newline
  $B_{0} a^{2} = 0.086 \pm 0.011$\newline
  $B_{1} a^{2} = 0.106 \pm 0.025$ & $0.958$ \\
  \hline 
  Bugg's parametrization & 
  $g_1^0 a = 0.0904 \pm 0.0037$ \newline 
  $(m_1 a)^2 = 0.15045 \pm 0.00086$ \newline
  $R_1 a^{-1} = 0 \pm 280$ \newline
  $G_0^0 a = 20 \pm 310$ \newline
  $(m_0 a)^2 = 5 \pm 88$ & $1.44$ \\
  \hline
  Effective-range expansion &  
  $g_1^0 a = 0.0898 \pm 0.0037$ \newline
  $(m_1 a)^2 = 0.15040 \pm 0.00086$ \newline
  $R_1 a^{-1} = 1.1 \pm 9.7$ \newline
  $c_{0} a = 0.173 \pm 0.030$ \newline
  $c_{1} a^{2} = -0.7 \pm 2.0$ & $0.926$ \\
  \hline
  Chung's parametrization & 
  $g_1^0 a = 0.0915 \pm 0.0037$ \newline
  $(m_1 a)^2 = 0.15048 \pm 0.00086$ \newline
  $R_1 a^{-1} = 1.5 \pm 7.0$ \newline
  $g_0^0 a  = 0.49 \pm 0.15$ \newline
  $(m_0 a)^2 = 0.36 \pm 0.12$ & $0.875$ \\
  \hline
  \end{tabularx}
  \caption{Like \cref{tab:C13fits}, but for the \texttt{D6} ensemble with $m_\pi\approx 176~\text{MeV}$.}\label{tab:D6fits}
  \end{table*}

  \FloatBarrier

\bibliographystyle{utphys-noitalics}

\begin{thebibliography}{100}

\bibitem{Lang:1978fk}
C.~B. Lang, ``{The $\pi K$ Scattering and Related Processes},''
\href{http://dx.doi.org/10.1002/prop.19780261002}{Fortsch. Phys. {\bfseries 26}
  (1978) 509}.

\bibitem{Bevan:2014iga}
{\bfseries BaBar, Belle} Collaboration, A.~Bevan {\em et~al.}, ``{The Physics
  of the $B$ Factories},''
  \href{http://dx.doi.org/10.1140/epjc/s10052-014-3026-9}{Eur. Phys. J. C
  {\bfseries 74} (2014) 3026}, \href{http://arxiv.org/abs/1406.6311}{{\ttfamily
  arXiv:1406.6311 [hep-ex]}}.

\bibitem{Kou:2018nap}
{\bfseries Belle-II} Collaboration, W.~Altmannshofer {\em et~al.}, ``{The Belle
  II Physics Book},'' \href{http://dx.doi.org/10.1093/ptep/ptz106}{PTEP
  {\bfseries 2019} no.~12, (2019) 123C01},
  \href{http://arxiv.org/abs/1808.10567}{{\ttfamily arXiv:1808.10567
  [hep-ex]}}. [Erratum: PTEP 2020, 029201 (2020)].

\bibitem{Bediaga:2018lhg}
{\bfseries LHCb} Collaboration, R.~Aaij {\em et~al.}, ``{Physics case for an
  LHCb Upgrade II - Opportunities in flavour physics, and beyond, in the HL-LHC
  era},'' \href{http://arxiv.org/abs/1808.08865}{{\ttfamily arXiv:1808.08865
  [hep-ex]}}.

\bibitem{Asner:2008nq}
D.~Asner {\em et~al.}, ``{Physics at BES-III},'' Int. J. Mod. Phys. A
  {\bfseries 24} (2009) S1--794,
  \href{http://arxiv.org/abs/0809.1869}{{\ttfamily arXiv:0809.1869 [hep-ex]}}.

\bibitem{Ablikim:2019hff}
M.~Ablikim {\em et~al.}, ``{Future Physics Programme of BESIII},''
  \href{http://dx.doi.org/10.1088/1674-1137/44/4/040001}{Chin. Phys. C
  {\bfseries 44} no.~4, (2020) 040001},
  \href{http://arxiv.org/abs/1912.05983}{{\ttfamily arXiv:1912.05983
  [hep-ex]}}.

\bibitem{Aaij:2014iva}
{\bfseries LHCb} Collaboration, R.~Aaij {\em et~al.}, ``{Measurements of $CP$
  violation in the three-body phase space of charmless $B^{\pm}$ decays},''
  \href{http://dx.doi.org/10.1103/PhysRevD.90.112004}{Phys. Rev. {\bfseries
  D90} no.~11, (2014) 112004},
\href{http://arxiv.org/abs/1408.5373}{{\ttfamily arXiv:1408.5373 [hep-ex]}}.

\bibitem{Aaij:2013iag}
{\bfseries LHCb} Collaboration, R.~Aaij {\em et~al.}, ``{Differential branching
  fraction and angular analysis of the decay $B^{0} \to K^{*0}
  \mu^{+}\mu^{-}$},'' \href{http://dx.doi.org/10.1007/JHEP08(2013)131}{JHEP
  {\bfseries 08} (2013) 131}, \href{http://arxiv.org/abs/1304.6325}{{\ttfamily
  arXiv:1304.6325 [hep-ex]}}.

\bibitem{Aaij:2013qta}
{\bfseries LHCb} Collaboration, R.~Aaij {\em et~al.}, ``{Measurement of
  Form-Factor-Independent Observables in the Decay $B^{0} \to K^{*0} \mu^+
  \mu^-$},''
  \href{http://dx.doi.org/10.1103/PhysRevLett.111.191801}{Phys.Rev.Lett.
  {\bfseries 111} (2013) 191801},
\href{http://arxiv.org/abs/1308.1707}{{\ttfamily arXiv:1308.1707 [hep-ex]}}.

\bibitem{Descotes-Genon:2013wba}
S.~Descotes-Genon, J.~Matias, and J.~Virto, ``{Understanding the $B \to
  K^*\mu^+\mu^-$ Anomaly},''
  \href{http://dx.doi.org/10.1103/PhysRevD.88.074002}{Phys.Rev. {\bfseries D88}
  (2013) 074002},
\href{http://arxiv.org/abs/1307.5683}{{\ttfamily arXiv:1307.5683 [hep-ph]}}.

\bibitem{Horgan:2013pva}
R.~R. Horgan, Z.~Liu, S.~Meinel, and M.~Wingate, ``{Calculation of $B^0 \to
  K^{*0} \mu^+ \mu^-$ and $B_s^0 \to \phi \mu^+ \mu^-$ observables using form
  factors from lattice QCD},''
  \href{http://dx.doi.org/10.1103/PhysRevLett.112.212003}{Phys. Rev. Lett.
  {\bfseries 112} (2014) 212003},
\href{http://arxiv.org/abs/1310.3887}{{\ttfamily arXiv:1310.3887 [hep-ph]}}.

\bibitem{Aaij:2017vbb}
{\bfseries LHCb} Collaboration, R.~Aaij {\em et~al.}, ``{Test of lepton
  universality with $B^{0} \rightarrow K^{*0}\ell^{+}\ell^{-}$ decays},''
  \href{http://dx.doi.org/10.1007/JHEP08(2017)055}{JHEP {\bfseries 08} (2017)
  055},
\href{http://arxiv.org/abs/1705.05802}{{\ttfamily arXiv:1705.05802 [hep-ex]}}.

\bibitem{Aebischer:2019mlg}
J.~Aebischer, W.~Altmannshofer, D.~Guadagnoli, M.~Reboud, P.~Stangl, and D.~M.
  Straub, ``{B-decay discrepancies after Moriond 2019},''
  \href{http://dx.doi.org/10.1140/epjc/s10052-020-7817-x}{Eur. Phys. J. C
  {\bfseries 80} no.~3, (2020) 252},
  \href{http://arxiv.org/abs/1903.10434}{{\ttfamily arXiv:1903.10434
  [hep-ph]}}.

\bibitem{Alguero:2019ptt}
M.~Algueró, B.~Capdevila, A.~Crivellin, S.~Descotes-Genon, P.~Masjuan,
  J.~Matias, M.~Novoa, and J.~Virto, ``{Emerging patterns of New Physics with
  and without Lepton Flavour Universal contributions},''
  \href{http://dx.doi.org/10.1140/epjc/s10052-019-7216-3}{Eur. Phys. J. C
  {\bfseries 79} no.~8, (2019) 714},
  \href{http://arxiv.org/abs/1903.09578}{{\ttfamily arXiv:1903.09578
  [hep-ph]}}.

\bibitem{Aaij:2020nrf}
{\bfseries LHCb} Collaboration, R.~Aaij {\em et~al.}, ``{Measurement of
  $C\!P$-averaged observables in the $B^{0}\rightarrow K^{*0}\mu^{+}\mu^{-}$
  decay},'' \href{http://arxiv.org/abs/2003.04831}{{\ttfamily arXiv:2003.04831
  [hep-ex]}}.

\bibitem{Estabrooks:1977xe}
P.~Estabrooks, R.~K. Carnegie, A.~D. Martin, W.~M. Dunwoodie, T.~A. Lasinski,
  and D.~W. G.~S. Leith, ``{Study of $K \pi$ Scattering Using the Reactions
  $K^{+-} p \to K^{+-} \pi^{+} n$ and $K^{+-} p \to K^{+-} \pi^{-} \Delta^{++}$
  at 13 GeV/c},''
\href{http://dx.doi.org/10.1016/0550-3213(78)90238-9}{Nucl. Phys. {\bfseries
  B133} (1978) 490--524}.

\bibitem{Aston:1987ir}
D.~Aston {\em et~al.}, ``{A Study of $K^- \pi^+$ Scattering in the Reaction
  $K^- p \to K^- \pi^+ n$ at 11 GeV/c},''
\href{http://dx.doi.org/10.1016/0550-3213(88)90028-4}{Nucl. Phys. {\bfseries
  B296} (1988) 493--526}.

\bibitem{Adeva:2014xtx}
{\bfseries DIRAC} Collaboration, B.~Adeva {\em et~al.}, ``{First πK atom
  lifetime and πK scattering length measurements},''
  \href{http://dx.doi.org/10.1016/j.physletb.2014.06.043}{Phys. Lett.
  {\bfseries B735} (2014) 288--294},
\href{http://arxiv.org/abs/1403.0845}{{\ttfamily arXiv:1403.0845 [nucl-ex]}}.

\bibitem{Gorchakov:2016thm}
O.~E. Gorchakov and L.~L. Nemenov, ``{The estimation of production rates of
  ${\pi }^{+}{K}^{-}$, ${\pi }^{-}{K}^{+}$ and ${\pi }^{+}{\pi }^{-}$ atoms in
  proton–nucleus interactions at 450 ${\rm GeV}c^{-1}$},''
\href{http://dx.doi.org/10.1088/0954-3899/43/9/095004}{J. Phys. {\bfseries G43}
  no.~9, (2016) 095004}.

\bibitem{Amaryan:2017ldw}
{\bfseries GlueX} Collaboration, S.~Adhikari {\em et~al.}, ``{Strange Hadron
  Spectroscopy with a Secondary $K_L$ Beam at GlueX},''
\href{http://arxiv.org/abs/1707.05284}{{\ttfamily arXiv:1707.05284 [hep-ex]}}.

\bibitem{Cherry:2000ut}
S.~N. Cherry and M.~R. Pennington, ``{There is no $\kappa(900)$},''
  \href{http://dx.doi.org/10.1016/S0375-9474(00)00587-X}{Nucl. Phys. {\bfseries
  A688} (2001) 823--841},
\href{http://arxiv.org/abs/hep-ph/0005208}{{\ttfamily arXiv:hep-ph/0005208
  [hep-ph]}}.

\bibitem{Tanabashi:2018oca}
{\bfseries Particle Data Group} Collaboration, M.~Tanabashi {\em et~al.},
  ``{Review of Particle Physics},''
  \href{http://dx.doi.org/10.1103/PhysRevD.98.030001}{Phys.\ Rev.\ D {\bfseries
  98} no.~3, (2018) 030001}.

\bibitem{Gasser:1984gg}
J.~Gasser and H.~Leutwyler, ``{Chiral Perturbation Theory: Expansions in the
  Mass of the Strange Quark},''
\href{http://dx.doi.org/10.1016/0550-3213(85)90492-4}{Nucl. Phys. {\bfseries
  B250} (1985) 465--516}.

\bibitem{Bernard:1990kw}
V.~Bernard, N.~Kaiser, and U.~G. Meissner, ``{$\pi K$ scattering in chiral
  perturbation theory to one loop},''
\href{http://dx.doi.org/10.1016/0550-3213(91)90461-6}{Nucl. Phys. {\bfseries
  B357} (1991) 129--152}.

\bibitem{Ishida:1997wn}
S.~Ishida, M.~Ishida, T.~Ishida, K.~Takamatsu, and T.~Tsuru, ``{Analysis of $K
  \pi$ scattering phase shift and existence of $\kappa (900)$ particle},''
  \href{http://dx.doi.org/10.1143/PTP.98.621}{Prog. Theor. Phys. {\bfseries 98}
  (1997) 621--629},
\href{http://arxiv.org/abs/hep-ph/9705437}{{\ttfamily arXiv:hep-ph/9705437
  [hep-ph]}}.

\bibitem{Oller:1997ng}
J.~A. Oller, E.~Oset, and J.~R. Pelaez, ``{Nonperturbative approach to
  effective chiral Lagrangians and meson interactions},''
  \href{http://dx.doi.org/10.1103/PhysRevLett.80.3452}{Phys. Rev. Lett.
  {\bfseries 80} (1998) 3452--3455},
\href{http://arxiv.org/abs/hep-ph/9803242}{{\ttfamily arXiv:hep-ph/9803242
  [hep-ph]}}.

\bibitem{Oller:1998hw}
J.~A. Oller, E.~Oset, and J.~R. Pelaez, ``{Meson meson interaction in a
  nonperturbative chiral approach},''
  \href{http://dx.doi.org/10.1103/PhysRevD.59.074001,
  10.1103/PhysRevD.60.099906, 10.1103/PhysRevD.75.099903}{Phys. Rev. {\bfseries
  D59} (1999) 074001}, \href{http://arxiv.org/abs/hep-ph/9804209}{{\ttfamily
  arXiv:hep-ph/9804209 [hep-ph]}}.
[Erratum: Phys. Rev.D60,099906(1999); Erratum: Phys. Rev.D75,099903(2007)].

\bibitem{Oller:1998zr}
J.~A. Oller and E.~Oset, ``{N/D description of two meson amplitudes and chiral
  symmetry},'' \href{http://dx.doi.org/10.1103/PhysRevD.60.074023}{Phys. Rev.
  {\bfseries D60} (1999) 074023},
\href{http://arxiv.org/abs/hep-ph/9809337}{{\ttfamily arXiv:hep-ph/9809337
  [hep-ph]}}.

\bibitem{Black:1998zc}
D.~Black, A.~H. Fariborz, F.~Sannino, and J.~Schechter, ``{Evidence for a
  scalar $\kappa(900)$ resonance in $\pi K$ scattering},''
  \href{http://dx.doi.org/10.1103/PhysRevD.58.054012}{Phys. Rev. {\bfseries
  D58} (1998) 054012},
\href{http://arxiv.org/abs/hep-ph/9804273}{{\ttfamily arXiv:hep-ph/9804273
  [hep-ph]}}.

\bibitem{Roessl:1999iu}
A.~Roessl, ``{Pion kaon scattering near the threshold in chiral SU(2)
  perturbation theory},''
  \href{http://dx.doi.org/10.1016/S0550-3213(99)00336-3}{Nucl. Phys. {\bfseries
  B555} (1999) 507--539},
\href{http://arxiv.org/abs/hep-ph/9904230}{{\ttfamily arXiv:hep-ph/9904230
  [hep-ph]}}.

\bibitem{Jamin:2000wn}
M.~Jamin, J.~A. Oller, and A.~Pich, ``{$S$ wave $K \pi$ scattering in chiral
  perturbation theory with resonances},''
  \href{http://dx.doi.org/10.1016/S0550-3213(00)00479-X}{Nucl. Phys. {\bfseries
  B587} (2000) 331--362},
\href{http://arxiv.org/abs/hep-ph/0006045}{{\ttfamily arXiv:hep-ph/0006045
  [hep-ph]}}.

\bibitem{GomezNicola:2001as}
A.~Gomez~Nicola and J.~R. Pelaez, ``{Meson meson scattering within one loop
  chiral perturbation theory and its unitarization},''
  \href{http://dx.doi.org/10.1103/PhysRevD.65.054009}{Phys. Rev. {\bfseries
  D65} (2002) 054009},
\href{http://arxiv.org/abs/hep-ph/0109056}{{\ttfamily arXiv:hep-ph/0109056
  [hep-ph]}}.

\bibitem{Bijnens:2004bu}
J.~Bijnens, P.~Dhonte, and P.~Talavera, ``{$\pi K$ scattering in three flavor
  ChPT},'' \href{http://dx.doi.org/10.1088/1126-6708/2004/05/036}{JHEP
  {\bfseries 05} (2004) 036},
\href{http://arxiv.org/abs/hep-ph/0404150}{{\ttfamily arXiv:hep-ph/0404150
  [hep-ph]}}.

\bibitem{Nebreda:2010wv}
J.~Nebreda and J.~R. Pelaez., ``{Strange and non-strange quark mass dependence
  of elastic light resonances from SU(3) Unitarized Chiral Perturbation Theory
  to one loop},'' \href{http://dx.doi.org/10.1103/PhysRevD.81.054035}{Phys.
  Rev. {\bfseries D81} (2010) 054035},
\href{http://arxiv.org/abs/1001.5237}{{\ttfamily arXiv:1001.5237 [hep-ph]}}.

\bibitem{Guo:2011pa}
Z.-H. Guo and J.~A. Oller, ``{Resonances from meson-meson scattering in U(3)
  CHPT},'' \href{http://dx.doi.org/10.1103/PhysRevD.84.034005}{Phys. Rev.
  {\bfseries D84} (2011) 034005},
\href{http://arxiv.org/abs/1104.2849}{{\ttfamily arXiv:1104.2849 [hep-ph]}}.

\bibitem{Magalhaes:2011sh}
P.~C. Magalhaes, M.~R. Robilotta, K.~S. F.~F. Guimaraes, T.~Frederico,
  W.~de~Paula, I.~Bediaga, A.~C.~d. Reis, C.~M. Maekawa, and G.~R.~S.
  Zarnauskas, ``{Towards three-body unitarity in $D^+ \to K^- \pi^+ \pi^+$},''
  \href{http://dx.doi.org/10.1103/PhysRevD.84.094001}{Phys. Rev. {\bfseries
  D84} (2011) 094001},
\href{http://arxiv.org/abs/1105.5120}{{\ttfamily arXiv:1105.5120 [hep-ph]}}.

\bibitem{Wolkanowski:2015jtc}
T.~Wolkanowski, M.~Sołtysiak, and F.~Giacosa, ``{$K_{0}^{\ast}(800)$ as a
  companion pole of $K_{0}^{\ast}(1430)$},''
  \href{http://dx.doi.org/10.1016/j.nuclphysb.2016.05.025}{Nucl. Phys.
  {\bfseries B909} (2016) 418--428},
\href{http://arxiv.org/abs/1512.01071}{{\ttfamily arXiv:1512.01071 [hep-ph]}}.

\bibitem{vanBeveren:1986ea}
E.~van Beveren, T.~A. Rijken, K.~Metzger, C.~Dullemond, G.~Rupp, and J.~E.
  Ribeiro, ``{A Low Lying Scalar Meson Nonet in a Unitarized Meson Model},''
  \href{http://dx.doi.org/10.1007/BF01571811}{Z. Phys. {\bfseries C30} (1986)
  615--620},
\href{http://arxiv.org/abs/0710.4067}{{\ttfamily arXiv:0710.4067 [hep-ph]}}.

\bibitem{vanBeveren:2006ua}
E.~van Beveren, D.~V. Bugg, F.~Kleefeld, and G.~Rupp, ``{The Nature of
  $\sigma$, $\kappa$, $a_0(980)$ and $f_0(980)$},''
  \href{http://dx.doi.org/10.1016/j.physletb.2006.08.051}{Phys. Lett.
  {\bfseries B641} (2006) 265--271},
\href{http://arxiv.org/abs/hep-ph/0606022}{{\ttfamily arXiv:hep-ph/0606022
  [hep-ph]}}.

\bibitem{Pelaez:2004xp}
J.~R. Pelaez, ``{Light scalars as tetraquarks or two-meson states from large
  $N_c$ and unitarized chiral perturbation theory},''
  \href{http://dx.doi.org/10.1142/S0217732304016160}{Mod. Phys. Lett.
  {\bfseries A19} (2004) 2879--2894},
\href{http://arxiv.org/abs/hep-ph/0411107}{{\ttfamily arXiv:hep-ph/0411107
  [hep-ph]}}.

\bibitem{Ledwig:2014cla}
T.~Ledwig, J.~Nieves, A.~Pich, E.~Ruiz~Arriola, and J.~Ruiz~de Elvira,
  ``{Large-$N_c$ naturalness in coupled-channel meson-meson scattering},''
  \href{http://dx.doi.org/10.1103/PhysRevD.90.114020}{Phys. Rev. {\bfseries
  D90} no.~11, (2014) 114020},
\href{http://arxiv.org/abs/1407.3750}{{\ttfamily arXiv:1407.3750 [hep-ph]}}.

\bibitem{Dobado:1992ha}
A.~Dobado and J.~R. Pelaez, ``{A Global fit of $\pi \pi$ and $\pi K$ elastic
  scattering in ChPT with dispersion relations},''
  \href{http://dx.doi.org/10.1103/PhysRevD.47.4883}{Phys. Rev. {\bfseries D47}
  (1993) 4883--4888},
\href{http://arxiv.org/abs/hep-ph/9301276}{{\ttfamily arXiv:hep-ph/9301276
  [hep-ph]}}.

\bibitem{Dobado:1996ps}
A.~Dobado and J.~R. Pelaez, ``{The Inverse amplitude method in chiral
  perturbation theory},''
  \href{http://dx.doi.org/10.1103/PhysRevD.56.3057}{Phys. Rev. {\bfseries D56}
  (1997) 3057--3073},
\href{http://arxiv.org/abs/hep-ph/9604416}{{\ttfamily arXiv:hep-ph/9604416
  [hep-ph]}}.

\bibitem{Buettiker:2003pp}
P.~Buettiker, S.~Descotes-Genon, and B.~Moussallam, ``{A new analysis of $\pi
  K$ scattering from Roy and Steiner type equations},''
  \href{http://dx.doi.org/10.1140/epjc/s2004-01591-1}{Eur. Phys. J. {\bfseries
  C33} (2004) 409--432},
\href{http://arxiv.org/abs/hep-ph/0310283}{{\ttfamily arXiv:hep-ph/0310283
  [hep-ph]}}.

\bibitem{DescotesGenon:2006uk}
S.~Descotes-Genon and B.~Moussallam, ``{The $K^{*0}(800)$ scalar resonance from
  Roy-Steiner representations of $\pi K$ scattering},''
  \href{http://dx.doi.org/10.1140/epjc/s10052-006-0036-2}{Eur. Phys. J.
  {\bfseries C48} (2006) 553},
\href{http://arxiv.org/abs/hep-ph/0607133}{{\ttfamily arXiv:hep-ph/0607133
  [hep-ph]}}.

\bibitem{Zhou:2006wm}
Z.~Y. Zhou and H.~Q. Zheng, ``{An improved study of the kappa resonance and the
  non-exotic $s$ wave $\pi K$ scatterings up to $\sqrt{s}=2.1$GeV of LASS
  data},'' \href{http://dx.doi.org/10.1016/j.nuclphysa.2006.06.170}{Nucl. Phys.
  {\bfseries A775} (2006) 212--223},
\href{http://arxiv.org/abs/hep-ph/0603062}{{\ttfamily arXiv:hep-ph/0603062
  [hep-ph]}}.

\bibitem{Pelaez:2020uiw}
J.~R. Peláez and A.~Rodas, ``{Determination of the lightest strange resonance
  $K_0^*(700)$ or $\kappa$, from a dispersive data analysis},''
\href{http://arxiv.org/abs/2001.08153}{{\ttfamily arXiv:2001.08153 [hep-ph]}}.

\bibitem{Ananthanarayan:2000cp}
B.~Ananthanarayan and P.~Buettiker, ``{Comparison of pion kaon scattering in
  SU(3) chiral perturbation theory and dispersion relations},''
  \href{http://dx.doi.org/10.1007/s100520100629}{Eur. Phys. J. {\bfseries C19}
  (2001) 517--528},
\href{http://arxiv.org/abs/hep-ph/0012023}{{\ttfamily arXiv:hep-ph/0012023
  [hep-ph]}}.

\bibitem{Pelaez:2016klv}
J.~R. Peláez, A.~Rodas, and J.~Ruiz~de Elvira, ``{Strange resonance poles from
  $K\pi $ scattering below 1.8 GeV},''
  \href{http://dx.doi.org/10.1140/epjc/s10052-017-4668-1}{Eur. Phys. J.
  {\bfseries C77} no.~2, (2017) 91},
\href{http://arxiv.org/abs/1612.07966}{{\ttfamily arXiv:1612.07966 [hep-ph]}}.

\bibitem{Pelaez:2016tgi}
J.~R. Pelaez and A.~Rodas, ``{Pion-kaon scattering amplitude constrained with
  forward dispersion relations up to 1.6 GeV},''
  \href{http://dx.doi.org/10.1103/PhysRevD.93.074025}{Phys. Rev. {\bfseries
  D93} no.~7, (2016) 074025},
\href{http://arxiv.org/abs/1602.08404}{{\ttfamily arXiv:1602.08404 [hep-ph]}}.

\bibitem{Bugg:2005xx}
D.~V. Bugg, ``{The Kappa in E791 data for $D \to K \pi \pi$},''
  \href{http://dx.doi.org/10.1016/j.physletb.2005.11.019}{Phys. Lett.
  {\bfseries B632} (2006) 471--474},
\href{http://arxiv.org/abs/hep-ex/0510019}{{\ttfamily arXiv:hep-ex/0510019
  [hep-ex]}}.

\bibitem{Bugg:2005ni}
D.~V. Bugg, ``{The Kappa in $J/\psi\to K^+ \pi^- K^- \pi^+$},''
  \href{http://dx.doi.org/10.1140/epja/i2005-10033-3}{Eur. Phys. J. {\bfseries
  A25} (2005) 107--114}, \href{http://arxiv.org/abs/hep-ex/0510026}{{\ttfamily
  arXiv:hep-ex/0510026 [hep-ex]}}.
[Erratum: Eur. Phys. J.A26,151(2005)].

\bibitem{Aitala:2002kr}
{\bfseries E791} Collaboration, E.~M. Aitala {\em et~al.}, ``{Dalitz Plot
  Analysis of the Decay $D^+ \to K^- \pi^+ \pi^+$ and the Study of the $K\pi$
  Scalar Amplitudes},''
  \href{http://dx.doi.org/10.1103/PhysRevLett.89.121801}{Phys. Rev. Lett.
  {\bfseries 89} (2002) 121801},
\href{http://arxiv.org/abs/hep-ex/0204018}{{\ttfamily arXiv:hep-ex/0204018
  [hep-ex]}}.

\bibitem{Bai:1994zm}
{\bfseries BES} Collaboration, J.~Z. Bai {\em et~al.}, ``{The BES detector},''
\href{http://dx.doi.org/10.1016/0168-9002(94)90081-7}{Nucl. Instrum. Meth.
  {\bfseries A344} (1994) 319--334}.

\bibitem{Bai:2001dw}
{\bfseries BES} Collaboration, J.~Z. Bai {\em et~al.}, ``{The BES upgrade},''
\href{http://dx.doi.org/10.1016/S0168-9002(00)00934-7}{Nucl. Instrum. Meth.
  {\bfseries A458} (2001) 627--637}.

\bibitem{Luscher:1990ux}
{M.~L\"uscher}, ``{Two particle states on a torus and their relation to the
  scattering matrix},''
\href{http://dx.doi.org/10.1016/0550-3213(91)90366-6}{Nucl. Phys. {\bfseries
  B354} (1991) 531--578}.

\bibitem{Rummukainen:1995vs}
K.~Rummukainen and S.~A. Gottlieb, ``{Resonance scattering phase shifts on a
  nonrest frame lattice},''
  \href{http://dx.doi.org/10.1016/0550-3213(95)00313-H}{Nucl. Phys. {\bfseries
  B450} (1995) 397--436},
\href{http://arxiv.org/abs/hep-lat/9503028}{{\ttfamily arXiv:hep-lat/9503028
  [hep-lat]}}.

\bibitem{Kim:2005gf}
C.~h. Kim, C.~T. Sachrajda, and S.~R. Sharpe, ``{Finite-volume effects for
  two-hadron states in moving frames},''
  \href{http://dx.doi.org/10.1016/j.nuclphysb.2005.08.029}{Nucl. Phys.
  {\bfseries B727} (2005) 218--243},
\href{http://arxiv.org/abs/hep-lat/0507006}{{\ttfamily arXiv:hep-lat/0507006
  [hep-lat]}}.

\bibitem{Davoudi:2011md}
Z.~Davoudi and M.~J. Savage, ``{Improving the Volume Dependence of Two-Body
  Binding Energies Calculated with Lattice QCD},''
  \href{http://dx.doi.org/10.1103/PhysRevD.84.114502}{Phys. Rev. D {\bfseries
  84} (2011) 114502}, \href{http://arxiv.org/abs/1108.5371}{{\ttfamily
  arXiv:1108.5371 [hep-lat]}}.

\bibitem{Fu:2011xz}
Z.~Fu, ``{Rummukainen-Gottlieb's formula on two-particle system with different
  mass},'' \href{http://dx.doi.org/10.1103/PhysRevD.85.014506}{Phys. Rev. D
  {\bfseries 85} (2012) 014506},
  \href{http://arxiv.org/abs/1110.0319}{{\ttfamily arXiv:1110.0319 [hep-lat]}}.

\bibitem{Hansen:2012tf}
M.~T. Hansen and S.~R. Sharpe, ``{Multiple-channel generalization of
  Lellouch-L\"uscher formula},''
  \href{http://dx.doi.org/10.1103/PhysRevD.86.016007}{Phys. Rev. {\bfseries
  D86} (2012) 016007},
\href{http://arxiv.org/abs/1204.0826}{{\ttfamily arXiv:1204.0826 [hep-lat]}}.

\bibitem{Leskovec:2012gb}
L.~Leskovec and S.~Prelovsek, ``{Scattering phase shifts for two particles of
  different mass and non-zero total momentum in lattice QCD},''
  \href{http://dx.doi.org/10.1103/PhysRevD.85.114507}{Phys. Rev. {\bfseries
  D85} (2012) 114507},
\href{http://arxiv.org/abs/1202.2145}{{\ttfamily arXiv:1202.2145 [hep-lat]}}.

\bibitem{Gockeler:2012yj}
M.~Gockeler, R.~Horsley, M.~Lage, U.~G. Meissner, P.~E.~L. Rakow, A.~Rusetsky,
  G.~Schierholz, and J.~M. Zanotti, ``{Scattering phases for meson and baryon
  resonances on general moving-frame lattices},''
  \href{http://dx.doi.org/10.1103/PhysRevD.86.094513}{Phys. Rev. {\bfseries
  D86} (2012) 094513},
\href{http://arxiv.org/abs/1206.4141}{{\ttfamily arXiv:1206.4141 [hep-lat]}}.

\bibitem{Briceno:2014oea}
R.~A. Briceno, ``{Two-particle multichannel systems in a finite volume with
  arbitrary spin},'' \href{http://dx.doi.org/10.1103/PhysRevD.89.074507}{Phys.
  Rev. {\bfseries D89} no.~7, (2014) 074507},
\href{http://arxiv.org/abs/1401.3312}{{\ttfamily arXiv:1401.3312 [hep-lat]}}.

\bibitem{Briceno:2017max}
R.~A. Briceno, J.~J. Dudek, and R.~D. Young, ``{Scattering processes and
  resonances from lattice QCD},''
  \href{http://dx.doi.org/10.1103/RevModPhys.90.025001}{Rev. Mod. Phys.
  {\bfseries 90} no.~2, (2018) 025001},
\href{http://arxiv.org/abs/1706.06223}{{\ttfamily arXiv:1706.06223 [hep-lat]}}.

\bibitem{Miao:2004gy}
C.~Miao, X.-i. Du, G.-w. Meng, and C.~Liu, ``{Lattice study on kaon pion
  scattering length in the I = 3/2 channel},''
  \href{http://dx.doi.org/10.1016/j.physletb.2004.05.073}{Phys. Lett.
  {\bfseries B595} (2004) 400--407},
\href{http://arxiv.org/abs/hep-lat/0403028}{{\ttfamily arXiv:hep-lat/0403028
  [hep-lat]}}.

\bibitem{Beane:2006gj}
S.~R. Beane, P.~F. Bedaque, T.~C. Luu, K.~Orginos, E.~Pallante, A.~Parreno, and
  M.~J. Savage, ``{$\pi K$ scattering in full QCD with domain-wall valence
  quarks},'' \href{http://dx.doi.org/10.1103/PhysRevD.74.114503}{Phys. Rev.
  {\bfseries D74} (2006) 114503},
\href{http://arxiv.org/abs/hep-lat/0607036}{{\ttfamily arXiv:hep-lat/0607036
  [hep-lat]}}.

\bibitem{Flynn:2007ki}
J.~M. Flynn and J.~Nieves, ``{Elastic $s$-wave $B \pi$, $D \pi$, $D K$ and $K
  \pi$ scattering from lattice calculations of scalar form-factors in
  semileptonic decays},''
  \href{http://dx.doi.org/10.1103/PhysRevD.75.074024}{Phys. Rev. {\bfseries
  D75} (2007) 074024},
\href{http://arxiv.org/abs/hep-ph/0703047}{{\ttfamily arXiv:hep-ph/0703047
  [hep-ph]}}.

\bibitem{Nagata:2008wk}
J.~Nagata, S.~Muroya, and A.~Nakamura, ``{Lattice study of $K \pi$ scattering
  in $I = 3/2$ and $1/2$},''
  \href{http://dx.doi.org/10.1103/PhysRevC.84.019904,
  10.1103/PhysRevC.80.045203}{Phys. Rev. {\bfseries C80} (2009) 045203},
  \href{http://arxiv.org/abs/0812.1753}{{\ttfamily arXiv:0812.1753 [hep-lat]}}.
[Erratum: Phys. Rev.C84,019904(2011)].

\bibitem{Fu:2011xb}
Z.-W. Fu, ``{Lattice calculation of $\kappa$ meson},''
  \href{http://dx.doi.org/10.1088/1674-1137/36/6/003}{Chin. Phys. {\bfseries
  C36} (2012) 489--497},
\href{http://arxiv.org/abs/1111.1835}{{\ttfamily arXiv:1111.1835 [hep-lat]}}.

\bibitem{Fu:2011xw}
Z.~Fu, ``{The preliminary lattice QCD calculation of $\kappa$ meson decay
  width},'' \href{http://dx.doi.org/10.1007/JHEP01(2012)017}{JHEP {\bfseries
  01} (2012) 017},
\href{http://arxiv.org/abs/1110.5975}{{\ttfamily arXiv:1110.5975 [hep-lat]}}.

\bibitem{Fu:2011wc}
Z.~Fu, ``{Lattice study on $\pi K $ scattering with moving wall source},''
  \href{http://dx.doi.org/10.1103/PhysRevD.85.074501}{Phys. Rev. {\bfseries
  D85} (2012) 074501},
\href{http://arxiv.org/abs/1110.1422}{{\ttfamily arXiv:1110.1422 [hep-lat]}}.

\bibitem{Fu:2013sua}
Z.~Fu, ``{Studying $\kappa$ meson with a MILC fine lattice},''
  \href{http://dx.doi.org/10.1142/S0217751X13500590}{Int. J. Mod. Phys.
  {\bfseries A28} (2013) 1350059},
\href{http://arxiv.org/abs/1305.4458}{{\ttfamily arXiv:1305.4458 [hep-lat]}}.

\bibitem{Lang:2012sv}
C.~B. Lang, L.~Leskovec, D.~Mohler, and S.~Prelovsek, ``{$K \pi$ scattering for
  isospin 1/2 and 3/2 in lattice QCD},''
  \href{http://dx.doi.org/10.1103/PhysRevD.86.054508}{Phys. Rev. {\bfseries
  D86} (2012) 054508},
\href{http://arxiv.org/abs/1207.3204}{{\ttfamily arXiv:1207.3204 [hep-lat]}}.

\bibitem{Sasaki:2013vxa}
{\bfseries PACS-CS} Collaboration, K.~Sasaki, N.~Ishizuka, M.~Oka, and
  T.~Yamazaki, ``{Scattering lengths for two pseudoscalar meson systems},''
  \href{http://dx.doi.org/10.1103/PhysRevD.89.054502}{Phys. Rev. {\bfseries
  D89} no.~5, (2014) 054502},
\href{http://arxiv.org/abs/1311.7226}{{\ttfamily arXiv:1311.7226 [hep-lat]}}.

\bibitem{Helmes:2018nug}
{\bfseries ETM} Collaboration, C.~Helmes, C.~Jost, B.~Knippschild,
  B.~Kostrzewa, L.~Liu, F.~Pittler, C.~Urbach, and M.~Werner, ``{Hadron-Hadron
  Interactions from $N_f=2+1+1$ Lattice QCD: $I=3/2$ $\pi K$ Scattering
  Length},'' \href{http://dx.doi.org/10.1103/PhysRevD.98.114511}{Phys. Rev.
  {\bfseries D98} no.~11, (2018) 114511},
\href{http://arxiv.org/abs/1809.08886}{{\ttfamily arXiv:1809.08886 [hep-lat]}}.

\bibitem{Alexandrou:2012rm}
C.~Alexandrou, J.~O. Daldrop, M.~Dalla~Brida, M.~Gravina, L.~Scorzato,
  C.~Urbach, and M.~Wagner, ``{Lattice investigation of the scalar mesons
  $a_0(980)$ and $\kappa$ using four-quark operators},''
  \href{http://dx.doi.org/10.1007/JHEP04(2013)137}{JHEP {\bfseries 04} (2013)
  137},
\href{http://arxiv.org/abs/1212.1418}{{\ttfamily arXiv:1212.1418 [hep-lat]}}.

\bibitem{Prelovsek:2010kg}
S.~Prelovsek, T.~Draper, C.~B. Lang, M.~Limmer, K.-F. Liu, N.~Mathur, and
  D.~Mohler, ``{Lattice study of light scalar tetraquarks with I=0,2,1/2,3/2:
  Are $\sigma$ and $\kappa$ tetraquarks?},''
  \href{http://dx.doi.org/10.1103/PhysRevD.82.094507}{Phys. Rev. {\bfseries
  D82} (2010) 094507},
\href{http://arxiv.org/abs/1005.0948}{{\ttfamily arXiv:1005.0948 [hep-lat]}}.

\bibitem{Guo:2013nja}
F.-K. Guo, L.~Liu, U.-G. Meissner, and P.~Wang, ``{Tetraquarks, hadronic
  molecules, meson-meson scattering and disconnected contributions in lattice
  QCD},'' \href{http://dx.doi.org/10.1103/PhysRevD.88.074506}{Phys. Rev.
  {\bfseries D88} (2013) 074506},
\href{http://arxiv.org/abs/1308.2545}{{\ttfamily arXiv:1308.2545 [hep-lat]}}.

\bibitem{Doring:2011nd}
M.~Doring and U.~G. Meissner, ``{Finite volume effects in pion-kaon scattering
  and reconstruction of the $\kappa$(800) resonance},''
  \href{http://dx.doi.org/10.1007/JHEP01(2012)009}{JHEP {\bfseries 01} (2012)
  009},
\href{http://arxiv.org/abs/1111.0616}{{\ttfamily arXiv:1111.0616 [hep-lat]}}.

\bibitem{Bernard:2010fp}
V.~Bernard, M.~Lage, U.~G. Meissner, and A.~Rusetsky, ``{Scalar mesons in a
  finite volume},'' \href{http://dx.doi.org/10.1007/JHEP01(2011)019}{JHEP
  {\bfseries 01} (2011) 019},
\href{http://arxiv.org/abs/1010.6018}{{\ttfamily arXiv:1010.6018 [hep-lat]}}.

\bibitem{Doring:2012eu}
M.~Doring, U.~G. Meissner, E.~Oset, and A.~Rusetsky, ``{Scalar mesons moving in
  a finite volume and the role of partial wave mixing},''
  \href{http://dx.doi.org/10.1140/epja/i2012-12114-6}{Eur. Phys. J. {\bfseries
  A48} (2012) 114},
\href{http://arxiv.org/abs/1205.4838}{{\ttfamily arXiv:1205.4838 [hep-lat]}}.

\bibitem{Xiao:2012vv}
C.~W. Xiao, F.~Aceti, and M.~Bayar, ``{The small $K \pi$ component in the $K^*$
  wave functions},'' \href{http://dx.doi.org/10.1140/epja/i2013-13022-y}{Eur.
  Phys. J. {\bfseries A49} (2013) 22},
\href{http://arxiv.org/abs/1210.7176}{{\ttfamily arXiv:1210.7176 [hep-ph]}}.

\bibitem{Doring:2013wka}
M.~Döring, U.-G. Meißner, and W.~Wang, ``{Chiral Dynamics and S-wave
  Contributions in Semileptonic B decays},''
  \href{http://dx.doi.org/10.1007/JHEP10(2013)011}{JHEP {\bfseries 10} (2013)
  011},
\href{http://arxiv.org/abs/1307.0947}{{\ttfamily arXiv:1307.0947 [hep-ph]}}.

\bibitem{Zhou:2014ana}
D.~Zhou, E.-L. Cui, H.-X. Chen, L.-S. Geng, and L.-H. Zhu, ``{$K\pi$
  interaction in finite volume and the $K^*$ resonance},''
  \href{http://dx.doi.org/10.1103/PhysRevD.91.094505}{Phys. Rev. {\bfseries
  D91} no.~9, (2015) 094505},
\href{http://arxiv.org/abs/1409.0178}{{\ttfamily arXiv:1409.0178 [hep-lat]}}.

\bibitem{Fu:2012tj}
Z.~Fu and K.~Fu, ``{Lattice QCD study on $K^\ast(892)$ meson decay width},''
  \href{http://dx.doi.org/10.1103/PhysRevD.86.094507}{Phys. Rev. {\bfseries
  D86} (2012) 094507},
\href{http://arxiv.org/abs/1209.0350}{{\ttfamily arXiv:1209.0350 [hep-lat]}}.

\bibitem{Prelovsek:2013ela}
S.~Prelovsek, L.~Leskovec, C.~B. Lang, and D.~Mohler, ``{$K$ $\pi$ Scattering
  and the $K^*$ Decay width from Lattice QCD},''
  \href{http://dx.doi.org/10.1103/PhysRevD.88.054508}{Phys. Rev. {\bfseries
  D88} no.~5, (2013) 054508},
\href{http://arxiv.org/abs/1307.0736}{{\ttfamily arXiv:1307.0736 [hep-lat]}}.

\bibitem{Bali:2015gji}
{\bfseries RQCD} Collaboration, G.~S. Bali, S.~Collins, A.~Cox, G.~Donald,
  M.~Göckeler, C.~Lang, and A.~Schäfer, ``{$\rho$ and $K^*$ resonances on the
  lattice at nearly physical quark masses and $N_f=2$},''
  \href{http://dx.doi.org/10.1103/PhysRevD.93.054509}{Phys. Rev. D {\bfseries
  93} no.~5, (2016) 054509}, \href{http://arxiv.org/abs/1512.08678}{{\ttfamily
  arXiv:1512.08678 [hep-lat]}}.

\bibitem{Wilson:2014cna}
D.~J. Wilson, J.~J. Dudek, R.~G. Edwards, and C.~E. Thomas, ``{Resonances in
  coupled $\pi K, \eta K$ scattering from lattice QCD},''
  \href{http://dx.doi.org/10.1103/PhysRevD.91.054008}{Phys. Rev. {\bfseries
  D91} no.~5, (2015) 054008},
\href{http://arxiv.org/abs/1411.2004}{{\ttfamily arXiv:1411.2004 [hep-ph]}}.

\bibitem{Dudek:2014qha}
{\bfseries Hadron Spectrum} Collaboration, J.~J. Dudek, R.~G. Edwards, C.~E.
  Thomas, and D.~J. Wilson, ``{Resonances in coupled $\pi K -\eta K$ scattering
  from quantum chromodynamics},''
  \href{http://dx.doi.org/10.1103/PhysRevLett.113.182001}{Phys. Rev. Lett.
  {\bfseries 113} no.~18, (2014) 182001},
\href{http://arxiv.org/abs/1406.4158}{{\ttfamily arXiv:1406.4158 [hep-ph]}}.

\bibitem{Brett:2018jqw}
R.~Brett, J.~Bulava, J.~Fallica, A.~Hanlon, B.~Hörz, and C.~Morningstar,
  ``{Determination of $s$- and $p$-wave $I=1/2$ $K\pi$ scattering amplitudes in
  $N_{\mathrm{f}}=2+1$ lattice QCD},''
  \href{http://dx.doi.org/10.1016/j.nuclphysb.2018.05.008}{Nucl. Phys.
  {\bfseries B932} (2018) 29--51},
\href{http://arxiv.org/abs/1802.03100}{{\ttfamily arXiv:1802.03100 [hep-lat]}}.

\bibitem{Wilson:2019wfr}
D.~J. Wilson, R.~A. Briceno, J.~J. Dudek, R.~G. Edwards, and C.~E. Thomas,
  ``{The quark-mass dependence of elastic $\pi K$ scattering from QCD},''
  \href{http://dx.doi.org/10.1103/PhysRevLett.123.042002}{Phys. Rev. Lett.
  {\bfseries 123} no.~4, (2019) 042002},
\href{http://arxiv.org/abs/1904.03188}{{\ttfamily arXiv:1904.03188 [hep-lat]}}.

\bibitem{Briceno:2014uqa}
R.~A. Briceño, M.~T. Hansen, and A.~Walker-Loud, ``{Multichannel 1
  $\rightarrow$ 2 transition amplitudes in a finite volume},''
  \href{http://dx.doi.org/10.1103/PhysRevD.91.034501}{Phys. Rev. D {\bfseries
  91} no.~3, (2015) 034501}, \href{http://arxiv.org/abs/1406.5965}{{\ttfamily
  arXiv:1406.5965 [hep-lat]}}.

\bibitem{Chung:1995dx}
S.~U. Chung, J.~Brose, R.~Hackmann, E.~Klempt, S.~Spanier, and C.~Strassburger,
  ``{Partial wave analysis in K matrix formalism},''
\href{http://dx.doi.org/10.1002/andp.19955070504}{Annalen Phys. {\bfseries 4}
  (1995) 404--430}.

\bibitem{Landau:1991wop}
L.~D. Landau and E.~M. Lifshits, {\em {Quantum Mechanics}}, vol.~v.3 of {\em
  Course of Theoretical Physics}.
\newblock Butterworth-Heinemann, Oxford,
1991.
\newblock

\bibitem{Krane:1987ky}
K.~Krane, {\em {INTRODUCTORY NUCLEAR PHYSICS}}.
\newblock 1987.

\bibitem{Bugg:2003kj}
D.~V. Bugg, ``{Comments on the sigma and kappa},''
  \href{http://dx.doi.org/10.1016/j.physletb.2004.06.050,
  10.1016/j.physletb.2003.07.078}{Phys. Lett. {\bfseries B572} (2003) 1--7}.
[Erratum: Phys. Lett.B595,556(2004)].

\bibitem{Bugg:2009uk}
D.~V. Bugg, ``{An Update on the Kappa},''
  \href{http://dx.doi.org/10.1103/PhysRevD.81.014002}{Phys. Rev. {\bfseries
  D81} (2010) 014002},
\href{http://arxiv.org/abs/0906.3992}{{\ttfamily arXiv:0906.3992 [hep-ph]}}.

\bibitem{Adler:1965ga}
S.~L. Adler, ``{Consistency conditions on the strong interactions implied by a
  partially conserved axial-vector current. II},''
  \href{http://dx.doi.org/10.1103/PhysRev.139.B1638}{Phys. Rev. {\bfseries 139}
  (1965) B1638--B1643}.
[,152(1965)].

\bibitem{Adler:1964um}
S.~L. Adler, ``{Consistency conditions on the strong interactions implied by a
  partially conserved axial vector current},''
  \href{http://dx.doi.org/10.1103/PhysRev.137.B1022}{Phys. Rev. {\bfseries 137}
  (1965) B1022--B1033}.
[,140(1964)].

\bibitem{Bessler:1974cb}
L.~Bessler and P.~T. Davies, ``{Necessary and sufficient conditions for the
  Adler zero},''
\href{http://dx.doi.org/10.1103/PhysRevD.9.2923}{Phys. Rev. {\bfseries D9}
  (1974) 2923--2925}.

\bibitem{VonHippel:1972fg}
F.~Von~Hippel and C.~Quigg, ``{Centrifugal-barrier effects in resonance partial
  decay widths, shapes, and production amplitudes},''
\href{http://dx.doi.org/10.1103/PhysRevD.5.624}{Phys. Rev. {\bfseries D5}
  (1972) 624--638}.

\bibitem{Link:2009ng}
{\bfseries FOCUS} Collaboration, J.~M. Link {\em et~al.}, ``{The $K^-\pi^+$
  S-wave from the $D^+ \to K^-\pi^+\pi^+$ decay},''
  \href{http://dx.doi.org/10.1016/j.physletb.2009.09.057}{Phys. Lett.
  {\bfseries B681} (2009) 14--21},
\href{http://arxiv.org/abs/0905.4846}{{\ttfamily arXiv:0905.4846 [hep-ex]}}.

\bibitem{Aitala:2005yh}
{\bfseries E791} Collaboration, E.~M. Aitala {\em et~al.}, ``{Model independent
  measurement of S-wave $K^-\pi^+$ systems using $D^+ \to K \pi \pi$ decays
  from Fermilab E791},'' \href{http://dx.doi.org/10.1103/PhysRevD.73.032004,
  10.1103/PhysRevD.74.059901}{Phys. Rev. {\bfseries D73} (2006) 032004},
  \href{http://arxiv.org/abs/hep-ex/0507099}{{\ttfamily arXiv:hep-ex/0507099
  [hep-ex]}}.
[Erratum: Phys. Rev.D74,059901(2006)].

\bibitem{Symanzik:1983pq}
K.~Symanzik, ``{Improved lattice actions for nonlinear sigma model and
  nonabelian gauge theory},'' in {\em {Workshop on Non-perturbative Field
  Theory and QCD Trieste, Italy, December 17-21, 1982}}, pp.~61--72.
\newblock 1983.
\newblock
[,61(1983)].

\bibitem{Symanzik:1983dc}
K.~Symanzik, ``{Continuum Limit and Improved Action in Lattice Theories. 1.
  Principles and $\phi^4$ Theory},''
\href{http://dx.doi.org/10.1016/0550-3213(83)90468-6}{Nucl. Phys. {\bfseries
  B226} (1983) 187--204}.

\bibitem{Symanzik:1983gh}
K.~Symanzik, ``{Continuum Limit and Improved Action in Lattice Theories. 2.
  O(N) Nonlinear Sigma Model in Perturbation Theory},''
\href{http://dx.doi.org/10.1016/0550-3213(83)90469-8}{Nucl. Phys. {\bfseries
  B226} (1983) 205--227}.

\bibitem{Luscher:1985zq}
{M.~L\"uscher and P.~Weisz}, ``{Computation of the Action for On-Shell Improved
  Lattice Gauge Theories at Weak Coupling},''
\href{http://dx.doi.org/10.1016/0370-2693(85)90966-9}{Phys. Lett. {\bfseries
  158B} (1985) 250--254}.

\bibitem{Wilson:1974sk}
K.~G. Wilson, ``{Confinement of Quarks},''
  \href{http://dx.doi.org/10.1103/PhysRevD.10.2445}{Phys. Rev. {\bfseries D10}
  (1974) 2445--2459}.
[,319(1974)].

\bibitem{Sheikholeslami:1985ij}
B.~Sheikholeslami and R.~Wohlert, ``{Improved Continuum Limit Lattice Action
  for QCD with Wilson Fermions},''
\href{http://dx.doi.org/10.1016/0550-3213(85)90002-1}{Nucl. Phys. {\bfseries
  B259} (1985) 572}.

\bibitem{Morningstar:2003gk}
C.~Morningstar and M.~J. Peardon, ``{Analytic smearing of SU(3) link variables
  in lattice QCD},'' \href{http://dx.doi.org/10.1103/PhysRevD.69.054501}{Phys.
  Rev. {\bfseries D69} (2004) 054501},
\href{http://arxiv.org/abs/hep-lat/0311018}{{\ttfamily arXiv:hep-lat/0311018
  [hep-lat]}}.

\bibitem{Davies:2009tsa}
{\bfseries HPQCD} Collaboration, C.~T.~H. Davies, E.~Follana, I.~D. Kendall,
  G.~P. Lepage, and C.~McNeile, ``{Precise determination of the lattice spacing
  in full lattice QCD},''
  \href{http://dx.doi.org/10.1103/PhysRevD.81.034506}{Phys. Rev. {\bfseries
  D81} (2010) 034506},
\href{http://arxiv.org/abs/0910.1229}{{\ttfamily arXiv:0910.1229 [hep-lat]}}.

\bibitem{Meinel:2010pv}
S.~Meinel, ``{Bottomonium spectrum at order $v^6$ from domain-wall lattice QCD:
  Precise results for hyperfine splittings},''
  \href{http://dx.doi.org/10.1103/PhysRevD.82.114502}{Phys. Rev. {\bfseries
  D82} (2010) 114502},
\href{http://arxiv.org/abs/1007.3966}{{\ttfamily arXiv:1007.3966 [hep-lat]}}.

\bibitem{Lepage:1992tx}
G.~P. Lepage, L.~Magnea, C.~Nakhleh, U.~Magnea, and K.~Hornbostel, ``{Improved
  nonrelativistic QCD for heavy quark physics},''
  \href{http://dx.doi.org/10.1103/PhysRevD.46.4052}{Phys. Rev. {\bfseries D46}
  (1992) 4052--4067},
\href{http://arxiv.org/abs/hep-lat/9205007}{{\ttfamily arXiv:hep-lat/9205007
  [hep-lat]}}.

\bibitem{Gusken:1989ad}
S.~Güsken, U.~Low, K.~H. Mutter, R.~Sommer, A.~Patel, and K.~Schilling,
  ``{Nonsinglet Axial Vector Couplings of the Baryon Octet in Lattice {QCD}},''
\href{http://dx.doi.org/10.1016/S0370-2693(89)80034-6}{Phys. Lett. {\bfseries
  B227} (1989) 266--269}.

\bibitem{Albanese:1987ds}
{\bfseries APE} Collaboration, M.~Albanese {\em et~al.}, ``{Glueball Masses and
  String Tension in Lattice QCD},''
\href{http://dx.doi.org/10.1016/0370-2693(87)91160-9}{Phys. Lett. {\bfseries
  B192} (1987) 163--169}.

\bibitem{Dresselhaus:2008}
{Mildred Dresselhaus, Gene Dresselhaus, Ado Jorio}, {\em {Group Theory:
  Application to the Physics of Condensed Matter}}.
\newblock {Springer-Verlag Berlin Heidelberg}, 1~ed., 2008.

\bibitem{Alexandrou:2017mpi}
C.~Alexandrou, L.~Leskovec, S.~Meinel, J.~Negele, S.~Paul, M.~Petschlies,
  A.~Pochinsky, G.~Rendon, and S.~Syritsyn, ``{$P$-wave $\pi\pi$ scattering and
  the $\rho$ resonance from lattice QCD},''
  \href{http://dx.doi.org/10.1103/PhysRevD.96.034525}{Phys. Rev. {\bfseries
  D96} no.~3, (2017) 034525},
\href{http://arxiv.org/abs/1704.05439}{{\ttfamily arXiv:1704.05439 [hep-lat]}}.

\bibitem{McNeile:2006bz}
{\bfseries UKQCD} Collaboration, C.~McNeile and C.~Michael, ``{Decay width of
  light quark hybrid meson from the lattice},''
  \href{http://dx.doi.org/10.1103/PhysRevD.73.074506}{Phys. Rev. {\bfseries
  D73} (2006) 074506},
\href{http://arxiv.org/abs/hep-lat/0603007}{{\ttfamily arXiv:hep-lat/0603007
  [hep-lat]}}.

\bibitem{Luscher:1990ck}
{M.~L\"uscher and U.~Wolff}, ``{How to Calculate the Elastic Scattering Matrix
  in Two-dimensional Quantum Field Theories by Numerical Simulation},''
\href{http://dx.doi.org/10.1016/0550-3213(90)90540-T}{Nucl. Phys. {\bfseries
  B339} (1990) 222--252}.

\bibitem{Blossier:2009kd}
B.~Blossier, M.~Della~Morte, G.~von Hippel, T.~Mendes, and R.~Sommer, ``{On the
  generalized eigenvalue method for energies and matrix elements in lattice
  field theory},'' \href{http://dx.doi.org/10.1088/1126-6708/2009/04/094}{JHEP
  {\bfseries 04} (2009) 094},
\href{http://arxiv.org/abs/0902.1265}{{\ttfamily arXiv:0902.1265 [hep-lat]}}.

\bibitem{Guo:2012hv}
P.~Guo, J.~Dudek, R.~Edwards, and A.~P. Szczepaniak, ``{Coupled-channel
  scattering on a torus},''
  \href{http://dx.doi.org/10.1103/PhysRevD.88.014501}{Phys. Rev. {\bfseries
  D88} no.~1, (2013) 014501},
\href{http://arxiv.org/abs/1211.0929}{{\ttfamily arXiv:1211.0929 [hep-lat]}}.

\bibitem{Briceno:2015dca}
R.~A. Briceno, J.~J. Dudek, R.~G. Edwards, C.~J. Shultz, C.~E. Thomas, and
  D.~J. Wilson, ``{The resonant $\pi^+\gamma\to\pi^+\pi^0$ amplitude from
  Quantum Chromodynamics},''
  \href{http://dx.doi.org/10.1103/PhysRevLett.115.242001}{Phys. Rev. Lett.
  {\bfseries 115} (2015) 242001},
  \href{http://arxiv.org/abs/1507.06622}{{\ttfamily arXiv:1507.06622
  [hep-ph]}}.

\bibitem{Alexandrou:2018jbt}
C.~Alexandrou, L.~Leskovec, S.~Meinel, J.~Negele, S.~Paul, M.~Petschlies,
  A.~Pochinsky, G.~Rendon, and S.~Syritsyn, ``{$\pi\gamma \to \pi\pi$
  transition and the $\rho$ radiative decay width from lattice QCD},''
  \href{http://dx.doi.org/10.1103/PhysRevD.98.074502}{Phys. Rev. {\bfseries
  D98} no.~7, (2018) 074502},
\href{http://arxiv.org/abs/1807.08357}{{\ttfamily arXiv:1807.08357 [hep-lat]}}.

\bibitem{Horgan:2013hoa}
R.~R. Horgan, Z.~Liu, S.~Meinel, and M.~Wingate, ``{Lattice QCD calculation of
  form factors describing the rare decays $B \to K^* \ell^+ \ell^-$ and $B_s
  \to \phi \ell^+ \ell^-$},''
  \href{http://dx.doi.org/10.1103/PhysRevD.89.094501}{Phys. Rev. {\bfseries
  D89} no.~9, (2014) 094501},
\href{http://arxiv.org/abs/1310.3722}{{\ttfamily arXiv:1310.3722 [hep-lat]}}.

\bibitem{XSEDE}
J.~{Towns}, T.~{Cockerill}, M.~{Dahan}, I.~{Foster}, K.~{Gaither},
  A.~{Grimshaw}, V.~{Hazlewood}, S.~{Lathrop}, D.~{Lifka}, G.~D. {Peterson},
  R.~{Roskies}, J.~R. {Scott}, and N.~{Wilkins-Diehr}, ``{XSEDE: Accelerating
  Scientific Discovery},'' Computing in Science Engineering {\bfseries 16}
  no.~5, (2014) 62--74. 

\end{thebibliography}
\providecommand{\href}[2]{#2}\begingroup\raggedright\endgroup

\end{document}